\documentclass[prl,twocolumn,aps,epsfig,subfig,prbbib]{revtex4}
\usepackage{graphicx,float}
\usepackage{hyperref}
\begin{document}

\title{Electronic properties of layered multicomponent wide-bandgap oxides: \\ a combinatorial approach}
\author{Altynbek Murat}
\author{Julia E. Medvedeva}\email{juliaem@mst.edu}

\affiliation{Department of Physics, Missouri University 
of Science \& Technology, Rolla, MO 65409, USA}

\begin{abstract}
The structural, electronic, and optical properties of twelve multicomponent oxides with layered structure, RAMO$_4$, 
where R$^{3+}$=In or Sc; A$^{3+}$=Al or Ga; and M$^{2+}$=Ca, Cd, Mg, or Zn, are investigated using first-principles density functional approach.
The compositional complexity of RAMO$_4$ leads to a wide range of band gap values varying from 2.45 eV for InGaCdO$_4$ to 6.29 eV for ScAlMgO$_4$.
Strikingly, despite the different band gaps in the oxide constituents, namely, 2-4 eV in CdO, In$_2$O$_3$, or ZnO; 5-6 for Ga$_2$O$_3$ or Sc$_2$O$_3$; and 7-9 eV in CaO, MgO, or Al$_2$O$_3$, the bottom of the conduction band in the multicomponent oxides is formed from the $s$-states of {\it all} cations and their neighboring oxygen $p$-states. We show that the hybrid nature of the conduction band in multicomponent oxides originates from the unusual five-fold atomic coordination of A$^{3+}$ and M$^{2+}$ cations which enables the interaction between the spatially-spread $s$-orbitals of adjacent cations via shared oxygen atoms. The effect of the local atomic coordination on the band gap,
the electron effective mass, the orbital composition of the conduction band, and the expected (an)isotropic character of the electron transport in layered RAMO$_4$ is thoroughly discussed.
\end{abstract}

\maketitle

\subsection{I. Introduction}
Transparent conducting oxides (TCOs) are unique materials that exhibit both low optical absorption in the visible region and nearly metallic electrical conductivity. 
Serving as a contact and a window layer simultaneously, TCOs are a vital part of many optoelectronic devices including solar cells, smart windows
and flat panel displays, and they also find application as heating, antistatic and optical coatings, for select reviews see Refs. 
\cite{Chopra,Thomas,MRS,TCO-optoelectr,Edwards,FMbook,TCOhandbook}.

Multicomponent TCOs -- complex oxides which contain a combination of post-transition metals, In, Zn, Ga, Cd or Sn, as well as light main-group metals such as Al or Mg -- have attracted wide attention due to a possibility to manipulate the optical, electronic, and thermal  properties via the chemical composition and, thus, to significantly broaden the application range of TCO materials \cite{Shannon,Chopra,Dawar,MRS,Unno,Phillips,spinel-review,Freeman,Mason-review,FMbook,TCOhandbook,Walsh}. 
To optimize the properties of a multicomponent TCOs, it is critical to understand the role played by each constituent oxide. 
For example, presence of lighter metals such as Ga, Al or Mg in multicomponent TCOs is attractive for achieving a broader optical transmission window associated with a wider band gap. At the same time, however, these cations are know to be detrimental for the electrical properties as they are believed to significantly suppress carrier concentration and transport.

Recent electronic band structure investigations of several main group metal oxides \cite{JMbook} reveal that the electronic configuration of the cations plays crucial role in the charge transport. It was shown that lighter metal cations (Ga, Ca, Al or Mg) have their empty $p$- or $d$-states near the conduction band bottom. The resulting strong (directional) hybridization of these anisotropic states with the $p$-states of the neighbor oxygen atoms result in significant charge localization (trapping) when extra electrons are introduced. This is in marked contrast to the conventional TCOs, In$_2$O$_3$, ZnO, SnO$_2$, or CdO, where the cation's $p$-states are deep in the conduction band (at about a few eV above the conduction band minimum \cite{JMbook}), and an extra charge is efficiently transported via a uniform network of the spatially-spread and spherically-symmetric metal $s$ orbitals connected by the oxygen $p$ states. 

In a multicomponent oxide containing the cations from both groups -- i.e., post-transition metals and light main-group metals -- the respective energy locations of the cations' states may not be the same as in single-oxide constituents due to the interaction between different cations via a shared oxygen neighbor.
Indeed, it was found \cite{JMprb} that the bottom of the conduction band in InGaZnO$_4$ is governed by the states of {\it all} cations despite the fact that the band gaps in the corresponding basis oxides differ significantly (2.9 eV for In$_2$O$_3$, 3.4 eV for ZnO, and 4.9 eV for Ga$_2$O$_3$).
Moreover, the electronic properties in a multicomponent oxide may significantly deviate from that expected from the electronic band structures of the single-cation (basis) oxides. This stems from the differences in the interatomic distances and the atomic coordination numbers in the complex oxide as compared to those in the bulk ground-state (lowest energy) structures of the constituent oxides.  

In this work, we systematically investigate the structural, electronic and optical properties of twelve RAMO$_4$ compounds with R$^{3+}$=In or Sc; A$^{3+}$=Al or Ga; and M$^{2+}$=Ca, Cd, Mg, or Zn. These materials possess the same layered crystal structure as the member of the homologous series InGaO$_3$(ZnO)$_m$ \cite{str,str-all} with $m$=1, where the chemically and structurally distinct layers -- the octahedrally coordinated RO$_{1.5}$ layer and wurtzite-like AMO$_{2.5}$ double-layer -- alternate along the crystallographic $z$ direction.
By comparing the calculated electronic properties of the set of multicomponent oxides, we determine 
how the composition affects (i) the nature of the conduction band bottom; (ii) the electron effective masses in the $ab$ plane (within the layers) and along the $z$ direction (across the layers); and (iii) the location of the cation(s) $p$-states with respect to the conduction band minimum.
For accurate determination of the electronic band structure of multicomponent oxides, we employ self-consistent screened-exchange LDA (sX-LDA) method which models the exchange-correlation hole within a {\it nonlocal} density scheme \cite{sxlda}.

The paper is organized as follows. 
First, details of the computational methods and approaches are given in Sec. II. In Sec. III, we discuss the  structural peculiarities of the investigated multicomponent compounds and compare them to the structural properties of the basis single-cation oxides. Specifically, we compare the cation-anion distances and the atomic coordination numbers in multicomponent and single-cation oxides in various structures. Further, the electronic properties of the basis, single-cation oxides are discussed in Sec. IV. We demonstrate how the electronic properties of the oxides, e.g., band gaps and the electron effective masses, vary upon changes in the interatomic distances and/or oxygen coordination by considering both the ground state and hypothetical structures of oxides. In Sec. V, the general electronic properties of multicomponent oxides are discussed first. Further, we thoroughly analyze the following: (A) how the atomic coordination affects the band gap formation in complex oxides; (B) what is the effect of chemical composition on (an)isotropy of conduction states in RAMO$_4$; (C) what is the orbital composition of the conduction band in RAMO$_4$ and the role he peculiar atomic coordination played in the respective energy location of cation's empty $s$, $p$, and $d$ orbitals in the conduction band; and (D) the electron effective masses within and accross the structural layers of different composition in RAMO$_4$. 
We give conclusions in Sec. VI.

\subsection{II. Methods and Approximations}
First-principles full-potential linearized augmented plane wave method (FLAPW) \cite{FLAPW,FLAPW1}
within the local density approximation is employed for the electronic band structure
investigations of twelve RAMO$_4$ compounds, R$^{3+}$=In or Sc, A$^{3+}$=Al, Ga, M$^{2+}$=Ca, Cd, Mg, and/or Zn,  \cite{kimizukaRAM} as well as their single-cation constituents,  MgO, CaO, ZnO, CdO, Sc$_2$O$_3$, In$_2$O$_3$, Al$_2$O$_3$, and Ga$_2$O$_3$. 
Cutoffs for the basis functions, 16.0 Ry, and the potential representation, 81.0 Ry,
and expansion in terms of spherical harmonics with $\ell \le$ 8 inside the muffin-tin spheres
were used. The muffin-tin radii of multi-cation and single-cation oxides are as follows: 2.3 to 2.6 a.u. for In, Sc, Cd, and Ca; 1.7 to 2.1 a.u.
for Ga, Mg, Zn, and Al; and 1.6 to 1.8 a.u. for O atoms.
Summations over the Brillouin zone were carried out using at least 23 special {\bf k} points in the irreducible wedge.

Because the local density approximation (LDA) underestimates the oxide band gaps and may give incorrect energy location of the states of different cations in the conduction band of multicomponent materials, 
we also employed the self-consistent screened-exchange LDA (sX-LDA) method \cite{sxlda,sxLDA1,sx-Asahi,sx-paper,sx-Kim} for more accurate
description of the band gap values and the valence/conduction band states. For the sX-LDA calculations, cutoff for the plane wave basis was 10.2 Ry and summations over the Brillouin zone were carried out using at least 14 special {\bf k} points in the irreducible wedge.
Ga and Zn 3d$^{10}$ states, which were treated as valence, were excluded from screening.


\subsection{III. Crystal Structure}

\begin{table*}
\centering
\caption{Lattice constants $a$ and $c$, in \AA; the range for the fractional z coordinates of A$^{3+}$=Al or Ga, and M$^{2+}$=Zn, Cd, Ca, or Mg atoms at the 6(c) positions of rhombohedral YbFe$_2$O$_4$ structure; and the average optimized cation-anion distances $\langle$D$_{R-O}\rangle$, the average planar $\langle$D$^{ab}_{A/M-O}\rangle$, nearest apical D$^c_{A/M-O}$, and next nearest apical distances D$^c_{A/M-O*}$ in \AA, for twelve multicomponent oxides. When available, the experimental lattice constants were used (from Refs. a, b, c and d given below), otherwise, the lattice parameters were obtained via the geometry optimization. The experimental data for the prototype structure YbFe$_2$O$_4$ is given for comparison.}
\label{tstr}
\begin{center}
\begin{tabular}{lccccccccccr}  \hline \hline
RAMO$_4$ &  $a$ &  $c$ &  z$_A$ &  z$_M$ & $\langle$D$_{R-O}\rangle$ & $\langle$D$^{ab}_{A-O}\rangle$ & D$^c_{A-O}$ & D$^c_{A-O*}$ & $\langle$D$^{ab}_{M-O}\rangle$ & D$^c_{M-O}$ & D$^c_{M-O*}$ \\ \hline
InAlCaO$_4$ & 3.34$^{ }$ & 27.25$^{ }$ & 0.228-0.230 & 0.215-0.216 & 2.20 & 1.77 & 1.78 & 2.17 & 2.20 & 2.20 & 2.60\\
InAlCdO$_4$ & 3.32$^{a}$ & 27.50$^{a}$ & 0.215-0.230 & 0.216-0.218 & 2.20 & 1.78 & 1.79 & 2.05 & 2.17 & 2.20 & 2.63 \\
InAlMgO$_4$ & 3.29$^{a}$ & 25.66$^{a}$ & 0.210-0.218 & 0.214-0.216 & 2.20 & 1.83 & 1.84 & 2.30 & 2.02 & 1.98 & 2.26\\
InAlZnO$_4$ & 3.31$^{b}$ & 26.33$^{b}$ & 0.211-0.221 & 0.216-0.217 & 2.21 & 1.84 & 1.84 & 2.14 & 2.05 & 2.00 & 2.38\\
InGaCaO$_4$ & 3.39$^{ }$ & 27.31$^{ }$ & 0.211-0.227 & 0.216-0.217 & 2.22 & 1.85 & 1.86 & 2.14 & 2.17 & 2.21 & 2.52\\
InGaCdO$_4$ & 3.38$^{ }$ & 27.16$^{ }$ & 0.215-0.226 & 0.217-0.219 & 2.21 & 1.86 & 1.89 & 2.31 & 2.15 & 2.17 & 2.61\\
InGaMgO$_4$ & 3.30$^{c}$ & 25.81$^{c}$ & 0.211-0.218 & 0.215-0.216 & 2.19 & 1.88 & 1.91 & 2.35 & 1.98 & 1.98 & 2.26\\
InGaZnO$_4$ & 3.29$^{c}$ & 26.07$^{c}$ & 0.213-0.217 & 0.217-0.218 & 2.21 & 1.88 & 1.92 & 2.35 & 1.98 & 1.97 & 2.38 \\
ScAlMgO$_4$ & 3.24$^{a}$ & 25.15$^{a}$ & 0.211-0.220 & 0.216-0.217 & 2.15 & 1.81 & 1.80 & 2.28 & 1.98 & 1.98 & 2.32\\
ScAlZnO$_4$ & 3.24$^{b}$ & 25.54$^{b}$ & 0.213-0.221 & 0.217-0.219 & 2.13 & 1.82 & 1.82 & 2.17 & 1.99 & 1.98 & 2.38\\
ScGaMgO$_4$ & 3.27$^{a}$ & 25.62$^{a}$ & 0.212-0.220 & 0.217-0.218 & 2.14 & 1.87 & 1.89 & 2.32 & 1.96 & 1.99 & 2.33\\
ScGaZnO$_4$ & 3.26$^{c}$ & 25.91$^{c}$ & 0.214-0.220 & 0.218-0.220 & 2.13 & 1.87 & 1.90 & 2.35 & 1.96 & 1.98 & 2.29\\ \hline
YbFe$_2$O$_4$ & 3.45$^{d}$ & 25.05$^{d}$ & \hspace{0.5cm}0.215 & \hspace{0.5cm}0.215 & 2.24 & 2.01 & 1.94 & 2.15 \\ \hline\hline
\end{tabular}
\end{center}
\footnotetext[1]{experimantal values from Ref. \cite{kimizukaRAM}} 
\footnotetext[2]{experimantal values from Ref. \cite{str-all}}
\footnotetext[3]{experimantal values from Ref. \cite{str}}
\footnotetext[4]{experimantal values from Ref. \cite{KimizukaRAM2}}

\end{table*}
\begin{table}
\centering
\caption{The cation-anion distances average, $\langle$D$\rangle$, and their ranges, in \AA\, in single-cation oxides as compared to the corresponding average cation-anion distances and ranges in multicomponent oxides. Also, the deviation in the ranges of distances in multicomponent oxides with respect to the distances in the corresponding single-cation oxide, in percent.}
\label{tcomp}
\begin{center}
\begin{tabular}{cc|c|c|c|c|c} \hline\hline
& & \multicolumn{2}{c|}{Basis oxide} & \multicolumn{3}{c}{RAMO$_4$} \\  \hline
& &  $\langle$D$\rangle$ & Range & $\langle$D$\rangle$ & Range & Deviation, \% \\ \hline
R$_2$O$_3$ & In-O & 2.17 & 2.12-2.21 & 2.21 & 2.13-2.37 & ---/+7 \\
           & Sc-O & 2.11 & 2.08-2.16 & 2.14 & 2.05-2.22 & --1/+3 \\\hline
A$_2$O$_3$ & Al-O & 1.91 & 1.86-1.97 & 1.85 & 1.71-2.30 & --8/+17 \\ 
           & Ga-O & 1.93 & 1.83-2.07 & 1.93 & 1.79-2.35 & --2/+14 \\\hline
MO         & Zn-O & 1.98 & 1.97-1.99 & 2.02 & 1.92-2.38 & --3/+20 \\
           & Mg-O & 2.08 &   2.08    & 2.04 & 1.92-2.33 & --8/+12 \\
           & Ca-O & 2.37 &   2.37    & 2.25 & 2.10-2.59 & --11/+9 \\
           & Cd-O & 2.35 &   2.35    & 2.38 & 2.09-2.63 & --11/+12 \\\hline\hline
\end{tabular}
\end{center}
\end{table}

The investigated multicomponent oxides have rhombohedral $R\bar{3}m$ layered crystal structure of YbFe$_2$O$_4$ type and belong to the homologous series RAO$_3$(MO)$_m$ with m=1 \cite{str,str-all,KimizukaRAM2}. In these compounds, R$^{3+}$ ions (In or Sc) have octahedral coordination with the oxygen atoms and reside in 3(a) position (Yb), whereas both A$^{3+}$ (Al or Ga) and M$^{2+}$ (Ca, Mg, Zn or Cd) ions reside in 6(c) position (Fe) and are distributed randomly \cite{Li}.
To model a random distribution, specifically, to avoid planes or chains of the same type of atoms, a 49-atom supercell was constructed with 
the lattice vectors (30$\bar{2}$), ($\bar{1}$12) and (02$\bar{1}$), given in the units of the rhombohedral primitive cell vectors \cite{my2epl}. Note, that the conventional rhombohedral unit cell of YbFe$_2$O$_4$ contains 21 atoms (Z=3), and the primitive, i.e., the smallest volume, cell contains 7 atoms (Z=1).

Because of the different ionic radii and the valence state of the cations in RAMO$_4$ compounds, the A$^{3+}$ and M$^{2+}$ atoms have different z component of the internal site position 6(c).
Since the exact internal positions of atoms are not known, we used those of the YbFe$_2$O$_4$\cite{str} as the starting values, and then optimized the internal positions of all atoms in the supercell via minimization of the total energy and the atomic forces. During the optimization the lattice parameters were fixed at the experimental values\cite{str,str-all,kimizukaRAM,KimizukaRAM2} except for InAlCaO$_4$, InGaCaO$_4$, and InGaCdO$_4$ where $a$ and $c$ were optimized since the experimental values are unavailable. Our optimized structural parameters for the latter compounds as well as the optimized $z$ values for every structure under consideration are given in Table \ref{tstr}. 

Next, we compare the local atomic structure in multicomponent oxides to that of the constituent basis oxides. First, the following ground state (lowest energy) structures of single-cation oxides were considered:
$Fm\bar{3}m$ (rocksalt) for MgO, CaO, and CdO; $Ia\bar{3}$ (bixbyite) for Sc$_2$O$_3$ and In$_2$O$_3$; $P6_3mc$ (wurtzite) for ZnO; $R\bar{3}c$ (corundum) for Al$_2$O$_3$; and $C2/m$ (monoclinic) for $\beta$-Ga$_2$O$_3$. For these structures, the lattice parameters were kept at the experimental values. The internal atomic positions for Sc$_2$O$_3$, In$_2$O$_3$, Al$_2$O$_3$, and Ga$_2$O$_3$ were optimized via the total energy and atomic forces minimization.
Additional phases for oxides of A and M metals were also calculated as explained in details below.

Our results show that the optimized cation-anion distances in multicomponent oxides correlate with the ionic radii of the cations, c.f., Tables \ref{tstr} and \ref{tcomp}. For the octahedrally coordinated R$^{3+}$ ions, i.e., In or Sc, the R-O distances in multicomponent oxides are close to those in the corresponding single-cation oxides, c.f., $\langle$D$_{R-O}\rangle$ in Table \ref{tstr} and $\langle$D$\rangle$ in Table \ref{tcomp}. The averaged In-O or Sc-O distance in RAMO$_4$ is only 0.03-0.04 \AA \, larger than that in In$_2$O$_3$ or Sc$_2$O$_3$. 
The largest deviations for one of the six In-O distances in the InO$_6$ octahedra (5-7\%) are found for Ca and Cd-containing compounds (i.e., InGaMO$_4$ and InAlMO$_4$ with M=Ca or Cd). These compounds represent the case of a large mismatch of the ionic radii of the A and M ions, which affects the In-O distances in the neighboring InO$_{1.5}$ layer. In other compounds, the In-O distances differ by only 1-2\% as compared to those in the bulk In$_2$O$_3$.

\begin{figure}[htbp]
\includegraphics[height=3.5cm]{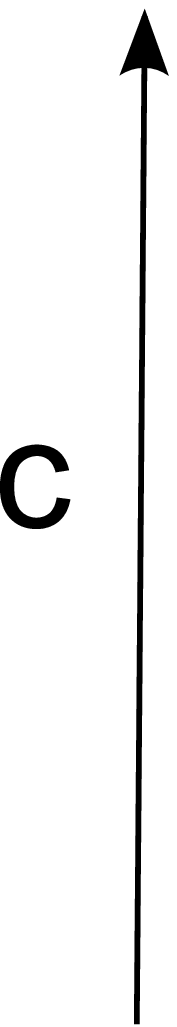}
\includegraphics[height=4cm]{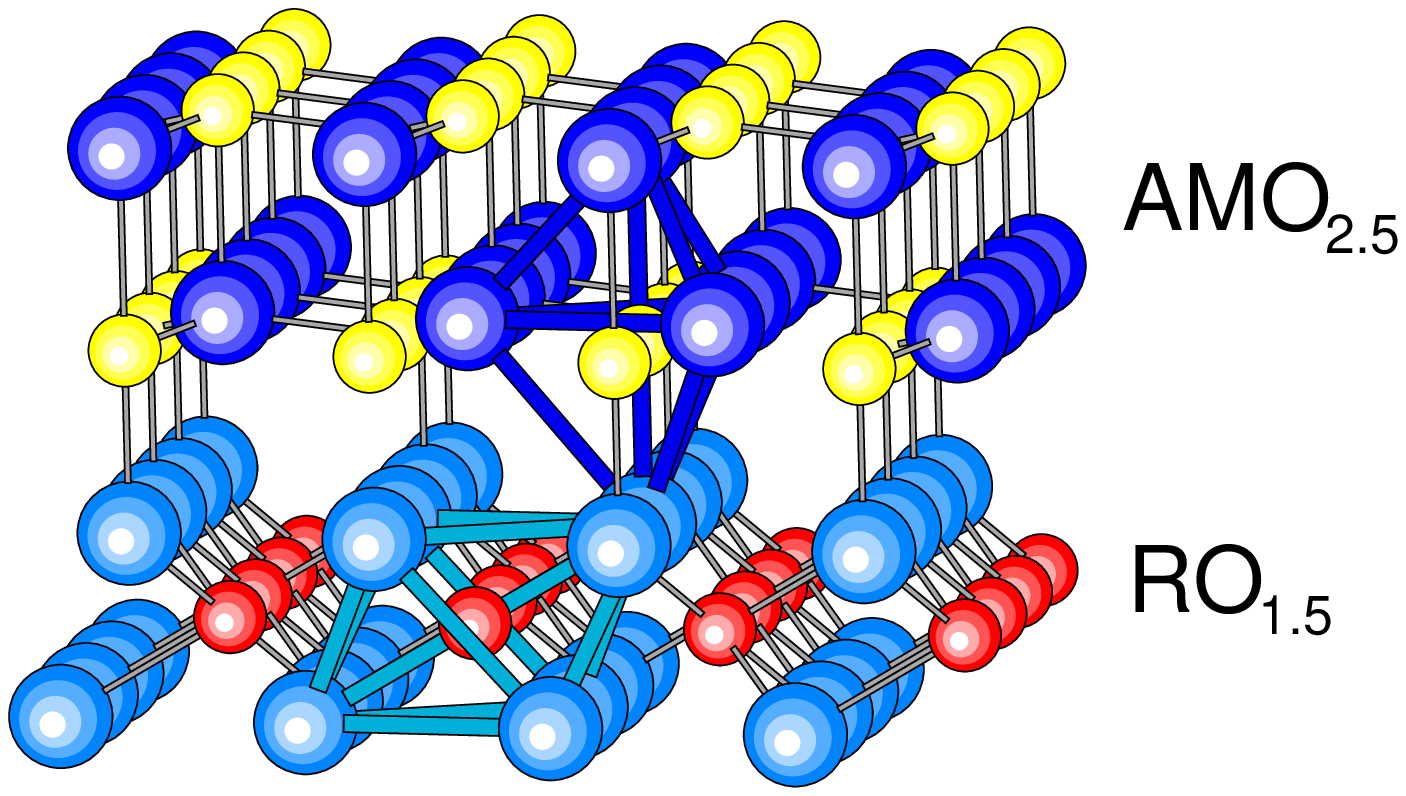}
\caption{Oxygen coordination of R=In or Sc (octahedra) and A=Al or Ga, and M=Zn, Cd, Ca, or Mg (bipyramid) in the single block of the unit cell of RAMO$_4$ compounds. The conventional unit cell of RAMO$_4$ consists of three similar blocks stacked along the $c$ direction.}
\label{str-block}
\end{figure}

\begin{figure}[htbp]
\includegraphics[height=7cm]{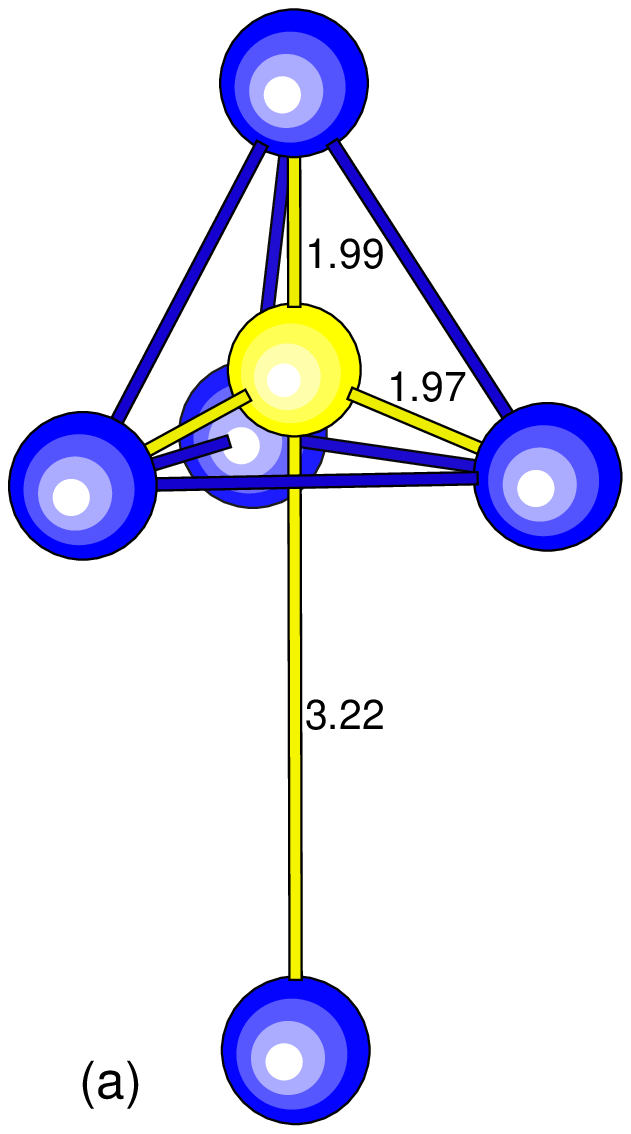}
\includegraphics[height=6.5cm]{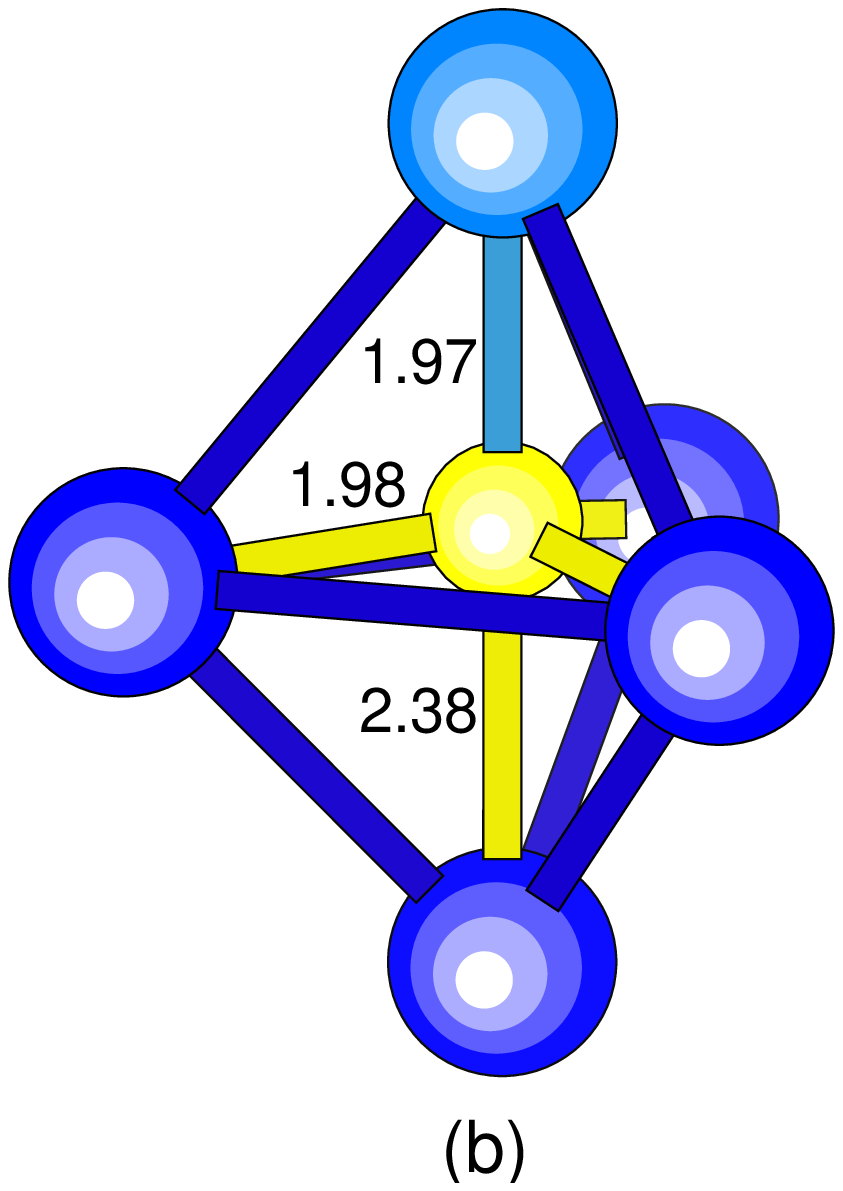}
\caption{Four-fold vs five-fold coordination of Zn with oxygen atoms in wurtzite ZnO (a) vs InGaZnO$_4$ (b). The cation-anion apical and planar distances are shown (in \AA). The corresponding charge densities are shown in Fig. \ref{ch-den}.}
\label{tetra-octa}
\end{figure}

The most important observation concerning the crystal structure in RAMO$_4$ compounds is that all A and M atoms are in five-fold coordination (bipyramid) with oxygen atoms, Fig. \ref{str-block}, and not in four-fold (tetrahedral) as it was previously assumed for decades. As one can see from Table \ref{tstr}, the A-O or M-O distance to the fifth atom (also called the second apical atom hereafter), denoted as $\langle$D$^{c}_{A/M-O*}\rangle$, is only $\sim$0.3-0.5 \AA \, longer than the distance to the nearest apical oxygen atom, denoted as $\langle$D$^{c}_{A/M-O}\rangle$.
For comparison, in wurtzite ZnO, the Zn-O distance to the next nearest oxygen atom (second apical O) is 3.22 \AA \, which is 1.23 \AA \, longer than the Zn-O distance to the nearest apical oxygen atom which belongs to the ZnO$_4$ tetrahedra, Fig. \ref{tetra-octa}(a). 

\begin{figure}[htbp]
\centering
\includegraphics[height=3.5cm]{uparrow.eps}
\includegraphics[height=6cm]{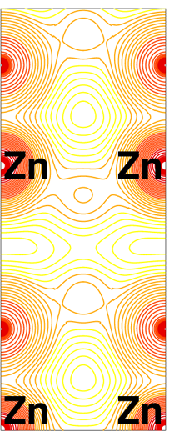}
\includegraphics[height=6cm]{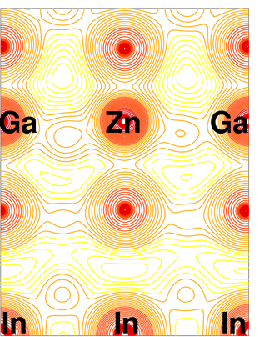}
\caption{Calculated total charge density distribution contour plots for wurtzite ZnO (left) and InGaZnO$_4$ (right). Zn as well as Ga have strong bonds with both apical oxygen atoms making them five-fold coordinated cations in the multicomponent oxide -- in marked contrast to wurtzite ZnO with four-fold oxygen coordination.}
\label{ch-den}
\end{figure}

The fact that Zn has five-fold oxygen coordination in RAZnO$_4$ is illustrated in Fig. \ref{ch-den} where we compare the calculated charge density distribution for InGaZnO$_4$ and wurtzite ZnO plotted in the (011) plane to include O-Zn-O bonds along the [0001] direction for both compounds. The strong bonding between Zn (as well as Ga) atom and the second apical oxygen atom in the multicomponent oxide is clearly seen from the charge density plot, Fig. \ref{ch-den}(b). In contrast, there is no overlap between Zn atom and its second apical oxygen atom in wurtzite ZnO, Fig. \ref{ch-den}(a). Thus, Zn atoms form five bonds with neighboring oxygen atoms in InGaZnO$_4$, whereas Zn has four bonds in the basis ZnO. 

Similar to Zn, all other M$^{2+}$ and all A$^{3+}$ cations in RAMO$_4$ compounds are five-fold coordinated with oxygen atoms. Strikingly, none of the A or M atoms possess five-fold coordination in the basis, single-cation oxides. The Ca, Cd, or Mg metals form rocksalt structure ($Fm\bar{3}m$) with octahedral oxygen coordination, whereas Al or Ga ions are in either four-fold or six-fold coordinations in corundum ($R\bar{3}c$) or monoclinic ($C2/m$) phases, respectively. 
(Other known phases of Al$_2$O$_3$, i.e., $\theta$- and $\kappa$-Al$_2$O$_3$ with $C2/m$ and $Pna2_1$ structures, respectively, also have four- and six-coordinated Al ions; $\alpha$-Ga$_2$O$_3$ has corundum structure, space group $R\bar{3}c$, with six-fold oxygen coordination of Ga.)
\begin{table}
\centering
\caption{Structural parameters for wurtzite-based hypothetical structures of M$^{2+}$O$^{2-}$ where metal-oxygen distances correspond to the average distances obtained for RAMO$_4$, Table \ref{tstr}. Lattice constants $a$ and $c$, internal paramater $u$ in \AA, as well as planar D$^{ab}_{M-O}$, nearest apical D$^c_{M-O}$, and next nearest apical distances D$^c_{M-O*}$ in \AA. To compare, in ground state wurtzite ZnO, a=3.25 \AA, c=5.21 \AA, u=0.3817: D$^{ab}_{Zn-O}$=1.97 \AA; D$^{c}_{Zn-O}$=1.99 \AA; D$^{c}_{Zn-O*}$=3.22 \AA.}
\label{t-phases}
\begin{center}
\begin{tabular*}{8.5cm}{p{1cm}  p{1cm} p{1cm}  p{1.2cm} | p{1cm} p{1cm}  p{1cm} } \hline \hline
& a & c & u & D$^{ab}_{M-O}$ & D$^{c}_{M-O}$ & D$^{c}_{M-O*}$ \\ \hline
ZnO & 3.44 & 4.34 & 0.4570 & 2.00 & 1.98 & 2.36 \\
MgO & 3.43 & 4.28 & 0.4639 & 1.98 & 1.98 & 2.29 \\
CaO & 3.77 & 4.76 & 0.4625 & 2.19 & 2.20 & 2.56 \\ 
CdO & 3.73 & 4.81 & 0.4557 & 2.16 & 2.19 & 2.62\\\hline\hline
\end{tabular*}
\end{center}
\end{table}
\begin{table*}
\centering
\caption{Structural parameters for hypothetical phases of Al$_2$O$_3$ and Ga$_2$O$_3$ in Al$_2$S$_3$-type (space group P6$_1$). The Ga-O and Al-O distances correspond to the average distances obtained in RAMO$_4$. Lattice constants $a$ and $c$ in \AA, positions for O(1) and O(2) in \AA, and planar D$^{ab}_{A-O}$, nearest apical D$^c_{A-O}$, and next nearest apical distances D$^c_{A-O*}$ in \AA \, for different sites. The internal atomic positions in Al$_2$S$_3$ are x=0.3417, y=0.3387 for O(1); and y=0.3501 for O(2).}
\label{t-phases2}
\begin{center}
\begin{tabular}{ccc|ccc|ccc|cccc|c|c} \hline\hline
& a & c & \multicolumn{3}{|c|}{O(1)} & \multicolumn{3}{|c|}{O(2)} & \multicolumn{4}{|c|}{D$^{ab}_{A-O}$} & D$^{c}_{A-O}$ & D$^{c}_{A-O*}$ \\ \hline
Al$_2$O$_3$ & 5.30 & 12.59 & x & y & 0.358 & 0.0239 & y & 0.2 & Al(1) & 1.78 & 1.81 & 1.82 & 1.79 & 2.30\\
      &      &         &   &   &        &        &   &     & Al(2) & 1.79 & 1.85 & 1.89 & 1.83 & 2.31\\\hline
Ga$_2$O$_3$ & 5.38 & 12.85 & x & y & 0.363 & 0.0239 & y & 0.2 & Ga(1) & 1.80 & 1.83 & 1.85 & 1.89 & 2.35   \\
      &      &         &   &   &        &        &   &     & Ga(2) & 1.82 & 1.88 & 1.92 & 1.87 & 2.38  \\\hline\hline
\end{tabular}
\end{center}
\end{table*}

The unusual five-fold coordination of A and M ions stabilized in RAMO$_4$ compounds is expected to manifest itself in the electronic properties of the complex oxides that differ from those for the basis oxides. Specifically, because the main features of the electronic band structure of oxides, such as the band gap value and the electron effective mass, are determined by the strong metal-oxygen interactions, direct comparison between the (averaged) values obtained for multicomponent oxides with those in the basis oxides in the ground state structures is invalid.

We stress here that the five-fold coordination of A and M atoms with the neighbor O atoms in the RAMO$_4$ compounds does not fall out of the fundamental principles governing the structure formation of multicomponent oxide systems.
As shown in the extensive works of Walsh et al. \cite{Walsh} (and references therein), the coordination environment is determined by satisfying the electronic octet rule for local charge neutrality as well as the material stoichiometry. 
The octahedral structure in the RO$_{1.5}$ layer which maximizes the atomic separation between the negatively charged O atoms, serves as a disruptive stacking fault to the wurtzite-like AMO$_{2.5}$ layer. At the same time, the A atoms, such as Al or Ga, do not have a strong preference for octahedral sites  \cite{Walsh}. Hence, while trying to accommodate the A and M atoms and obey the electronic octet rule, changes must occur in the AMO$_{2.5}$ layer leading to the formation of five-fold trigonal bipyramid structures \cite{Walsh}.

To determine how the local atomic coordination affects the electronic properties of oxides, we performed calculations for the hypothetical phases with five-fold oxygen coordination of A and M cations.
Moreover, we set the lattice parameters as well as the internal atomic positions in the hypothetical phases so that the metal-oxygen distances are similar to those in the corresponding multicomponent RAMO$_4$ oxides (given in Table \ref{tstr}). This will allow us to compare the band gap value calculated for each RAMO$_4$ compound with the value obtained via averaging over the band gaps in the corresponding single-cation oxides with the same local atomic coordination and bond-lengths. For ZnO, MgO, CdO, and MgO, i.e., for M$^{2+}$O$^{2-}$ compounds, we performed calculations for wurtzite-based structures where the second nearest apical oxygen atom is located close enough to the metal ion to make it a five-fold coordination, Table \ref{t-phases}. Similarly, for Al$_2$O$_3$ and Ga$_2$O$_3$, we used Al$_2$S$_3$-type structure, space group $P6_1$, and modified the lattice parameters and the internal atomic positions to obtain A-O distances similar to those in the corresponding RAlMO$_4$ or RGaMO$_4$ compounds, Table \ref{t-phases2}.
Note, that the In and Sc are octahedrally coordinated with oxygen atoms both in the basis oxides and in RAMO$_4$. The In-O or Sc-O distances in the multicomponent oxides are slightly larger than those in the basis oxides, c.f., Tables \ref{tstr} and \ref{tcomp}. 

In the next section, we begin our discussions with the electronic properties of single-cation oxides and how the atomic coordination affects their electronic band structure.

\begin{table*}
\centering
\caption{ 
The averaged electron effective mass in m$_e$, for single-cation oxides within both LDA and sX-LDA are
given for the basis oxides in the ground state phase, $\langle$m$^g\rangle$ 
, and in the hypothetical phase, $\langle$m$^h\rangle$. The effective mass anisotropy $\delta$, which is defined as 
$\delta=(m^{[100]}+m^{[010]})/2m^{[001]}$. Also, the band gap values (in eV) obtained within both LDA and sX-LDA are given for the basis oxides in the ground state phase, E$_g^{g}$, and in the hypothetical phase, E$_g^{h}$, with the bond lengths and oxygen coordination resembling those in the corresponding RAMO$_4$ compounds. The fundamental band gaps as well as optical, i.e., direct, band gaps (in parenthesis) are given.}
\label{tsingleband}
\begin{tabular*}{13.2cm}{p{1cm} p{1cm} | p{1cm} p{1cm} p{1.8cm} p{1cm}| p{1cm} p{1.5cm} p{1cm} p{1.5cm}}  \hline\hline
\multicolumn{2}{c}{} & \multicolumn{4}{c}{LDA} & \multicolumn{4}{c}{sX-LDA} \\ \hline
            &     & $\langle$m$^g\rangle$ & $\delta$ & E$_g^g$ & E$_g^h$ & $\langle$m$^g\rangle$ & E$_g^g$ &$\langle$m$^h\rangle$ & E$_g^h$\\ \hline
R$_2$O$_3$  & In$_2$O$_3$ & 0.18 & 1.00 & 1.16        & 0.85 & 0.28 &  2.90(3.38) & 0.28 & 2.61(3.07) \\
            & Sc$_2$O$_3$ & 1.12 & 1.00 & 3.66        & 3.61 & 1.19 &  6.06       & 1.19 & 5.98 \\\hline
A$_2$O$_3$  & Al$_2$O$_3$ & 0.39 & 1.00 & 6.27        & 3.86 & 0.45 &  9.08       & 0.52 & 6.80 \\
            & Ga$_2$O$_3$ & 0.26 & 1.17 & 2.32        & 2.42 & 0.34 &  4.86(4.91) & 0.43 & 4.82 \\\hline
MO          & ZnO         & 0.17 & 1.09 & 0.81        & 1.14 & 0.35 &  3.41       & 0.36 & 3.63 \\\
            & MgO         & 0.38 & 1.00 & 4.76        & 3.44 & 0.46 &  7.55       & 0.52 & 6.50 \\
            & CaO         & 0.37 & 1.00 & 3.45(4.42)  & 3.52 & 0.42 &  5.95(7.15) & 0.53 & 6.51 \\
            & CdO         & 0.15 & 1.00 & -0.51(0.92) & 0.00 & 0.23 &  0.50(2.29) & 0.31 & 1.01 \\  \hline\hline
\end{tabular*}
\end{table*}

\subsection{IV. Electronic properties of single-cation oxides}

\subsubsection{A. Ground state structures}
The investigated basis oxides of post-transition and light main group metals possess qualitatively similar electronic band structure: the valence band is formed from non-bonding and bonding 2$p$ states of oxygen, whereas the highly dispersed conduction band arises from the metal $s$ states and the anti-bonding O-2$p$ states. Strong metal-oxygen interaction 
is responsible for wide band gaps and small electron effective masses in these oxides, Table \ref{tsingleband}.
Note that, as expected, LDA underestimates the band gap values as well as the electron effective masses. The nonlocal density scheme of the sX-LDA method corrects the LDA failure and gives an excellent agreement between the calculated, Table \ref{tsingleband}, and experimental band gaps for both the semiconductor-like materials with band gap of $\sim$2.3-3.4 eV (CdO, In$_2$O$_3$, ZnO) and the insulators with band gaps of $\sim$6-9 eV (CaO, MgO, Al$_2$O$_3$, Sc$_2$O$_3$). 

\begin{figure}
\includegraphics[height=6.5cm]{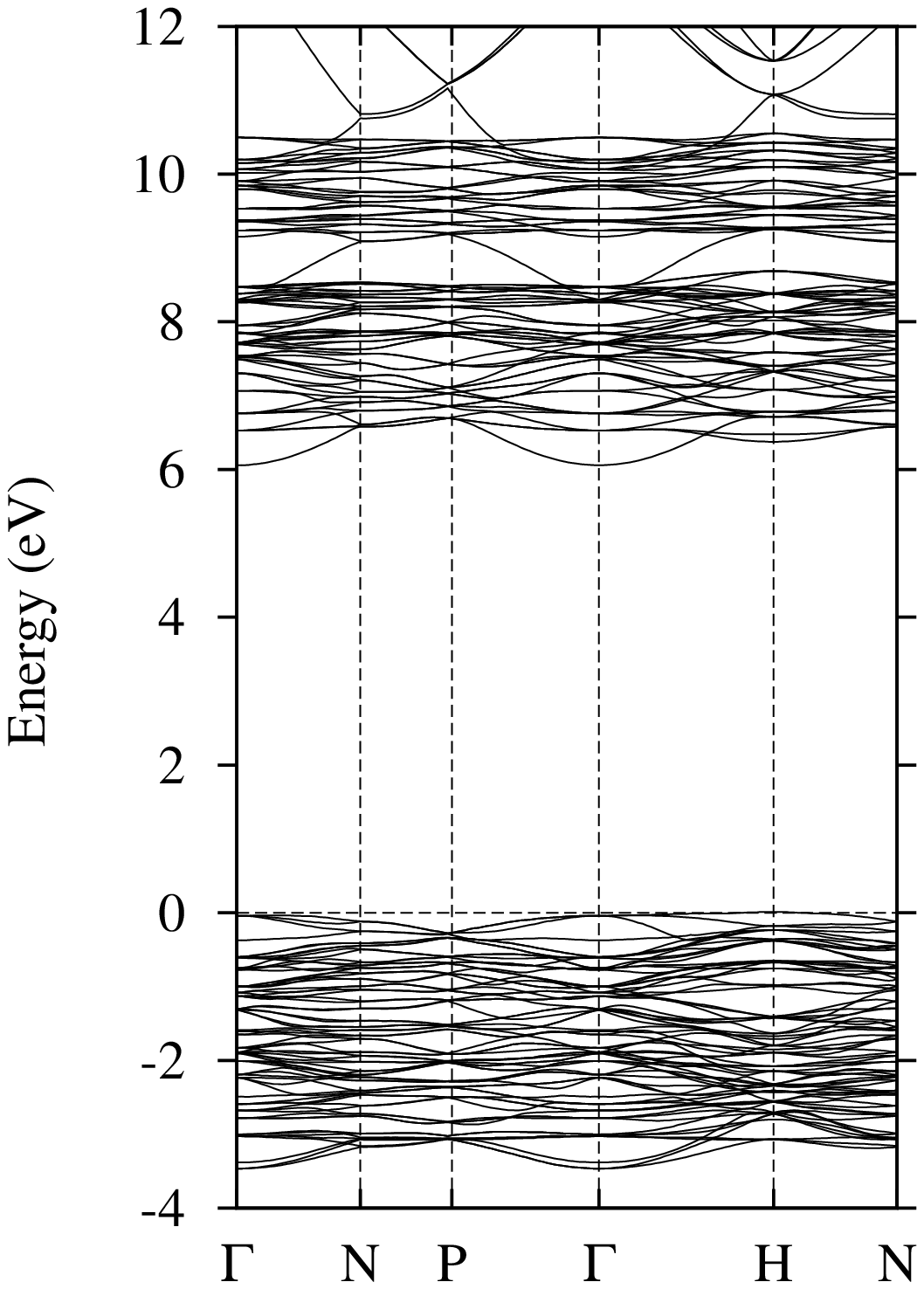}
\includegraphics[height=6.5cm]{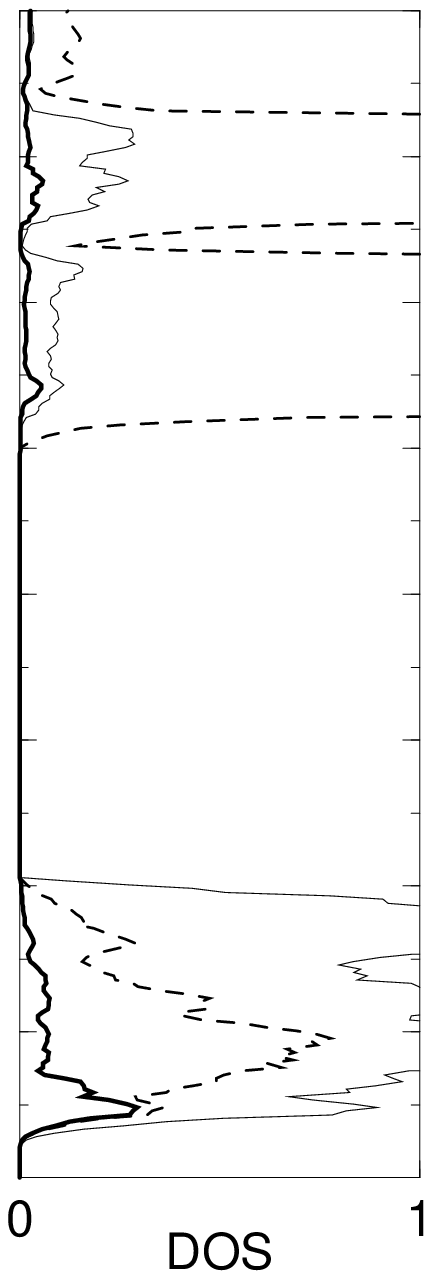}
\caption{Band structure and partial DOS of Sc$_2$O$_3$. The thin, dashed, and thick lines in the partial DOS plots represent the metal d, oxygen p, and metal s states, respectively.}
\label{sc2o3}
\end{figure}

The sX-LDA calculated electronic band structures and partial density of states of all single-cation oxides studied in this work have been published earlier  \cite{JMprb,JMbook} except for Sc$_2$O$_3$. The bottom of the conduction band in scandium oxide is governed by the localized Sc $d$-states, Fig. \ref{sc2o3}, and, as a result of the low dispersion of the conduction band, the electron effective mass in Sc$_2$O$_3$ is the largest among the oxides and is greater than the mass of the free electron, Table \ref{tsingleband}.

Recent comparative investigations of main group metal oxides \cite{JMprb,JMbook} have revealed that the fundamental differences in the electronic properties of the conventional TCO hosts (In$_2$O$_3$, ZnO and CdO) and the light metal oxides (Al$_2$O$_3$, CaO and MgO) originate from the different energy location of the cation's empty $p$ or $d$ states with respect to the conduction band bottom. In the former oxides, the cation $p$ band is well above its $s$ band which is a prerogative for a good charge transport via a uniform network formed by the spherically symmetric metal $s$ orbitals and the neighboring oxygen $p$ orbitals in degenerately doped materials. In striking contrast to the post-transition metal oxides, the light metal $p$ or $d$ band almost coincides (as in Al$_2$O$_3$ or MgO) or is even below its $s$ band (as in CaO or Sc$_2$O$_3$) in the classical insulators. The proximity of the $p$ or $d$ states to the bottom of the conduction band and the resulting strong directional interaction of these anisotropic orbitals with the $p$ orbitals of the neighboring oxygen atoms have three consequences: (1) wide band gaps of 6-9 eV; (2) the electron effective masses which are at least twice larger than those in the conventional TCO hosts, Table \ref{tsingleband}; and (3) charge localization (widely known as an F-center or colour center) of extra electrons near an electron-donor defect. The deep defect states are unable to produce electrical conductivity in these oxides.

We note here that Ga$_2$O$_3$ does not belong to either of the two groups of oxides but rather represents an intermediate case, c.f., Table \ref{tsingleband}, illustrating that, naturally, the transition between the oxide groups is not abrupt. The Ga $p$ band is located relatively close to the metal $s$ band but does not coincide with it as, for example, 
in Al$_2$O$_3$ or MgO. This leads to a considerable but not dominant contributions from the Ga $p$ states near the bottom of the conduction band.
Consequently, in oxygen deficient Ga$_2$O$_3$, extra electrons induced by the oxygen vacancy are not fully localized near the defect as in light metal oxides,
yet, the electron group velocity is nearly an order of magnitude smaller than that in TCOs, e.g., In$_2$O$_3$ \cite{JMprb}. This explains why Ga$_2$O$_3$ is not a viable TCO itself, nonetheless, Ga-containing multicomponent TCOs are common.

In section V.B, we will come back to the discussion of the proximity of the cation's $p$ or $d$ states to the conduction band in multicomponent oxides.

\subsubsection{B. Hypothetical phases with five-fold coordination}
As mentioned above, the main features in the electronic band structure of oxides are determined by the nature and degree of the metal-oxygen interaction.
Here we discuss how the electronic properties, in particular, the band gap values of single-cation oxides vary when the metal-oxygen distances and oxygen coordination are changed to resemble those in the RAMO$_4$ multicomponent oxides.

First, we note that In and Sc are octahedrally coordinated with oxygen atoms both in the basis oxides and in RAMO$_4$. The In-O or Sc-O distances in the basis oxides are slightly smaller than those in the multicomponent oxides, c.f., Tables \ref{tstr} and \ref{tcomp}. To reproduce the R-O distances found in the multicomponent oxides, we increased the lattice parameter $a$ from 10.12 to 10.26 \AA \, and from 9.81 to 9.90 \AA \, for cubic In$_2$O$_3$ and Sc$_2$O$_3$, respectively. As expected from a smaller nearest neighbor orbital overlap associated with longer metal-oxygen distances, we obtained smaller band gaps for indium and scandium oxides, cf., Table \ref{tsingleband}.

For A$_2$O$_3$ and MO oxides, we considered hypothetical structures with five-fold coordination and metal-oxygen distances that resemble those obtained in multicomponent oxides (see section III for details). The band gap values calculated within sX-LDA methods for the hypothetical structures are given in Table \ref{tsingleband}.
For Al$_2$O$_3$ and MgO with five-coordinated Al and Mg cations, the gap becomes smaller by 2.2 eV and 1.0 eV, respectively, as compared to the ground state phases (corundum and rocksalt, respectively) with six-fold coordination. In the hypothetical CaO and CdO with five-fold coordinated Ca and Cd, the band gap becomes direct and its value decreases by 0.6 eV and 1.3 eV, respectively, as compared to the optical, direct band gap of rocksalt CaO and CdO with octahedral coordination of cations, Tabel \ref{tsingleband}. (Note, the case of Cd represents the largest coordination-induced change in the band gap, namely, 56\%). Accordingly, the band gap in hypothetical ZnO with five-fold coordinated Zn slightly increases (by $\sim$0.2 eV) with respect to four-coordinated Zn in wurtzite ZnO. Finally, there is a negligible change in the band gap of $\beta$-Ga$_2$O$_3$ which has four-fold and six-fold coordinated Ga atoms in the ground state monoclinic phase as opposed to the five-fold coordination of Ga in the hypothetical Al$_2$S$_3$-type structure. 

Thus, we find that lower coordination number leads to a smaller band gap. We must stress here that this conclusion should not be generalized to other coordinations. For example, we do not expect the band gap to increase further for structures with 8-fold coordination (e.g., as in CsCl-type structure) with respect to the six-fold coordination. We believe that octahedral coordination provides a largest band gap because it corresponds to the largest overlap between the metal orbitals and the $p_x$, $p_y$, and $p_z$ orbitals of the neighboring oxygen atoms\cite{JMbook}. Therefore, with respect to the six-coordinated case, higher- and lower-coordinated structures are expected to produce a smaller band gap. Variations in the metal-oxygen distances (c.f., ranges in Table \ref{tcomp}) may further affect the orbital overlap and, hence, the band gap values, but perhaps to a lesser extent compared to the changes caused by the different atomic coordination.

In the next section we will demonstrate that the band gap values of multicomponent RAMO$_4$ compounds can be reproduced via averaging over those obtained for the single-cation oxides in the hypothetical structures, i.e., with the corresponding atomic coordination and interatomic distances.

\subsection{V. Electronic properties of multicomponent oxides}
\subsubsection{A. Role of atomic coordination in band gap formation}
The electronic band structure of 12 multicomponent oxides, RAMO$_4$, is similar to that of the single-cation oxides: the valence band is formed from the oxygen 2$p$ states, whereas the conduction band arises from the antibonding oxygen $2p$ states and the metal $s$, $p$ or $d$ states.  

\begin{figure*}
\centering
\includegraphics[height=7cm]{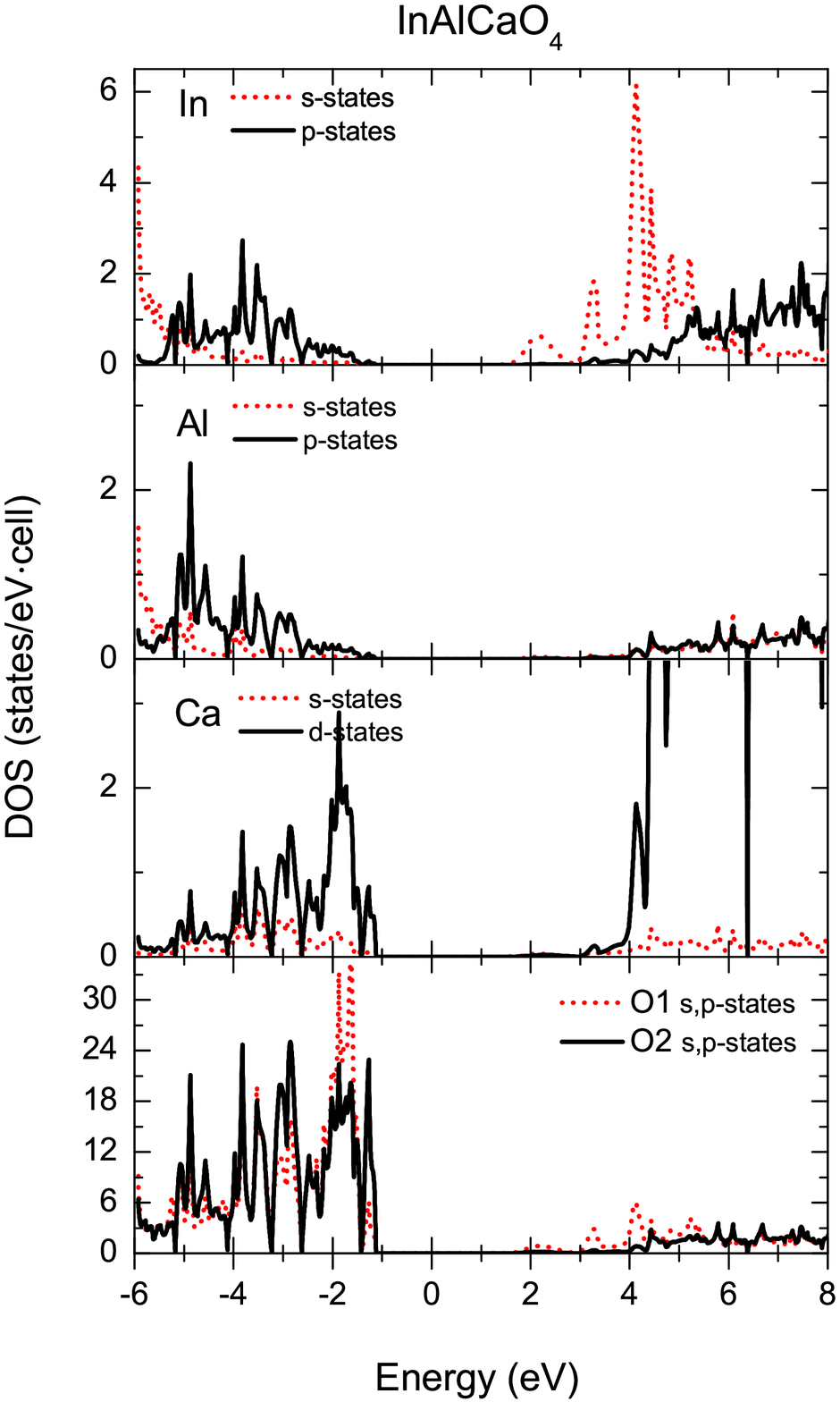}
\includegraphics[height=7cm]{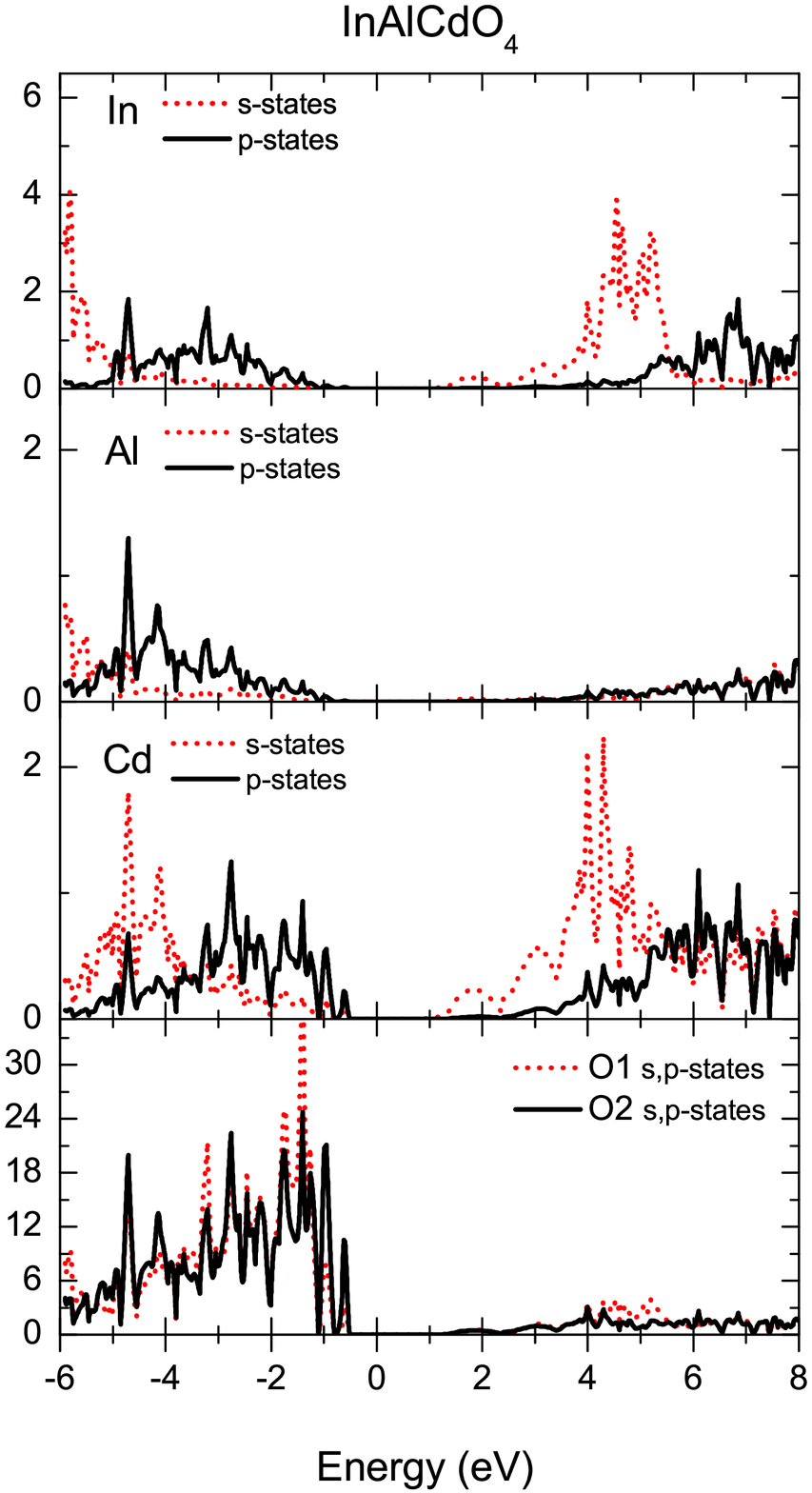}
\includegraphics[height=7cm]{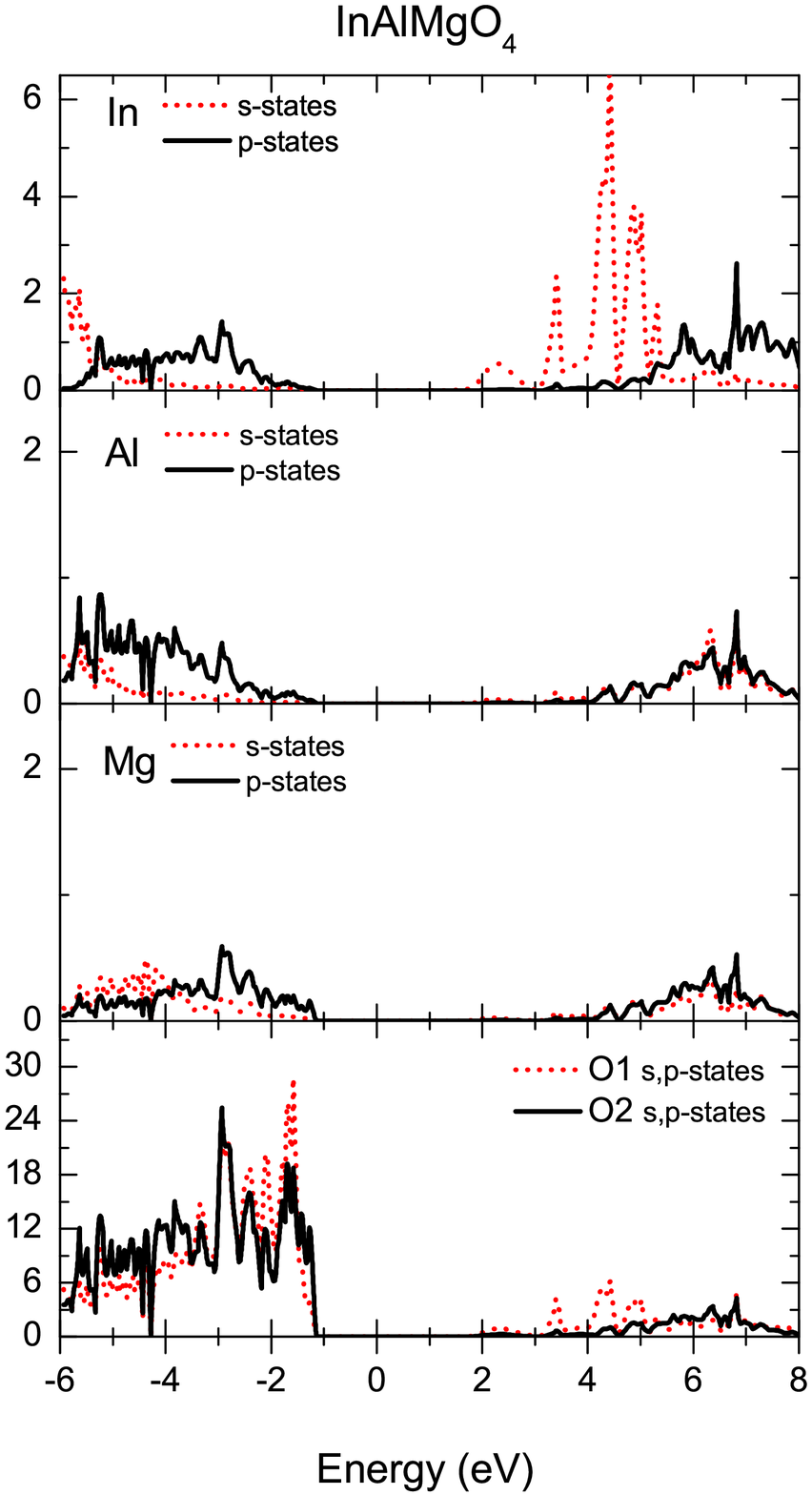}
\includegraphics[height=7cm]{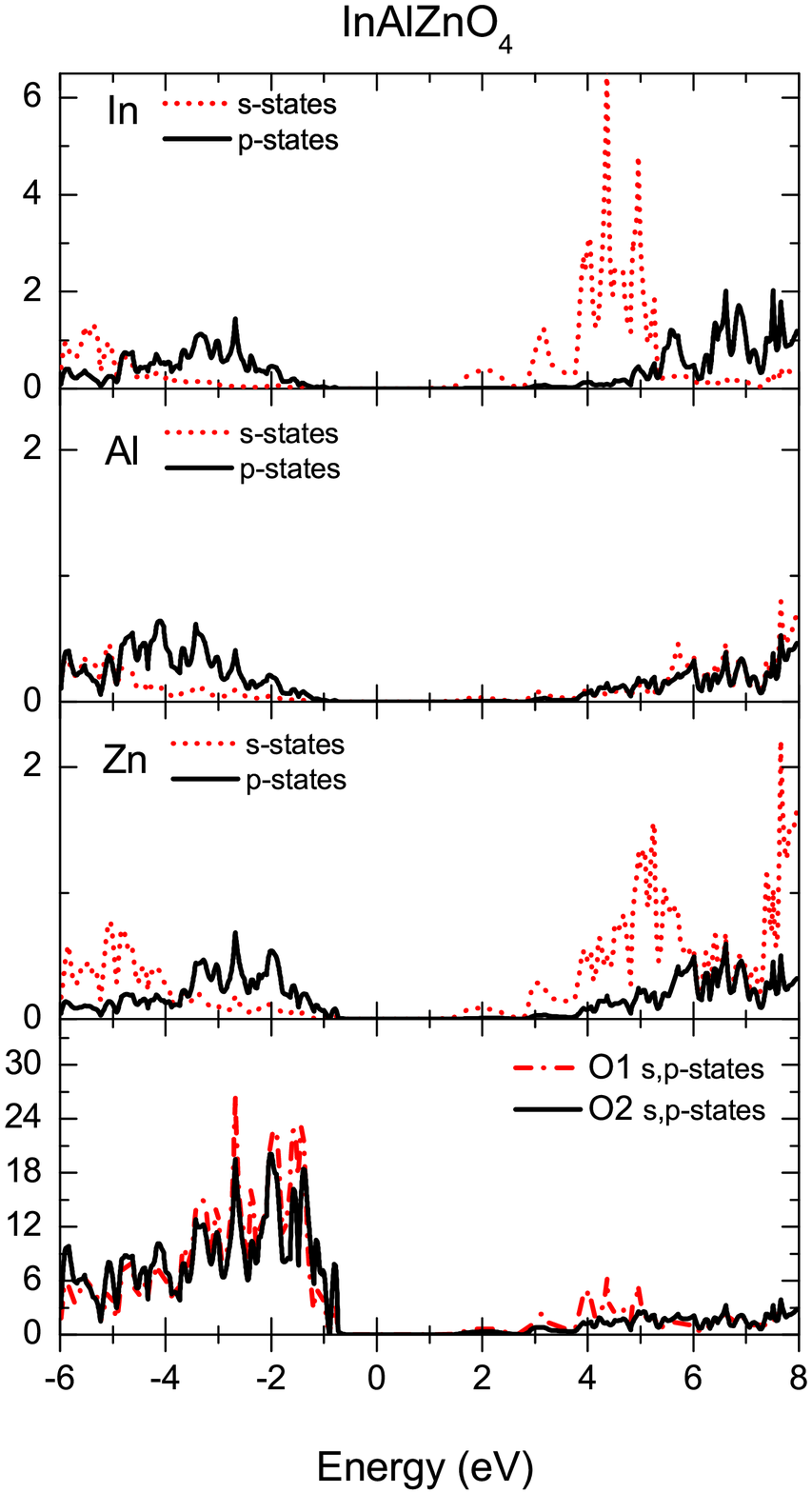}
\vspace{0.5cm}

\includegraphics[height=7cm]{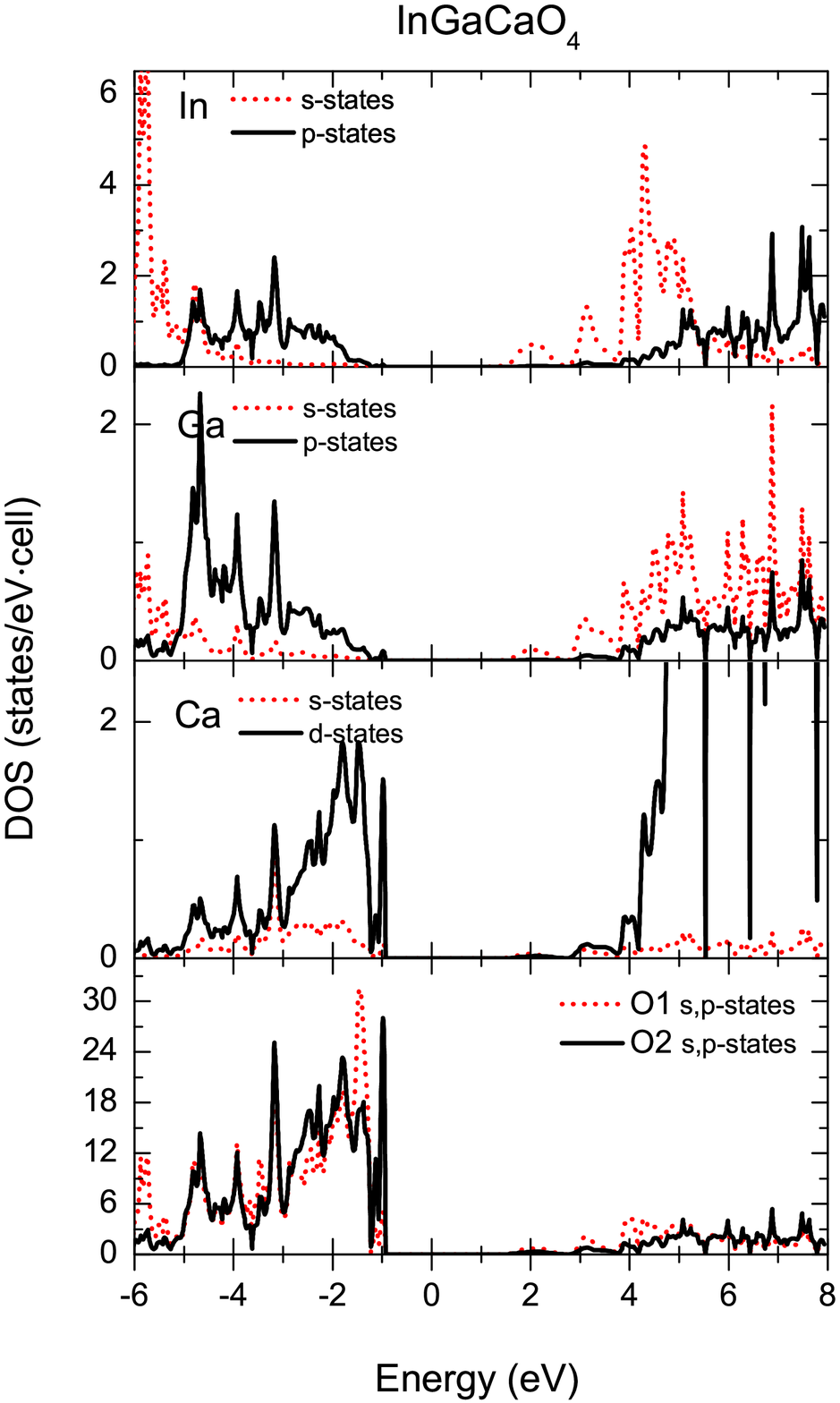}
\includegraphics[height=7cm]{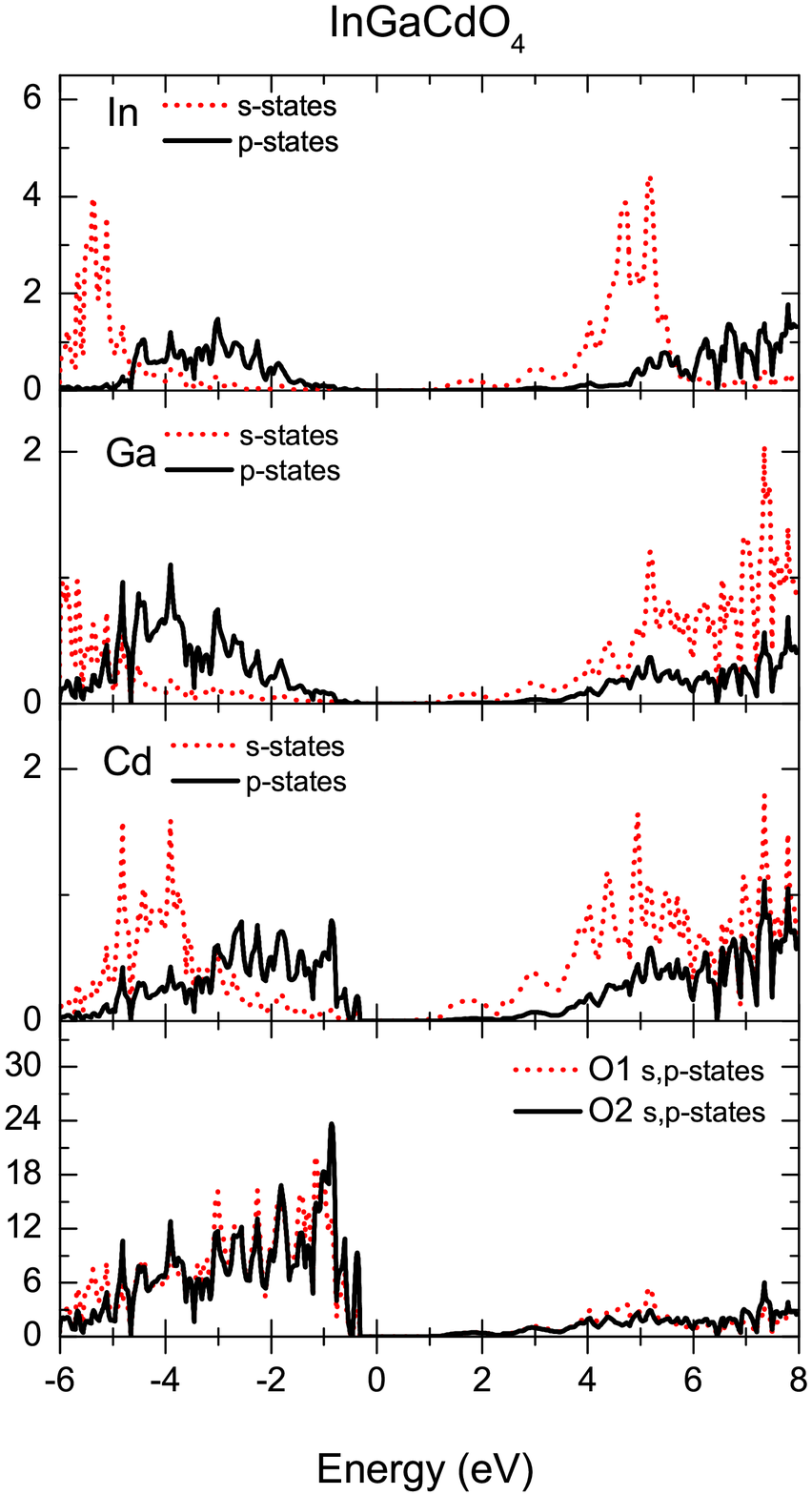}
\includegraphics[height=7cm]{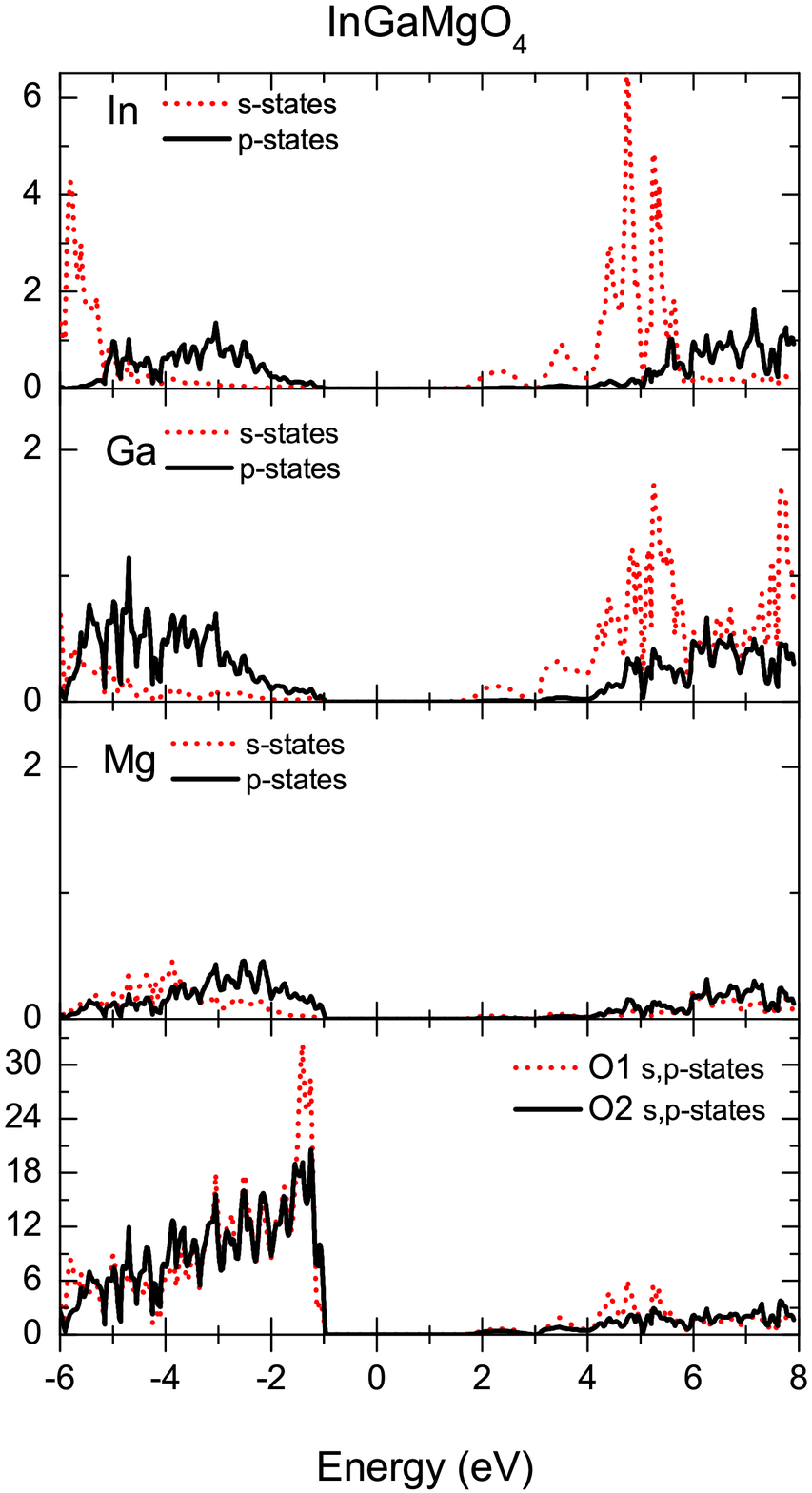}
\includegraphics[height=7cm]{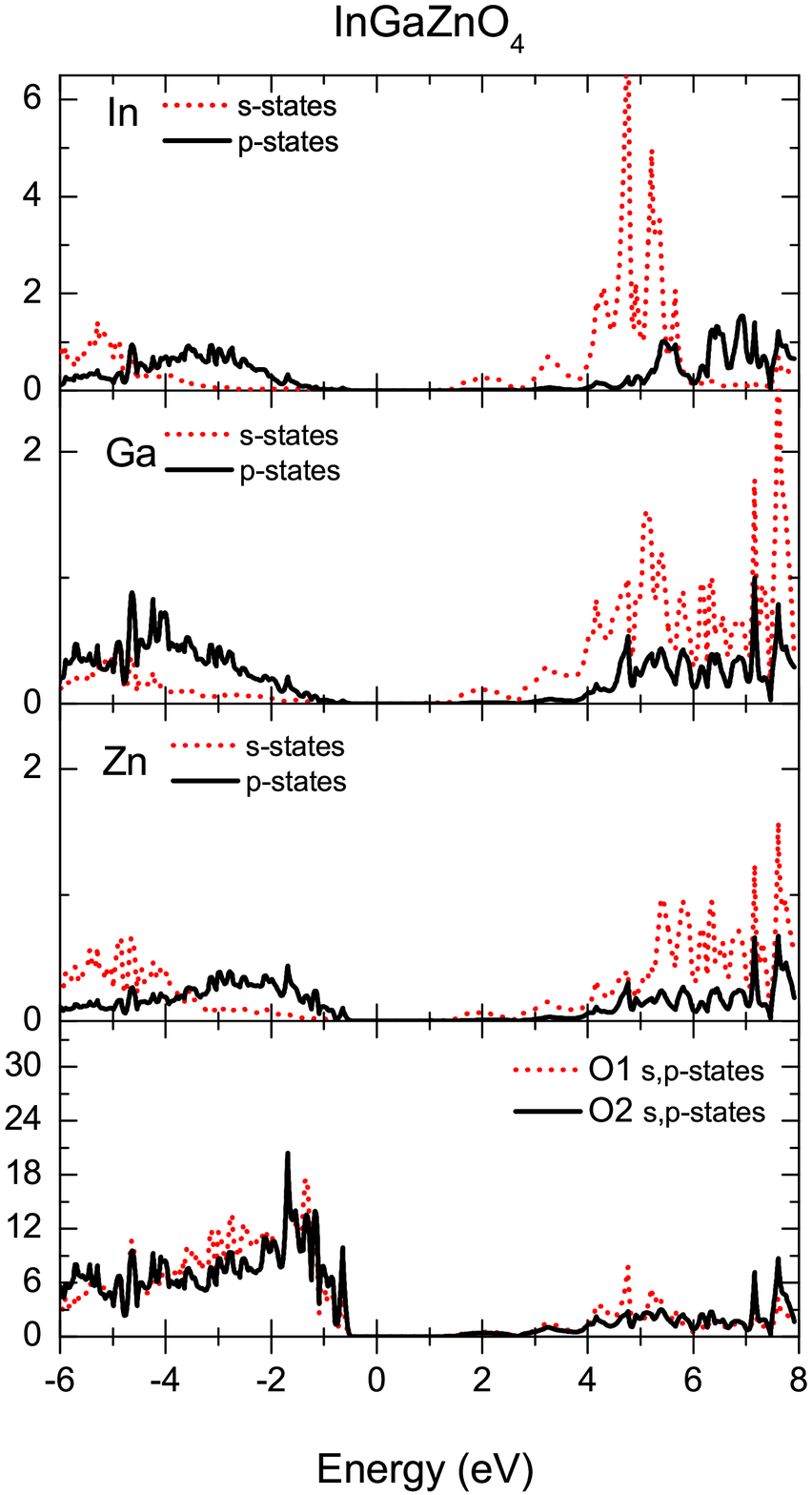}
\vspace{0.5cm}

\includegraphics[height=7cm]{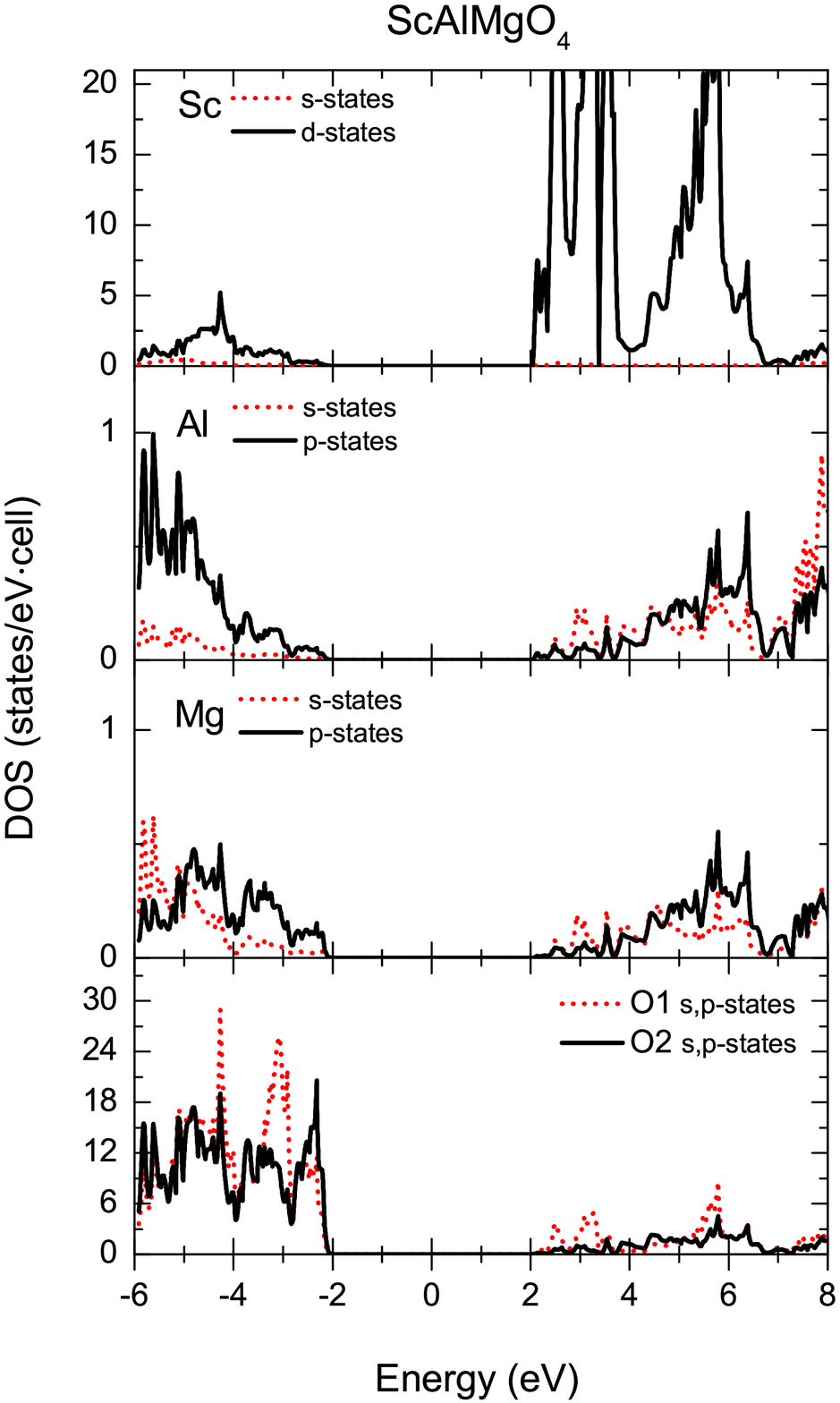}
\includegraphics[height=7cm]{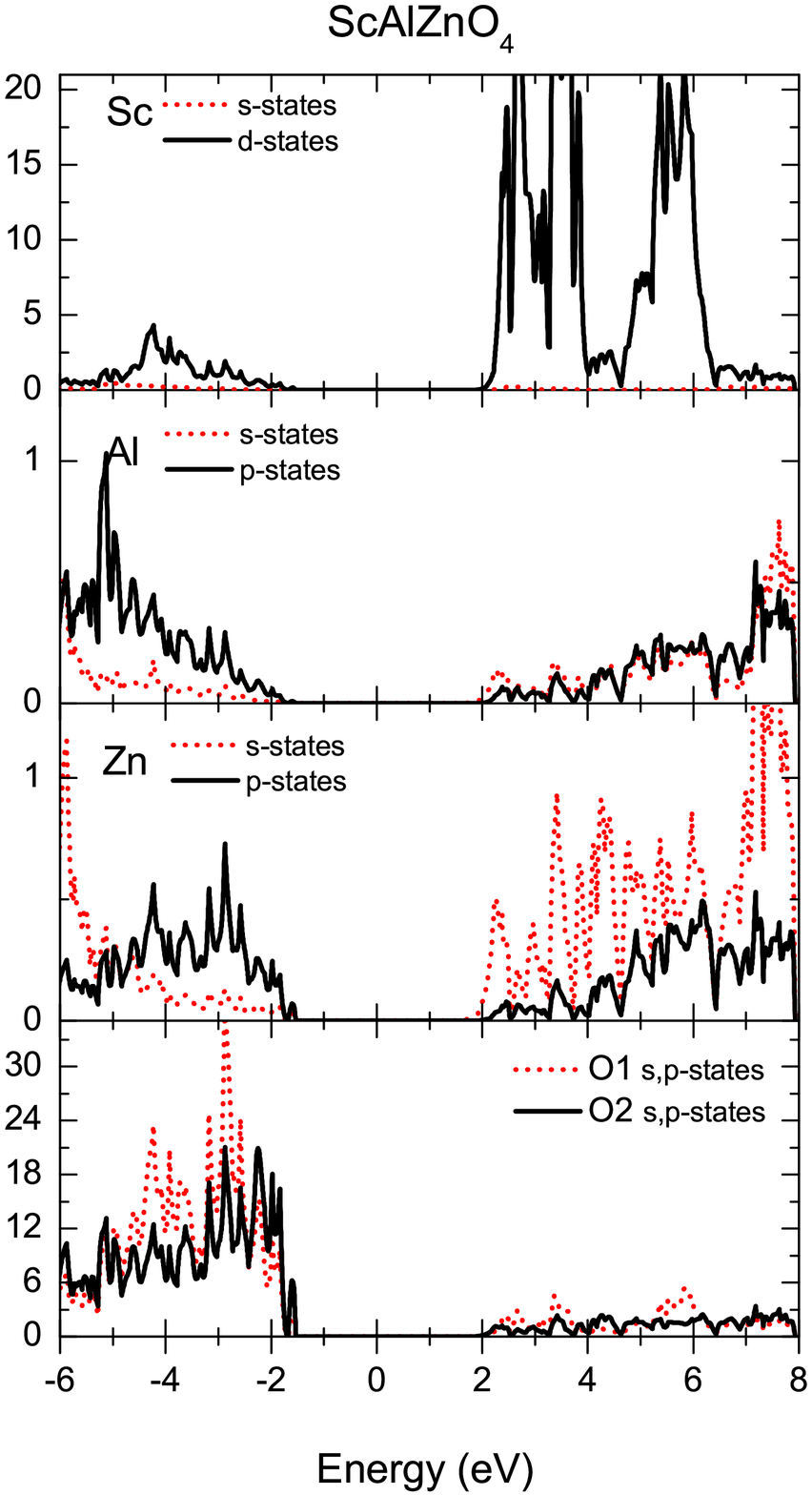}
\includegraphics[height=7cm]{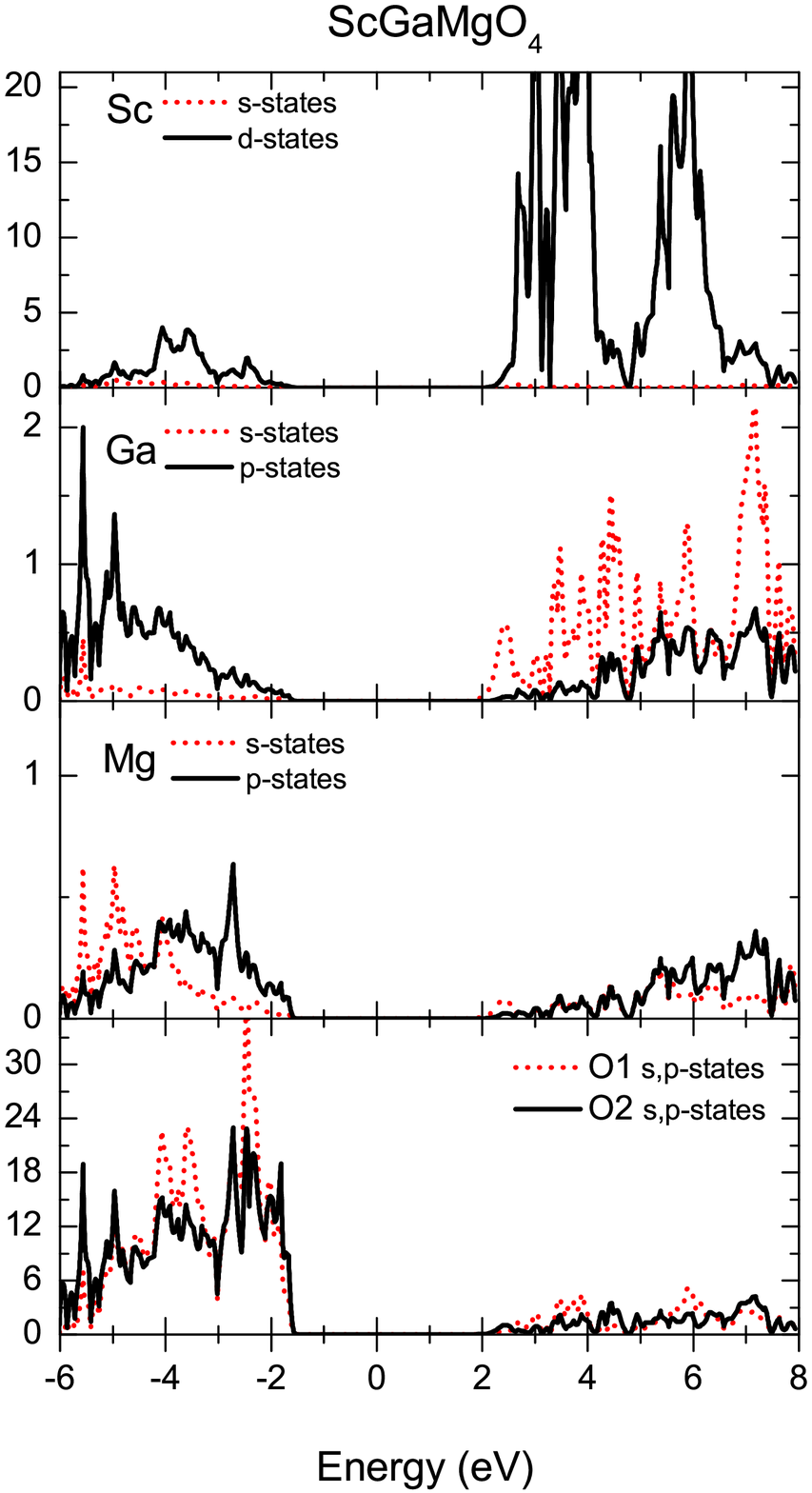}
\includegraphics[height=7cm]{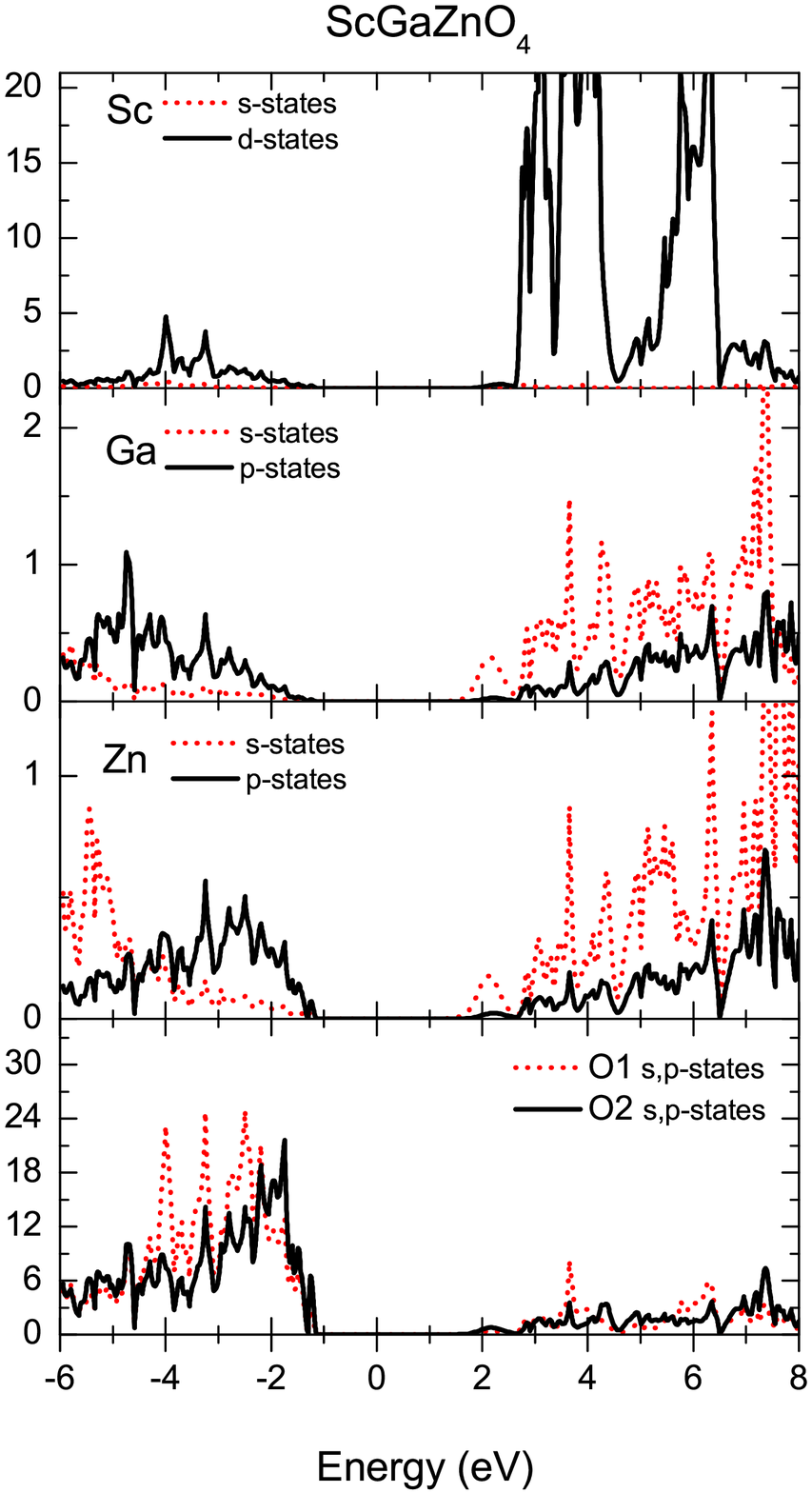}
\caption{Partial density of states in multicomponent RAMO$_4$ compounds as obtained from LDA calculations.}
\label{pdos}
\end{figure*}

In the valence band, both types of the oxygen atoms, O(1) and O(2), give comparable contributions, Fig. \ref{pdos}. However, at the very top of the valence band, the contributions from O(2), i.e., the oxygen that belongs to the AMO$_{2.5}$ double layer, are at least two times larger except for ScAlMgO$_4$ and ScGaMgO$_4$ where the oxygen contributions are similar. The average width of the valence band is about 6.4 eV for all compounds with the largest value of 7.5 eV obtained for ScGaZnO$_4$.

\begin{figure*}
\centering
\includegraphics[height=5.8cm]{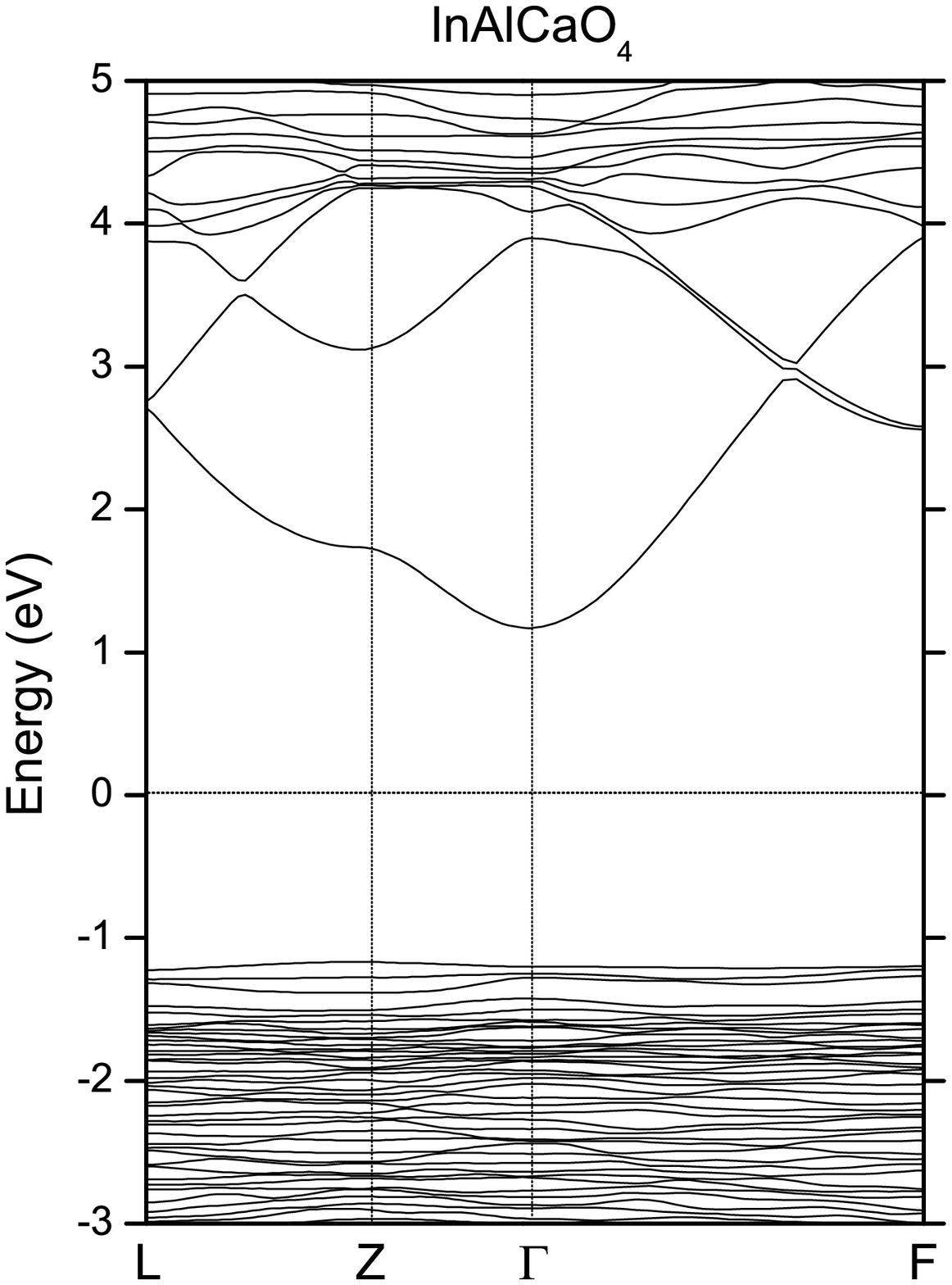}
\includegraphics[height=5.8cm]{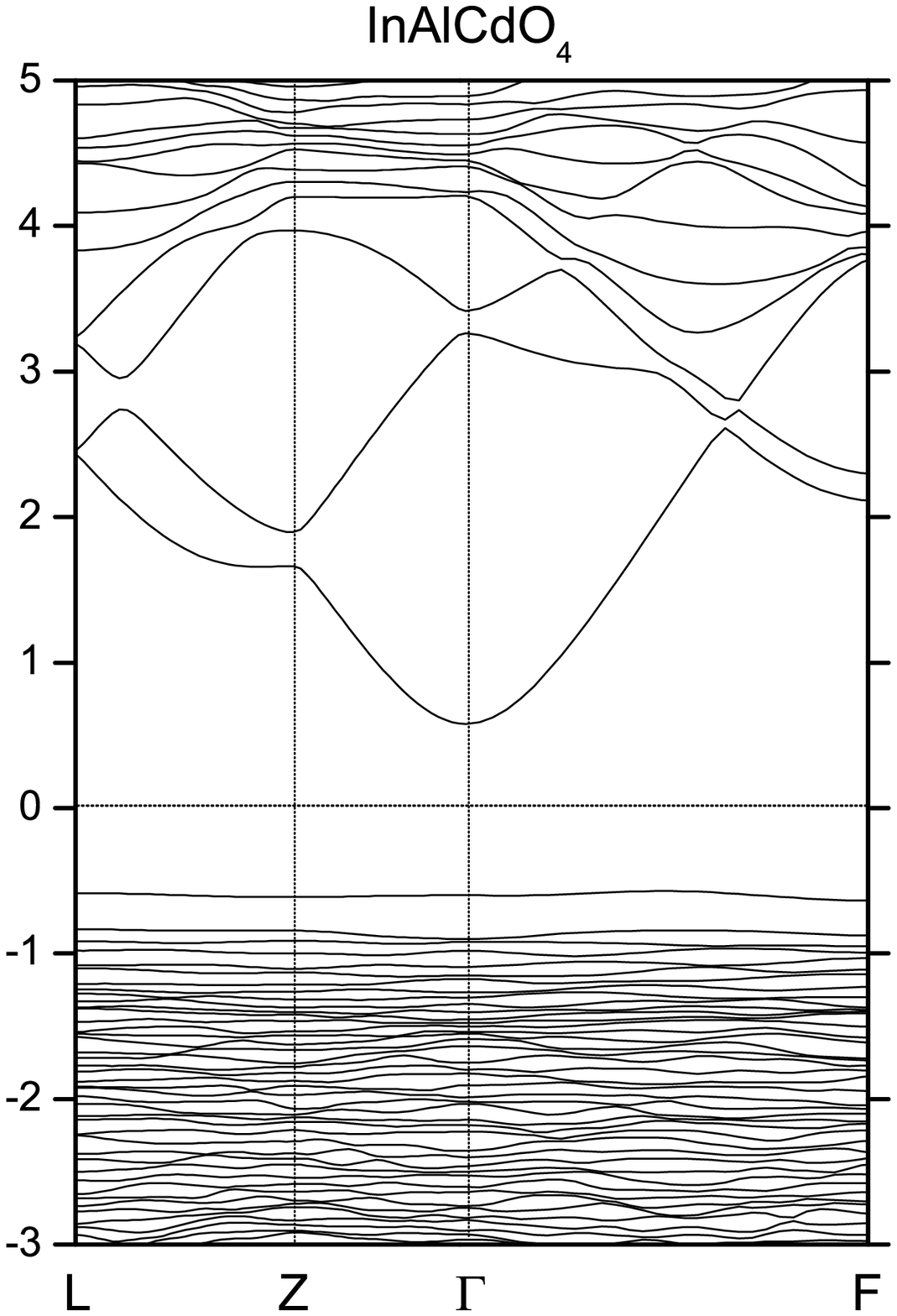}
\includegraphics[height=5.8cm]{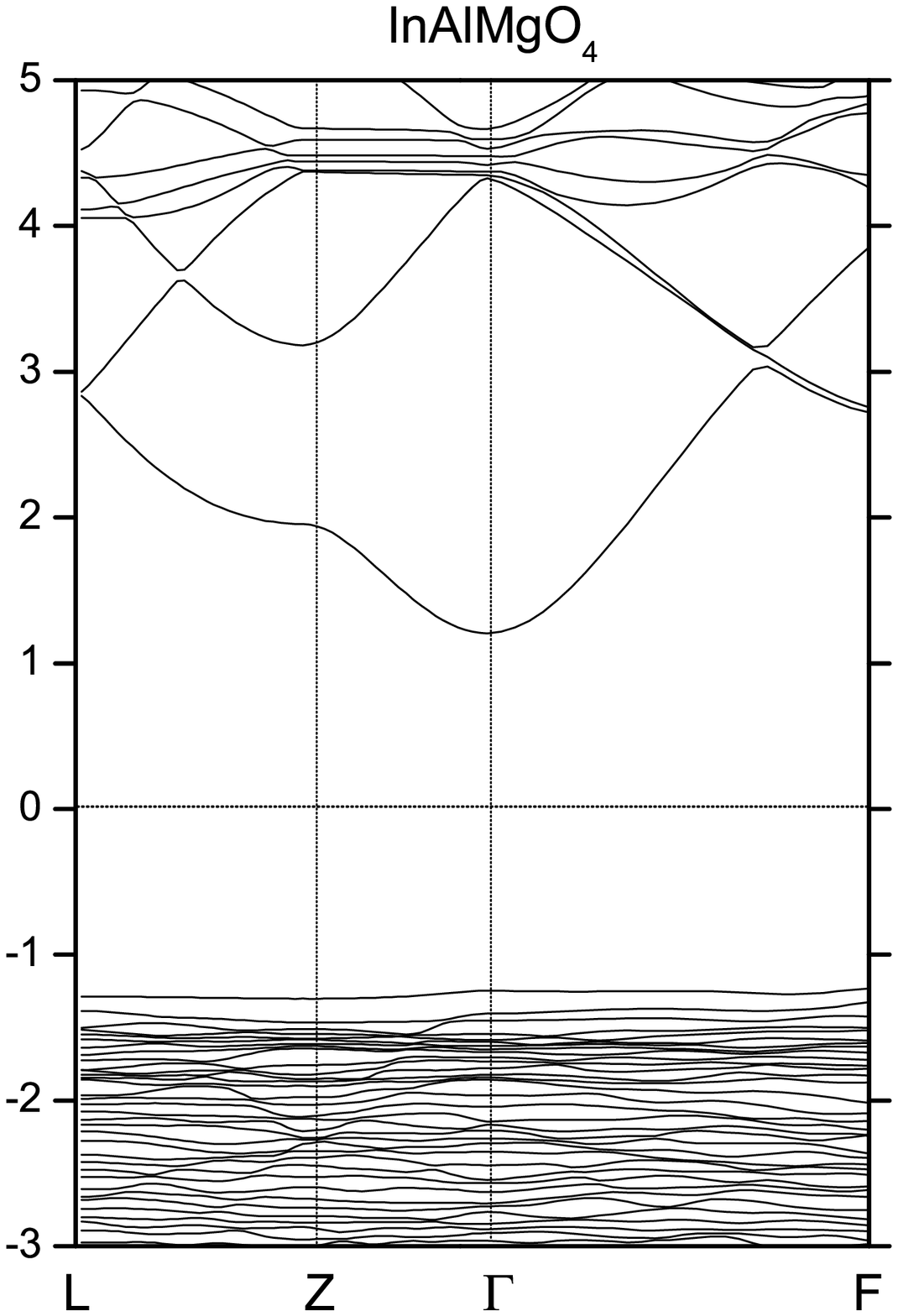}
\includegraphics[height=5.8cm]{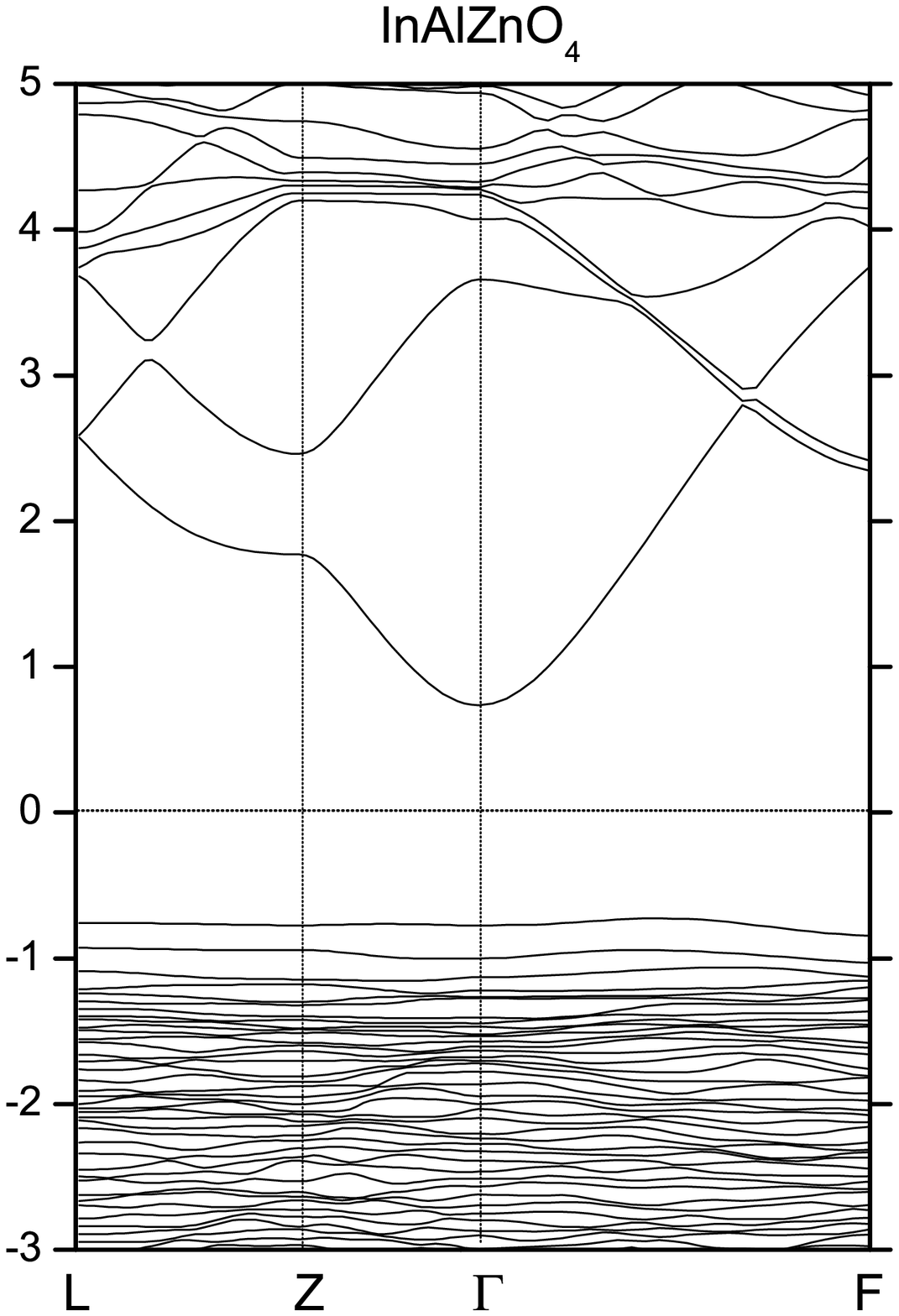}
\vspace{0.5cm}

\includegraphics[height=5.8cm]{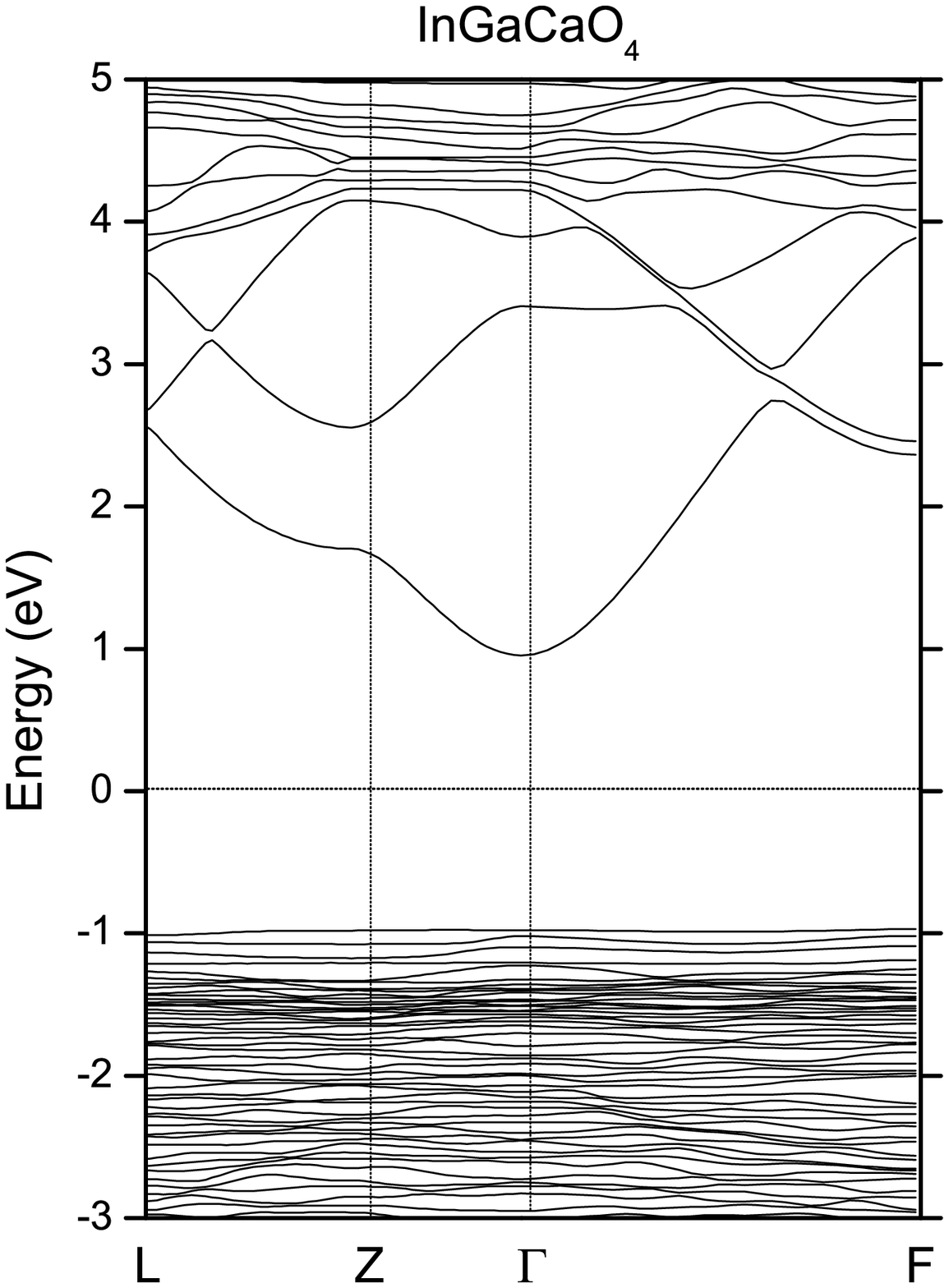}
\includegraphics[height=5.8cm]{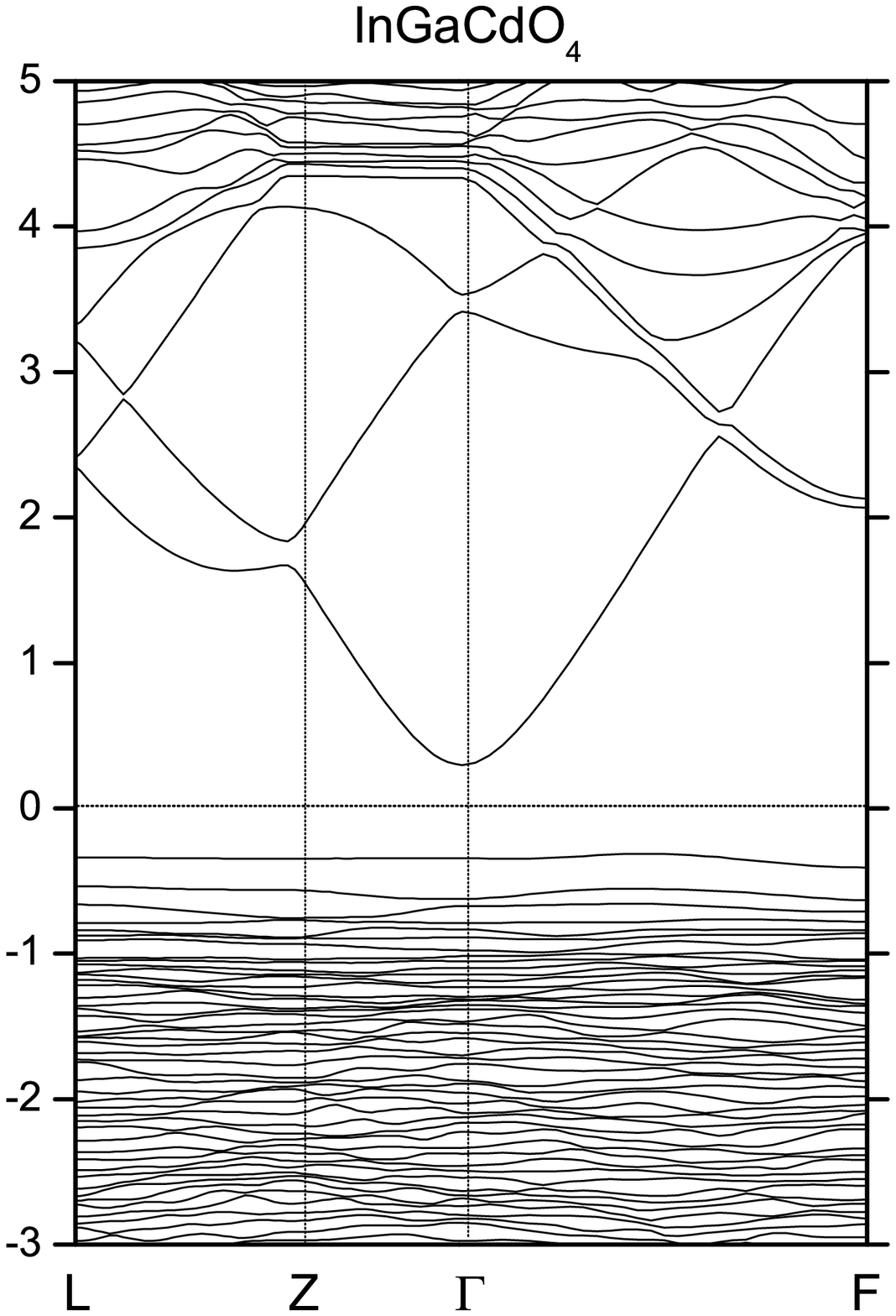}
\includegraphics[height=5.8cm]{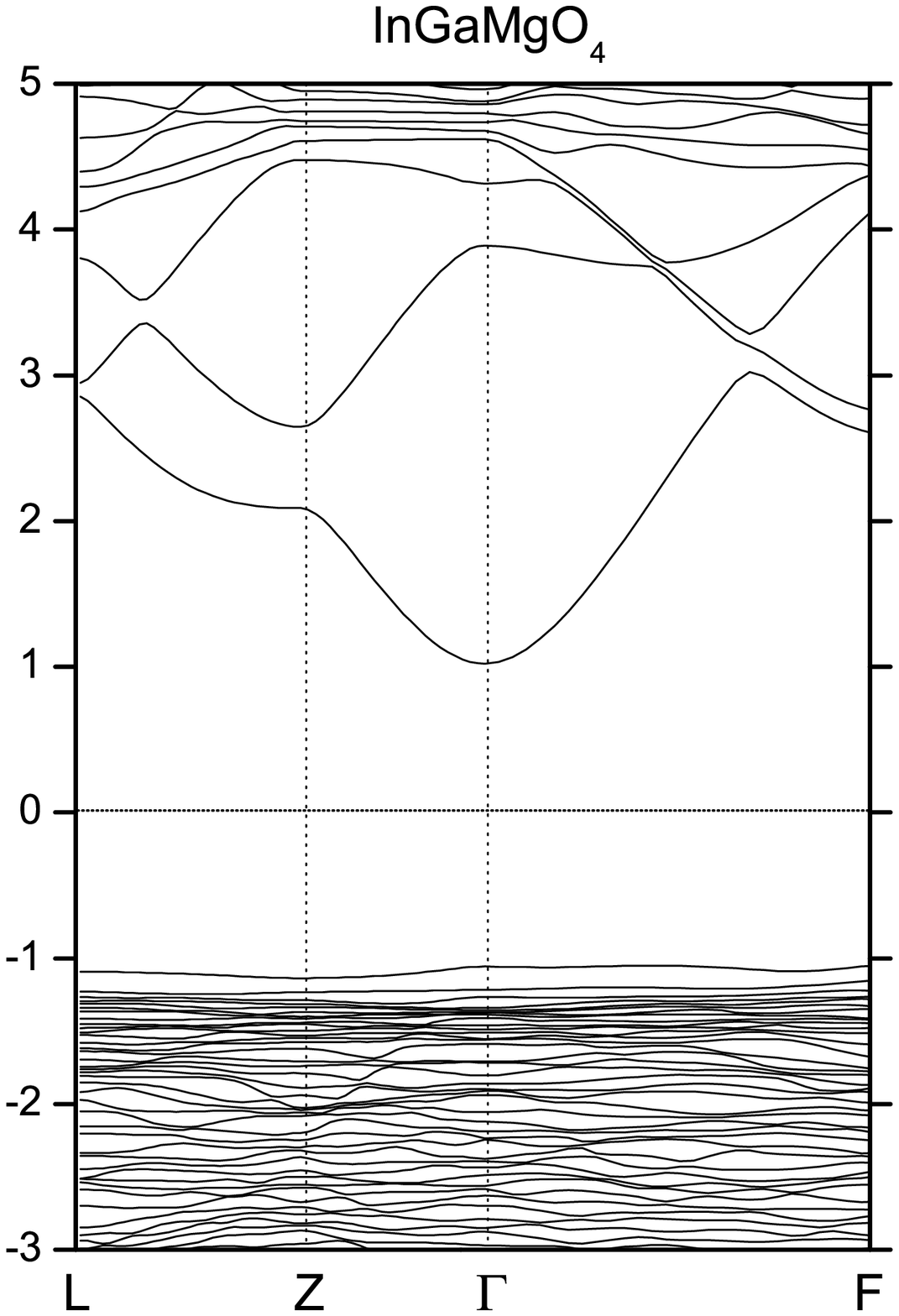}
\includegraphics[height=5.8cm]{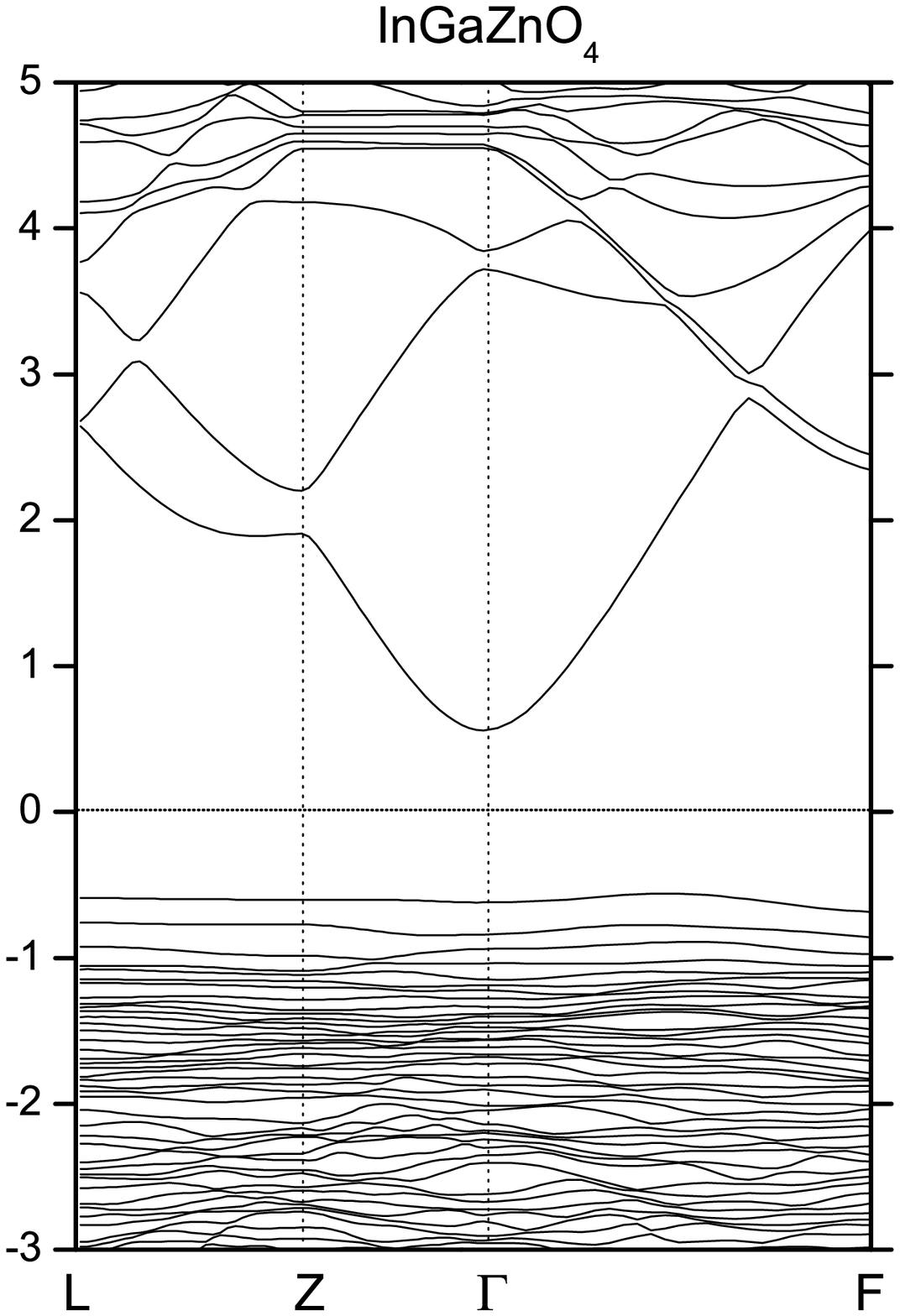}
\vspace{0.5cm}

\includegraphics[height=5.8cm]{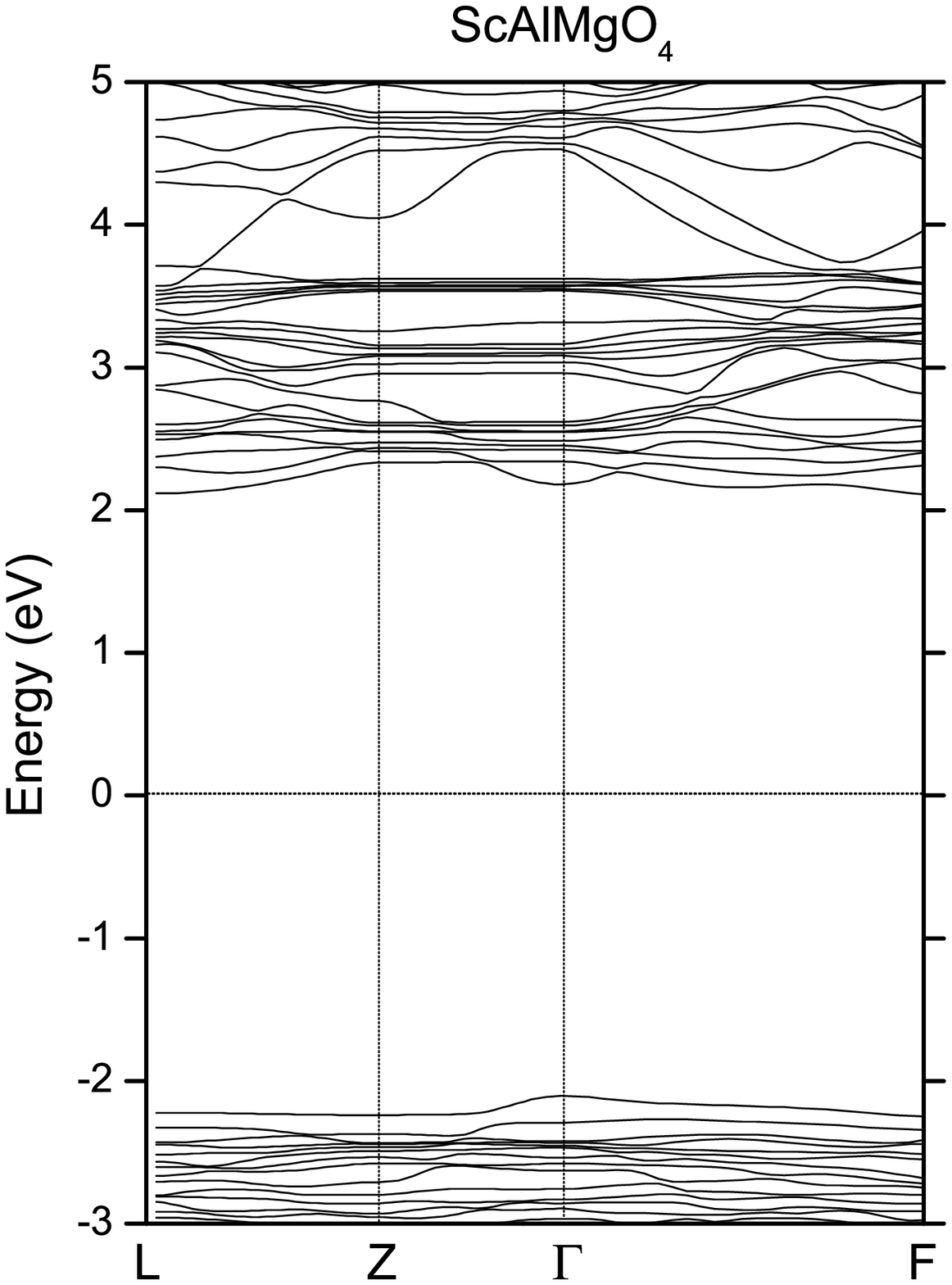}
\includegraphics[height=5.8cm]{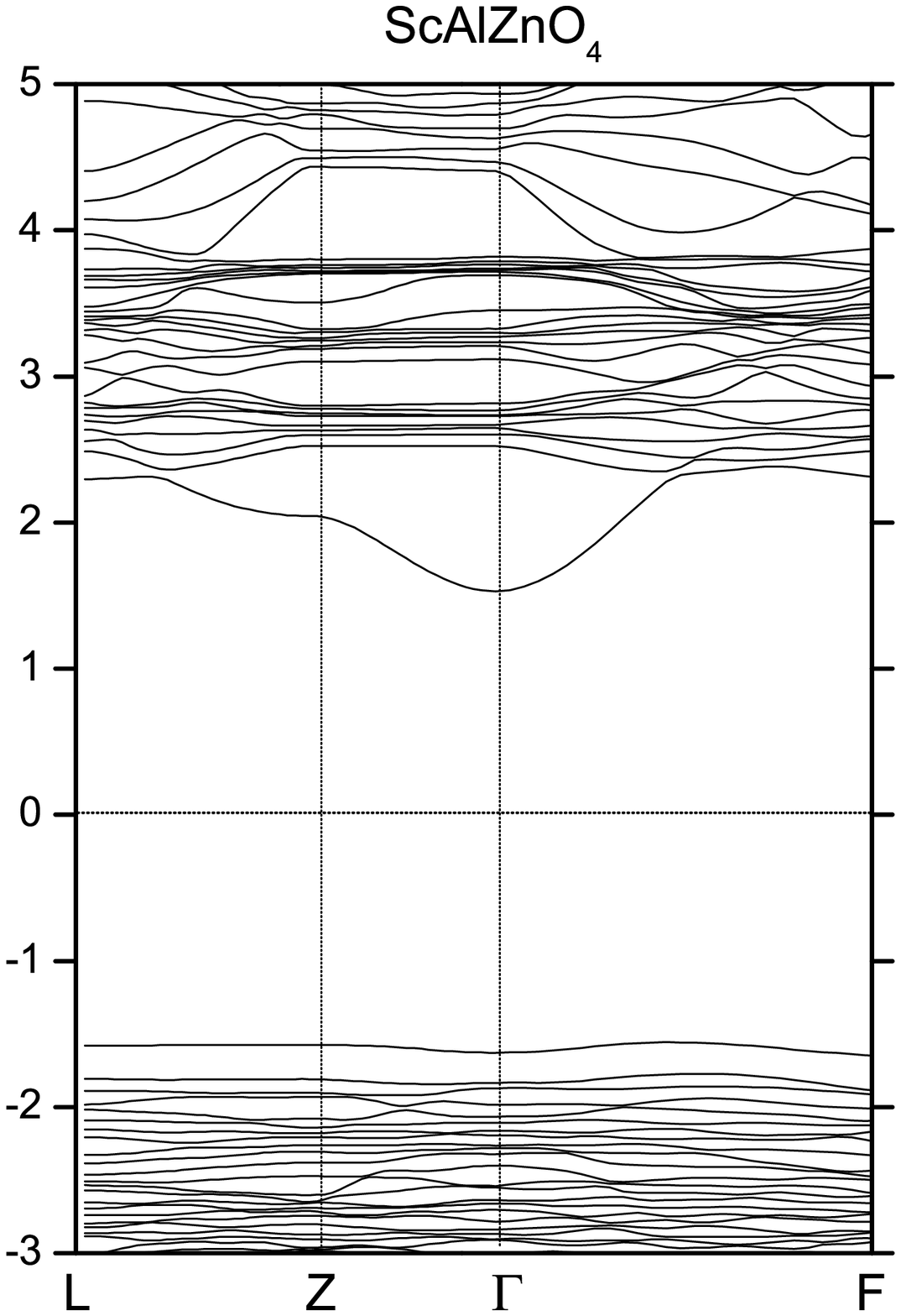}
\includegraphics[height=5.8cm]{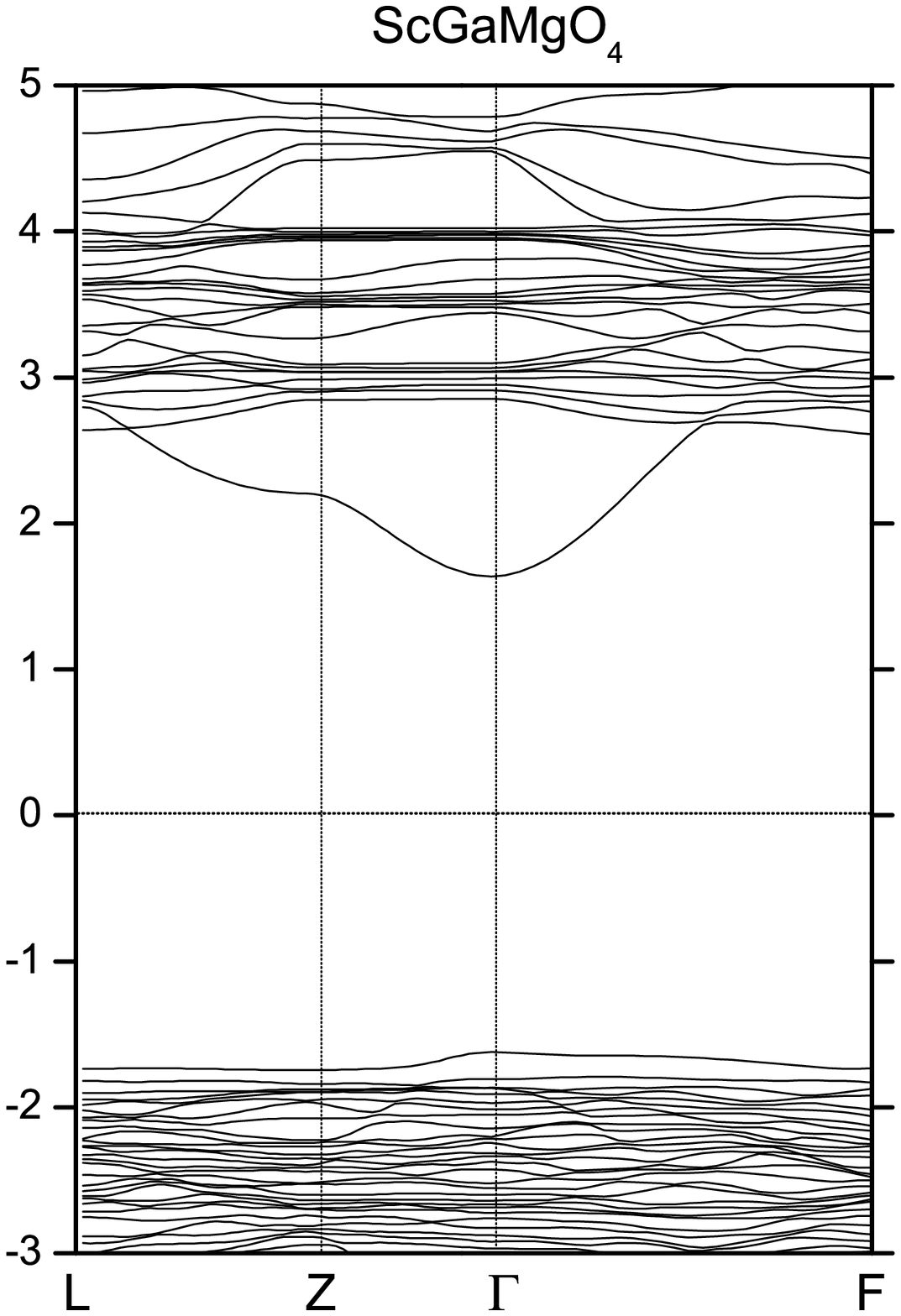}
\includegraphics[height=5.8cm]{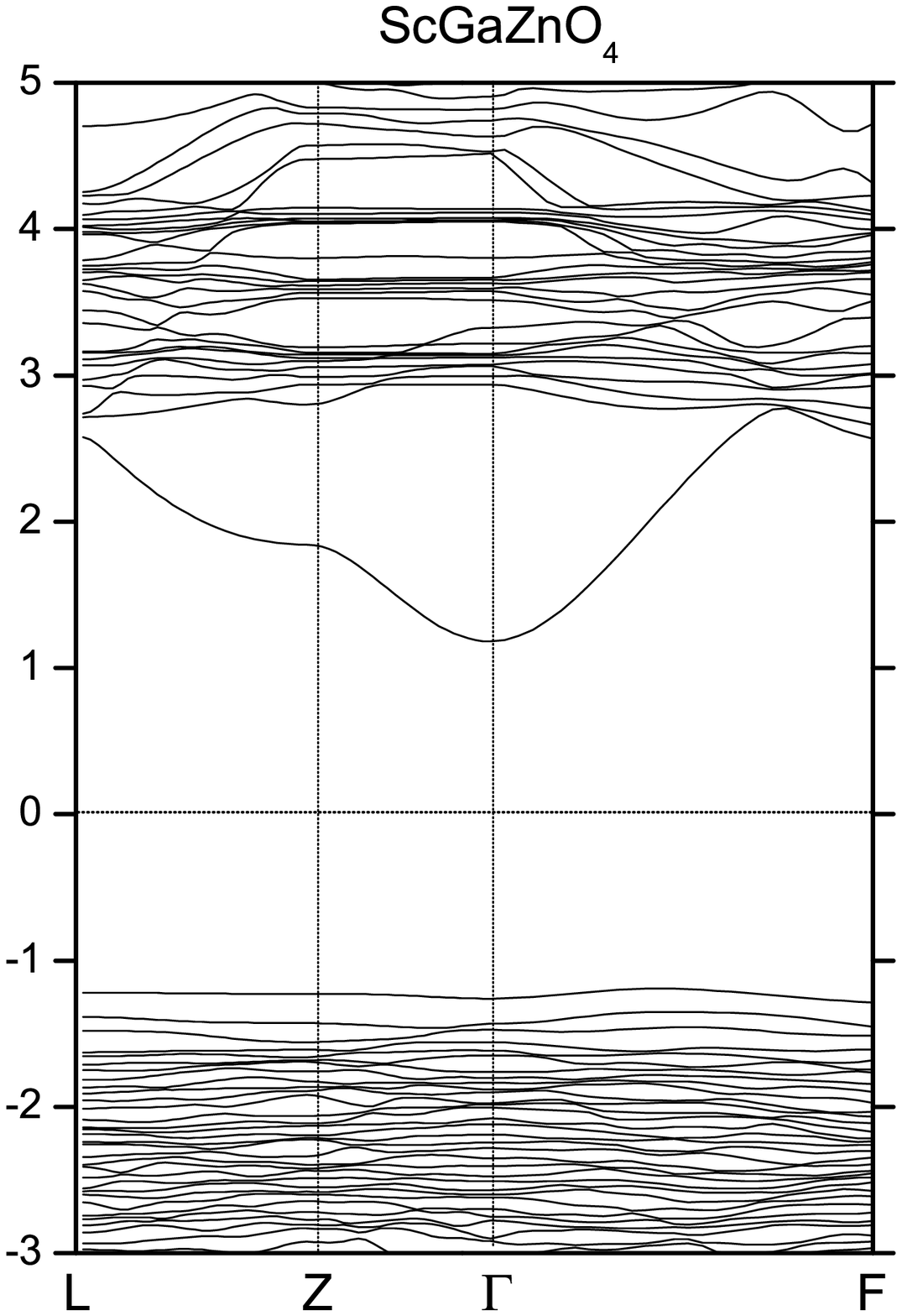}
\caption{Electronic band structure of twelve multicomponent RAMO$_4$ compounds as obtained from LDA calculations.}
\label{band-str}
\end{figure*}

Metal-oxygen interactions result in a band gap between the valence and the conduction bands which varies significantly with composition. From the LDA calculations, the smallest gap of 0.64 eV is found for InGaCdO$_4$, and the largest, 4.35 eV, for ScAlMgO$_4$, Table \ref{tband}. We also employ sX-LDA method to accurately calculate the band gap of the multicomponent oxides, and obtained the range of 2.45 eV to 6.29 eV, respectively.  We note that independent of the composition of the complex oxides, the sX-LDA band gap values are larger by about 2 eV (more precisely, by 1.7--2.5 eV) than the LDA values for all compounds investigated. 

\begin{table*}
\centering
\caption{LDA and sX-LDA calculated band gaps E$_g$ (in eV) in RAMO$_4$ compounds and the band gap averages obtained using the band gaps of the corresponding single-cation oxides in the ground state, $\langle$E$_g^g$$\rangle$, or the hypothetical phases, $\langle$E$_g^h$$\rangle$, cf., Table \ref{tsingleband}, with equal weights. In addition, weighted averages, $\langle$E$_g^g$$\rangle ^w$ and $\langle$E$_g^h$$\rangle ^w$, calculated based on the respective contributions of the cations to the bottom of the conduction band, Fig. \ref{fcontrib}, are given.}
\label{tband}
\begin{center}
\begin{tabular*}{15.3cm}{p{2.0cm}| p{1.2cm} p{1.2cm} p{1.2cm} p{1.2cm} p{1.2cm} | p{1.2cm} p{1.2cm} p{1.2cm}  p{1.2cm} p{1.2cm} p{1.2cm} }  \hline\hline
\multicolumn{1}{c}{} & \multicolumn{5}{c}{LDA} & \multicolumn{5}{c}{sX-LDA} \\ \hline 
RAMO$_4$    & E$_g$ &  $\langle$E$_g^g$$\rangle$ & $\langle$E$_g^g$$\rangle ^w$ & $\langle$E$_g^h$$\rangle$ & $\langle$E$_g^h$$\rangle ^w$ & 
E$_g$ &  $\langle$E$_g^g$$\rangle$ & $\langle$E$_g^g$$\rangle ^w$ & $\langle$E$_g^h$$\rangle$ & $\langle$E$_g^h$$\rangle ^w$  \\ \hline
InAlZnO$_4$ & 1.51 & 2.75 & 2.41 & 1.95 & 1.73 & 3.48 & 5.13 & 4.68 & 4.35 & 4.01\\
InAlCaO$_4$ & 2.37 & 3.63 & 3.02 & 2.74 & 2.21 & 4.87 & 5.98 & 5.21 & 5.31 & 4.54\\
InAlMgO$_4$ & 2.45 & 4.06 & 3.20 & 2.72 & 2.15 & 4.62 & 6.51 & 5.43 & 5.30 & 4.47\\
InAlCdO$_4$ & 1.18 & 2.31 & 1.88 & 1.57 & 1.32 & 2.87 & 4.16 & 3.62 & 3.47 & 3.11\\ \hline
InGaZnO$_4$ & 1.18 & 1.43 & 1.41 & 1.47 & 1.43 & 3.29 & 3.72 & 3.67 & 3.69 & 3.62\\ 
InGaCaO$_4$ & 1.93 & 2.31 & 2.10 & 2.26 & 2.02 & 4.08 & 4.57 & 4.28 & 4.65 & 4.29\\
InGaMgO$_4$ & 2.08 & 2.75 & 2.54 & 2.24 & 2.07 & 4.31 & 5.10 & 4.83 & 4.64 & 4.40\\
InGaCdO$_4$ & 0.64 & 0.99 & 0.85 & 1.09 & 0.99 & 2.45 & 2.75 & 2.54 & 2.81 & 2.65\\ \hline
ScGaZnO$_4$ & 2.44 & 2.26 & 1.93 & 2.39 & 2.10 & 4.45 & 4.78 & 4.47 & 4.81 & 4.53\\
ScAlZnO$_4$ & 3.16 & 3.58 & 3.00 & 2.87 & 2.49 & 5.52 & 6.18 & 5.61 & 5.47 & 5.08\\
ScGaMgO$_4$ & 3.26 & 3.58 & 3.55 & 3.16 & 3.09 & 5.76 & 6.16 & 6.15 & 5.77 & 5.73\\
ScAlMgO$_4$ & 4.35 & 4.90 & 4.12 & 3.64 & 3.64 & 6.29 & 7.56 & 6.60 & 6.43 & 6.13\\ \hline\hline
\end{tabular*}
\end{center}
\end{table*}

The band gaps of multicomponent oxides seem to follow the general trend expected from the band gap values of the basis oxide constituents, i.e., the incorporation of ligher metals results in a band gap increase. However, the increase is not the same in otherwise similar compounds: for example, when Ga is replaced by Al in InAMO$_4$ compounds, the gap does not increase by the same amount for the four compounds, i.e., those with M=Zn, Ca, Mg, or Cd. Rather, the increase is about 0.2 eV, 0.8 eV, 0.3 eV, or 0.4 eV, respectively, Table \ref{tband}, as obtained within sX-LDA calculations.
A thorough analysis of the obtained trends in the band gap values and a comparison with those in the corresponding basis oxides allow us to make the following important conclusions:

(i) The band gap in a multicomponent oxide is not governed by the smallest-gap basis oxide constituent. For example, for two Cd-containing complex oxides, the sX-LDA band gaps are 2.5 eV and 2.9 eV which are larger than the CdO band gap, Table \ref{tsingleband}. For InAMO$_4$ compounds excluding those with Cd, the band gap values vary from 3.3 eV to 4.9 eV, Table \ref{tband} -- despite the fact that In$_2$O$_3$ has the band gap of 2.90 eV (from sX-LDA), Table \ref{tsingleband}. 

(ii) The band gap in the multicomponent oxides is affected by the presence of {\it all} oxide constituents disregarding the differences in the band gaps of the basis oxides. In other words, not only the post-transition metal oxides (smaller-gap constituents) but also the light metal oxides (large-gap constituents) contribute to the formation of the band gap (for example, compare the band gaps of InGaMO$_4$ with M=Cd, Zn, Ca or Mg, or other sets of compounds). This arises from the close interaction between the alternating cations via shared oxygen atoms in mixed A and M or neighbor R-layers, and points to a hybrid nature of the conduction band, as discussed in the next section.

(iii) An equal-weight average, $\langle$E$_g^g\rangle$, over the band gaps of the basis oxides in their ground state phases (c.f., Table \ref{tsingleband}) correlates with the calculated band gaps for corresponding multicomponent oxides but gives significantly overestimated values in most cases (one exception is ScGaZnO$_4$ where the LDA calculated band gap is greater than the one obtained via averaging), Table \ref{tband}. 

(iv) An equal-weight average, $\langle$E$_g^h\rangle$, over the band gaps of the basis oxides in the hypothetical phases (c.f., Table \ref{tsingleband}) provides a better guess but still overestimates the value of the band gap in multicomponent oxides, Table \ref{tband}. 

(v) Weighted average over the band gaps of the basis oxides (in either the ground state phase or the hypothetical phase) with weights taken as the percent contributions from the cations states to the lowest conduction band wave-function at the $\Gamma$ point yields underestimated band gap values with respect to those calculated for multicomponent oxides (these values are not given in the Table \ref{tband}). For the RAMO$_4$ compounds with two or more light metal oxide constituents, the underestimation is significant, of $\sim$30\%. This suggests that the states located above the conduction band minimum (such as the states of the light metals) play an important role and must be taken into account.

(vi) Weighted average, $\langle$E$_g^h\rangle ^w$, over the band gaps of the basis oxides in the hypothetical phase with weights taken as the relative cation contributions to the conduction band within an energy range, Fig. \ref{fcontrib}, provides a closest match to the calculated band gap values in multicomponent oxides, Table \ref{tband}. The energy range at the bottom of the conduction band which is used to determine the cations contributions, represents the Fermi energy displacement, or the so-called Burstein-Moss (BM) shift, which corresponds to an extra electron concentration of 1$\times$10$^{21}$ cm$^{-3}$ in each compound. 
Due to the high energy dispersion of the conduction band in InAMO$_4$ compounds, Fig. \ref{band-str}, the BM shift is large, of 1.0-1.5 eV. In ScAMO$_4$, the presence of the Sc $d$-states near the bottom of the conduction band result in a high density of states, and hence, the BM shift is significantly smaller, e.g., 0.05 eV for ScAlMgO$_4$ and $\sim$0.7 eV for ScAlZnO$_4$ and ScGaMgO$_4$.

Thus, the local atomic structure in multicomponent oxides which differs from that of the basis oxides in the ground state (see sections III and IV), plays an important role in determining the resulting electronic properties and must be taken into account for accurate predictions.
We note here that an improved agreement between the calculated and the averaged band gaps is expected when the metal-oxygen distances in the hypothetical oxide phases closely correspond to the distances in particular multicomponent oxide, Table \ref{tstr}. In our calculations for the hypothetical single-cation phases we used the metal-oxygen distances averaged over all RAMO$_4$ for each particular metal, $\langle$D$\rangle$ in Table \ref{tcomp}, while the actual distances in each RAMO$_4$ may differ essentially, c.f., deviations of the ranges in Table \ref{tcomp}. For example, the 15 \% overestimation of the band gap average in InAlZnO$_4$ is due to the fact that the Al-O and Zn-O distances in this compound, $\langle$D$_{Zn-O}^{ab}$$\rangle$=2.05 \AA, $\langle$D$_{Zn-O}^{c}$$\rangle$=2.00 \AA, $\langle$D$_{Al-O}^{ab}$$\rangle$=1.84 \AA \, and $\langle$D$_{Al-O}^{c}$$\rangle$=1.84 \AA, Table \ref{tstr}, are larger than those in the hypothetical ZnO phase, $\langle$D$_{Al-O}^{ab}$$\rangle$=2.00 \AA \, and $\langle$D$_{Zn-O}^{c}$$\rangle$=1.98 \AA, Table \ref{t-phases}, and the hypothetical $P6_1$ phase of Al$_2$O$_3$, $\langle$D$_{Al-O}^{ab}$$\rangle$=1.825 \AA \, and $\langle$D$_{Al-O}^{c}$$\rangle$=1.81 \AA, Table \ref{t-phases2}. Increased distances in the hypothetical oxide phases will result in smaller band gaps for these compounds, bringing the average band gap closer to the calculated one in InAlZnO$_4$. Conversely, the 7\% underestimation of the band gap average in InAlCaO$_4$ is due to the smaller Al-O distances in the multicomponent oxide ($\langle$D$_{Al-O}^{ab}$$\rangle$=1.77 \AA \, and $\langle$D$_{Al-O}^{c}$$\rangle$=1.78 \AA, Table \ref{tstr}) as compared to those in the hypothetical Al$_2$O$_3$, Table \ref{t-phases2}.

\subsubsection{B. Nature of the conduction band in RAMO$_4$}

The nature of the conduction band in a complex TCO host is of primary interest since the charge transport in degenerately doped material will occur through 
the states which form the conduction band. One of the reasons that the oxides of homologues series (In,Ga)$_2$O$_3$(ZnO)$_m$, $m$=integer, have attracted wide attention was a common assumption that the conduction band in these complex oxides is formed from the In $s$-states. Based on this assumption it was  suggested that these layered materials offer a possibility to spatially separate carrier donors located within non-conducting layers and the conducting layers which transfer the carriers effectively, i.e., without charge scattering on the impurities, that would lead to an increased conductivity\cite{spinel-review}. 

From the density of states (DOS) plots, Fig. \ref{pdos}, it may appear that the In states solely govern the conduction band in all InAMO$_4$ compounds. However, analysis of the DOS plots alone may provide a misleading picture of the nature of the conduction bands for three reasons.
First, due to the high energy dispersion at the bottom of the conduction band in the oxides under consideration, the corresponding density of states is small. This tail in the DOS should not be neglected.
Second, one should compare the relative contributions from different atoms within a rather narrow energy range at the bottom of the conduction band which corresponds to a Fermi level displacement associated with introduction of a particular electron concentration upon degenerate doping of the material. Usually, the extra electron concentrations are of the order of 10$^{19}-$10$^{21}$ cm$^{-3}$. Third, the partial DOS is commonly calculated within thse muffin-tin spheres and, therefore, the interstitial region which may give a significant contribution owing to the spatial distribution of the metal $s$-orbitals, is not taken into account. 
\begin{figure*}
\includegraphics[height=4.2cm]{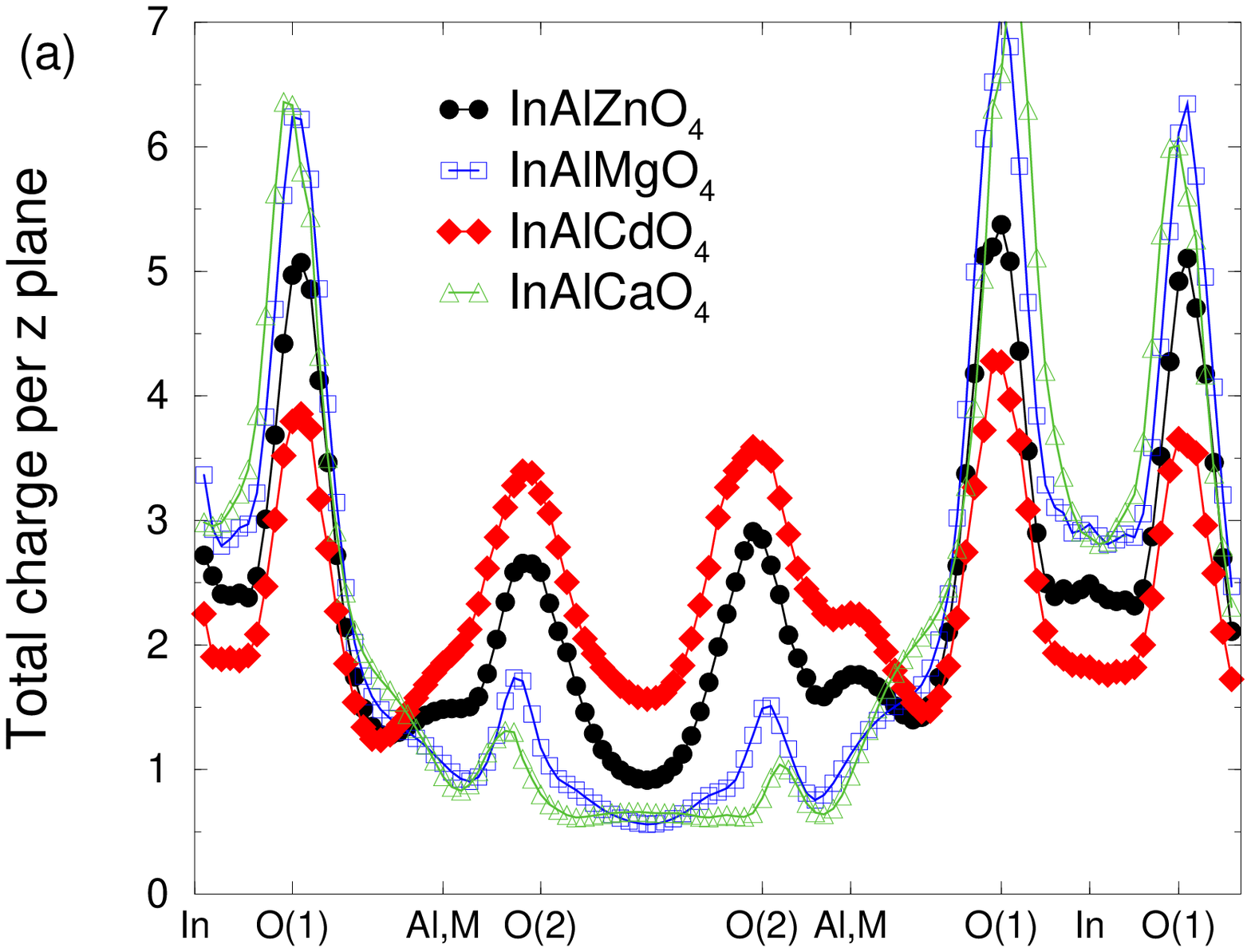}
\includegraphics[height=4.2cm]{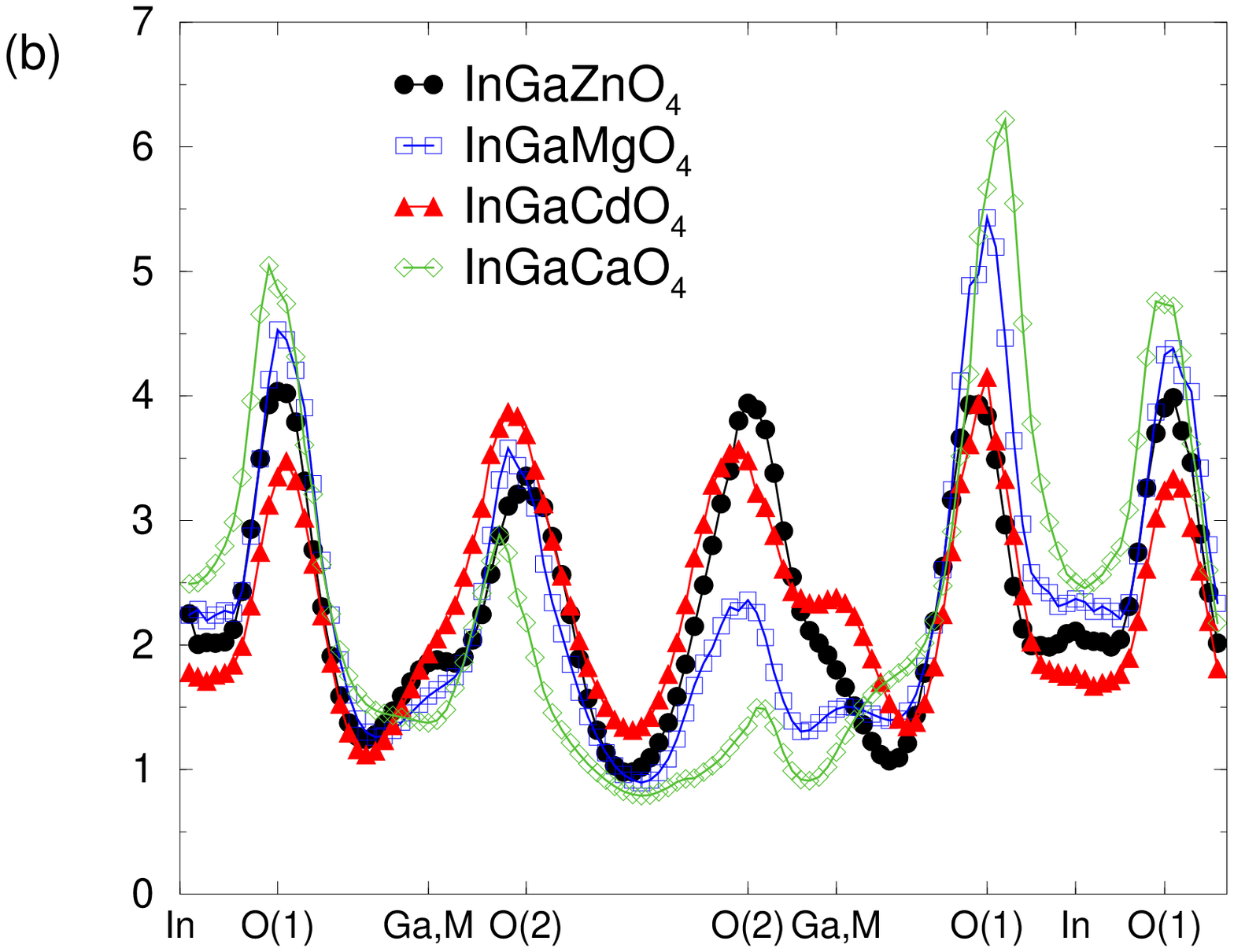}
\includegraphics[height=4.2cm]{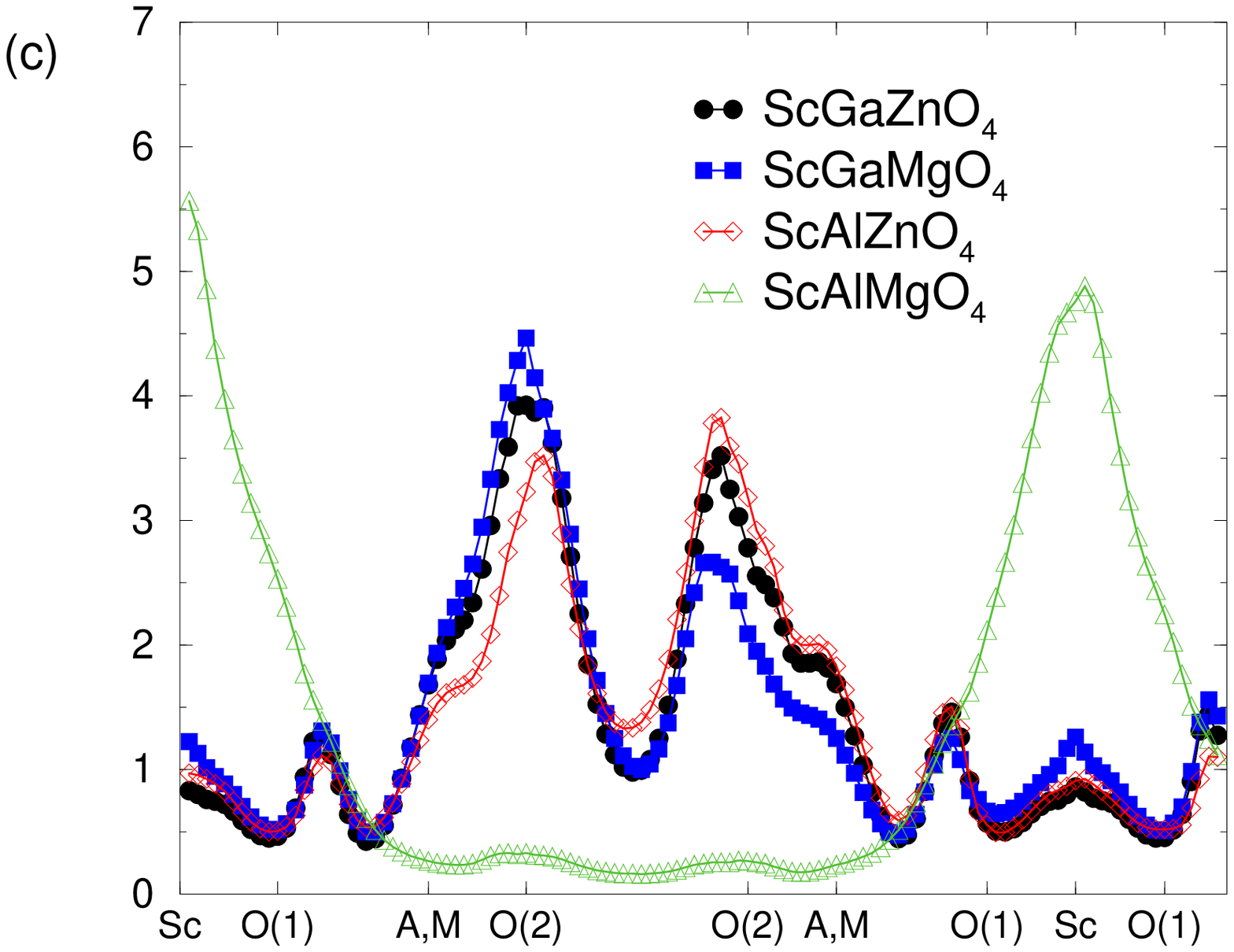}
\caption{(Color online) Total charge densities calculated within one unit cell and at the bottom of the conduction band for the energy window that represents $\sim$1x10$^{21}$ cm$^{-3}$ extra electrons in each RAMO$_4$.}
\label{fcontrib}
\end{figure*}

To obtain a more reliable description of the conduction states in multicomponent oxides, we calculated the charge density distribution within an energy range at the bottom of the conduction band. For each RAMO$_4$ compound, the energy range was chosen to correspond to an extra electron concentration of 1.0-1.3$\times$10$^{21}$ cm$^{-3}$. The resulting Fermi energy displacement depends on the density of states at the bottom of the conduction band: a small density of states (i.e., high energy dispersion of the conduction band bottom) leads to a pronounced E$_F$ shift, while  the Fermi level rises slow with electron concentration in the case of a large density of states. Specifically, we find that in InAMO$_4$ compounds, Fig. \ref{band-str}, the E$_F$ shift is large: it is 1.5 eV for InAlCdO$_4$, 0.9-1.0 eV for InACaO$_4$, and 1.1-1.3 eV for all other InAMO$_4$ compounds. In ScAMO$_4$, the presence of the Sc $d$-states near the bottom of the conduction band result in a high density of states, and hence, the E$_F$ shift is significantly smaller, namely, 0.05 eV for ScAlMgO$_4$, $\sim$0.7 eV for ScAlZnO$_4$ and ScGaMgO$_4$, and 0.9 for ScGaZnO$_4$.

The charge density distributions calculated within the specified energy ranges are obtained for the full conventional unit cell of RAMO$_4$ to include both layers, RO$_{1.5}$ and AMO$_{2.5}$, and the interstitial region between the layers. We summed up the charge within each [0001] plane, Fig. \ref{fcontrib}, in order to compare the contributions from the two structurally and chemically different layers. We found that:

(1) Different layer contributions to the conduction band are nearly identical in InGaZnO$_4$, InGaCdO$_4$, and InAlCdO$_4$. Hence, both layers, InO$_{1.5}$ and AMO$_{2.5}$, are expected to participate in the charge transport once degenerate doping is achieved.

(2) In InAlZnO$_4$, InGaMgO$_4$, and InGaCaO$_4$, contributions from the In-O layer are larger, yet comparable to those from the A-M-O layers. Together with the compounds in the above case (1), these oxides possess two post-transition metals (In, Zn, Cd and/or Ga) and one light metal cation (Al, Mg or Ca). These results suggest that the AMO$_{2.5}$ layers where post-transition and light metals are mixed, will serve as conducting path for extra electrons in degenerately doped materials.

(3) If the AMO$_{2.5}$ layer consists of two light metal cations, as in InAlMgO$_4$ or InAlCaO$_4$, its contribution to the charge density is low, yet it is not zero as, for example, in ScAlMgO$_4$, Fig. \ref{fcontrib}(c). Similarly, the Sc-O layer contributions are negligible if the AMO$_{2.5}$ layer contains one or two post-transition metals, as in ScGaZnO$_4$, ScGaMgO$_4$, or ScAlZnO$_4$.

(4) In ScAlMgO$_4$, the Al-Mg-O layers have zero contributions, while the charge is localized within the Sc-O layer. Hence, if extra electrons are introduced, the AMO$_{2.5}$ layers would be non-conducting.

Thus, despite well-defined crystal lattice anisotropy and presence of a light metal cation in the AMO$_{2.5}$ layer, several RAMO$_4$ compounds are capable of giving rise to a nearly isotropic conductivity (i.e., within and across the structural layers) when properly doped. The role of light metal cations in carrier generation in these multicomponent oxides, i.e., the effect of these cations on the formation of native electron-donor and electron-``killer'' defects, should be investigated further.

\subsubsection{C. Role of atomic coordination on the conduction states in RAMO$_4$}

As mentioned in the introduction, the proximity of the cations empty $p$- or $d$-states to the bottom of the conduction band may help predict the degree of electron localization in the oxides upon doping. Specifically, it was found \cite{JMbook} that in oxides of light metals, such as Ga$_2$O$_3$, CaO, Al$_2$O$_3$, or MgO, the Ga, Mg or Al $p$-states or Ca $d$-states are energetically compatible with the $s$-states of cations in the conduction band. Upon electron doping, extra charge becomes trapped on the anisotropic $p$ or $d$-orbitals which form strong covalent metal-oxygen bonds around defect, leading to the charge confinement (known as a color or F center). Now, we want to determine the energy location of the detrimental $p$- or $d$-states of cations in the conduction band of multicomponent oxides. Our goal is to understand how the $p$- or $d$-states location with respect to the conduction band bottom is affected by the local atomic coordination, i.e., the five-fold coordination in RAMO$_4$ vs the four- or six-fold coordinations in the ground state structures of basis oxides.

\begin{figure}
\includegraphics[height=6cm]{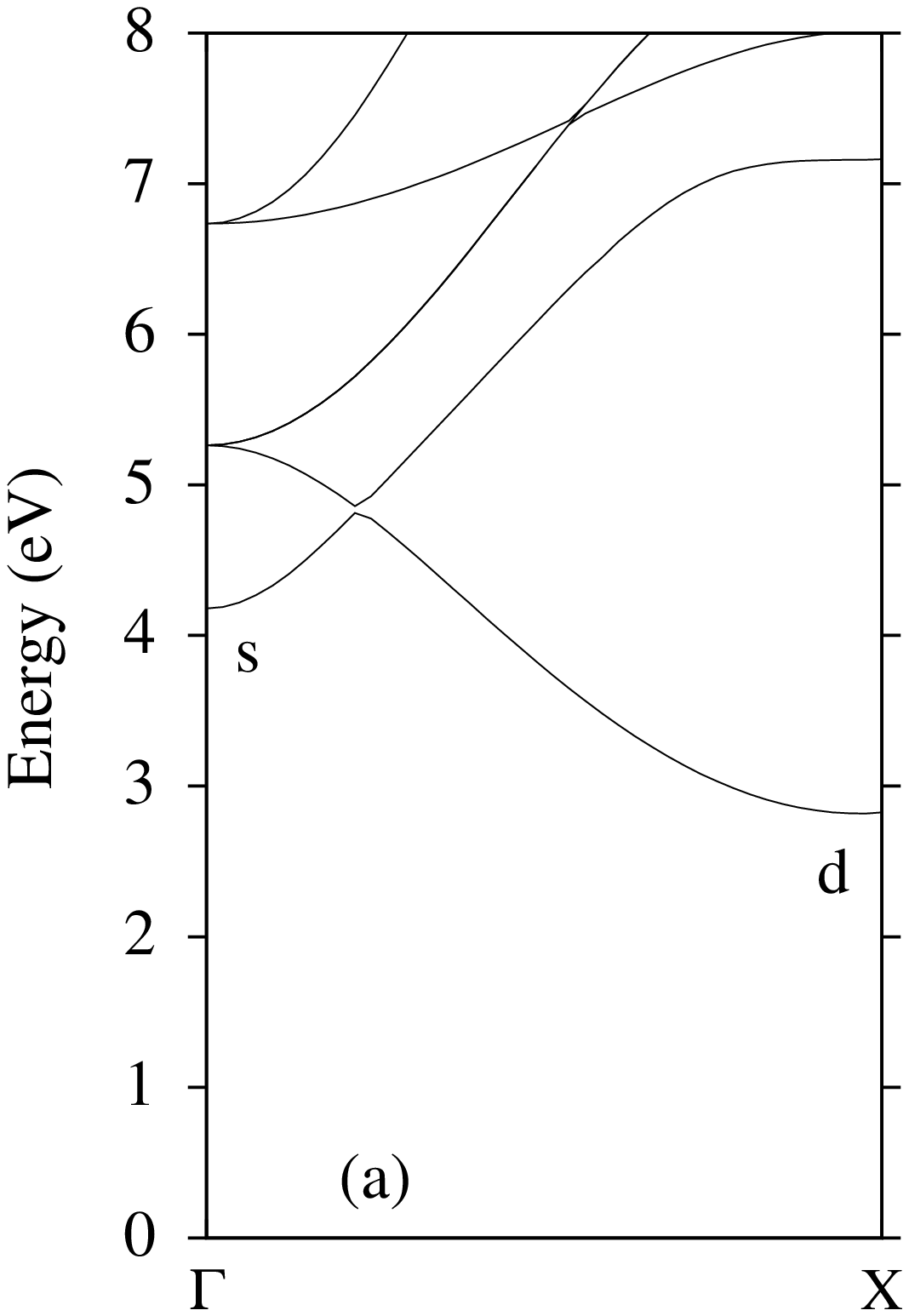}
\includegraphics[height=6cm]{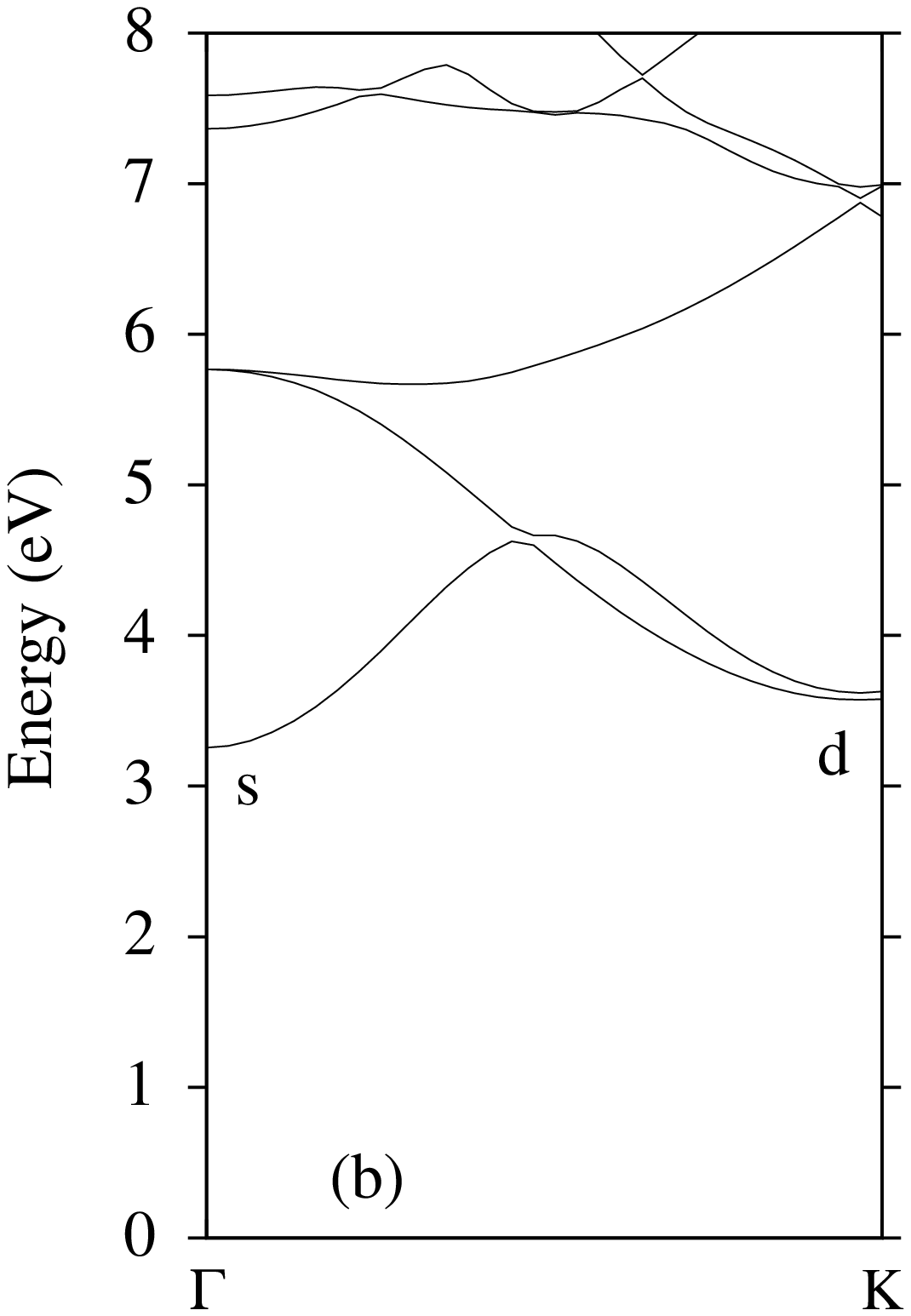}
\caption{Electronic band structure of (a) rocksalt CaO, and (b) hypothetical wurtzite CaO with the interatomic distances matching those in RACaO$_4$. Only the bottom of the conduction band is shown. The calculations are performed within sX-LDA.}
\label{fcao}
\end{figure}

First, we find that the local structural variations significantly affect the conduction bands of oxides -- in addition to the band gap value discussed in the section V.A above. Specifically, in rocksalt CaO with six-fold atomic coordination, the charge-trapping $d$-states of Ca govern the bottom of the conduction band being about 1.2 eV below the Ca $s$ states, Fig. \ref{fcao}(a). In marked contrast to the ground-state CaO, we find that in hypothetical wurtzite CaO with five-fold coordinated Ca, the Ca $d$-states are pushed into the conduction band and are {\it above} the $s$-states, resulting in a direct band gap, Fig. \ref{fcao}(b). This occurs since the octahedral symmetry favors strong directional interaction between the $d$ states of Ca and the $p$ states of oxygen neighbors, whereas the $s$-$p$ interaction is prefered when the symmetry is broken, as in five-fold coordinated Ca. Therefore, low-symmetry coordination helps diminish the detrimental effect of the anisotropic $d$-states on the oxide transport properties by promoting the $s$-character of the bottom of the conduction band. 

Further, from the calculated density of states for InAlCaO$_4$ or InGaCaO$_4$, Fig. \ref{pdos}, we find that the Ca $d$-states are well above the bottom of the conduction band formed from the $s$-states of the constituent cations. Similarly, the empty $p$ band of Al, Mg or Ga in RAMO$_4$ are located at a higher energy, i.e., deep inside the conduction band. We conclude that not only the unusual five-fold coordination of the A and M cations but also the hybridization between the spatially extended $s$ states of the cations (via shared oxygen atoms) are the reasons for a deeper cation's $p$ and $d$ bands in RAMO$_4$. Because of the interaction of cations (e.g., in the mixed AMO$_{2.5}$ layers) and due to the difference in the band gaps of the constituent oxides, namely, 2.3-3.4 eV in CdO, In$_2$O$_3$, or ZnO, and 7-9 eV in CaO, MgO, or Al$_2$O$_3$, the bottom of the hybrid $s$-like conduction band of complex oxides is {\it driven away} from the Ga, Al, and Mg $p$ states or Ca $d$ states. The fact that the Ga, Al, Mg, or Ca atoms do contribute their states (which are the $s$-states) to the conduction band bottom is clearly illustrated by the calculated charge densities within different layers, Fig. \ref{fcontrib}. Hence, those atoms are expected to participate in change transport upon degenerate doping.

Here we stress the importance of the five-fold coordination in the formation of the hybrid $s$-like conduction band in all considered RAMO$_4$ except those containing Sc. Because the Sc coordination is the same in ScAMO$_4$ and Sc$_2$O$_3$, i.e., octahedral, the Sc $d$-states remain below its $s$-states in all the oxides. As a result, the interaction between the Sc and other cations in a Sc-containing multicomponent oxide is very weak, and the bottom of the conduction band is formed by the states of the basis oxides with smaller band gap, i.e., Sc $d$-states in ScAlMgO$_4$ or the $s$-states of A and M atoms in ScGaZnO$_4$, ScGaMgO$_4$, or ScAlZnO$_4$. This leads to a clear separation of the particular layers (Sc-O layers in the former case and AMO$_{2.5}$ layers in the latter cases) into potentially conducting and non-conducting, Fig. \ref{fcontrib}.

\begin{table*}
\centering
\caption{Electron effective masses m, in $m_e$, calculated within LDA and sX-LDA along the specified crystallographic directions in RAMO$_4$ compounds. The components of the electron effective-mass tensor, $m_{a,b}$, $m_z$, and weighted $m_{a,b}^w$, $m_z^w$, calculated for both the ground state and hypothetical phases using the effective masses of the corresponding single-cation oxides from Table \ref{tsingleband}.}
\label{tmass}
\begin{center}
\begin{tabular}{c|ccc|ccc|cccc|cccccccc} \hline \hline
    & \multicolumn{3}{c|}{LDA} & \multicolumn{11}{c}{sX-LDA}   \\  \hline
\multicolumn{1}{c|}{} & \multicolumn{3}{c|}{Calculated} & \multicolumn{3}{c|}{Calculated} &  \multicolumn{8}{c}{Predicted} \\  \hline
 RAMO$_4$ & m$_{[100]}$ & m$_{[010]}$ & m$_{[001]}$ & m$_{[100]}$ & m$_{[010]}$ & m$_{[001]}$  & m$_{ab}^g$ & m$_z^g$ & (m$_{ab}^g)^w$ & (m$_z^g)^w$ &  m$_{ab}^h$ & m$_z^h$ & (m$_{ab}^h)^w$ & (m$_z^h)^w$ \\ \hline

InAlZnO$_4$ & 0.25 & 0.25 & 0.26 & 0.39 &       0.38 & 0.38&    0.35&   0.36    &0.34   &0.34   &0.37   &0.39   &0.35&  0.37 \\
InAlCaO$_4$ & 0.35 & 0.36 & 0.34 & 0.49 & 0.50 & 0.46&  0.37&   0.38&   0.34&   0.35&   0.41&   0.44&   0.36&   0.40  \\
InAlMgO$_4$ & 0.31 & 0.30 & 0.30 & 0.46 & 0.47 & 0.44&  0.38&   0.40&   0.34&   0.36&   0.40&   0.44&   0.36&   0.39\\
InAlCdO$_4$ & 0.25 & 0.24 & 0.25 & 0.38 &       0.38&   0.38&   0.32&   0.32&   0.32&   0.31&   0.36&   0.37&   0.35&   0.36 \\ \hline
InGaZnO$_4$ & 0.21 & 0.20 & 0.21 & 0.34 & 0.34  &0.34&  0.32&   0.32&   0.32&   0.32&   0.35&   0.36&   0.34    &0.35 \\
InGaCaO$_4$ & 0.30 & 0.30 & 0.30 & 0.43 & 0.44& 0.42&   0.34&   0.35&   0.33&   0.33&   0.39&   0.41&   0.37&   0.39  \\
InGaMgO$_4$ & 0.26 & 0.26 & 0.27 & 0.41 & 0.41  &0.40&  0.35    &0.36&  0.34    &0.35& 0.39    &0.41   &0.37   &0.39 \\
InGaCdO$_4$ & 0.18 & 0.17 & 0.19 & 0.33 &       0.34&   0.33&   0.28    &0.28&  0.28&   0.28    &0.33   &0.34   &0.34   &0.34 \\ \hline
ScGaZnO$_4$ & 0.32 & 0.33 & 0.33 & 0.44 & 0.45  &0.43   &0.45   &0.63&  0.39&   0.49&   0.51&   0.66&   0.45&   0.53  \\ 
ScAlZnO$_4$ & 0.37 & 0.42 & 0.40 & 0.48 & 0.51  &0.48   &0.51&  0.66&   0.48&   0.59&   0.56&   0.69&   0.52&   0.61\\
ScGaMgO$_4$ & 0.39 & 0.40 & 0.39 & 0.53 & 0.54  &0.52&  0.51&   0.66&   0.48&   0.59&   0.59&   0.71&   0.56&   0.65\\
ScAlMgO$_4$ & 0.45 & 0.47 & 0.46 & 0.78 & 0.69  &0.64   &0.57   &0.70&  0.90    &1.04   &0.64&  0.74&   0.95&   1.06 \\ \hline\hline
\end{tabular}
\end{center}
\end{table*}

\subsubsection{D. Electron effective mass in RAMO$_4$}

The electron effective masses calculated along the [100], [010], and [001] crystallographic directions  in the multicomponent oxides RAMO$_4$ are given in Table \ref{tmass}. LDA underestimates the effective mass values which are in the range of 0.2-0.5 m$_e$, and the sX-LDA gives larger values -- as expected from larger band gaps, Table \ref{tband}. Within the sX-LDA, the smallest electron effective mass, 0.33 m$_e$, is found in InGaCdO$_4$, and the largest, 0.78 m$_e$, are in ScAlMgO$_4$. The trend in the effective mass values of RAMO$_4$ compounds follows the one in the calculated band gaps, c.f., Table \ref{tband}. Significantly, we find that both LDA and sX-LDA yield isotropic electron effective masses, i.e., the $m$ values along and across the structural layers are nearly identical in every RAMO$_4$ compound except for ScAlMgO$_4$.
This is in agreement with the hybrid nature of the conduction band and the similar contributions from the R-O and A-M-O layers to the electron density, as discussed in sections V.B and V.C. 

In section V.A, we demonstrated that the band gap in RAMO$_4$ compounds can be predicted via averaging over the values obtained for the single-cation oxide constituents with corresponding local atomic structure. Here, we perform similar analysis for the electron effective masses. The results are given in Table \ref{tmass}, where the LDA and sX-LDA values calculated for RAMO$_4$ compounds are given along with those obtained via averaging over the masses of the bases sinlge-cation oxides. The $ab$ and $z$ components of the average effective-mass tensors are found according to \cite{my2epl}.
We find that:

1) Equal-weight or weighted averaging over the electron effective masses of the single-cation oxides in their ground state structures, c.f., m$^g$ and (m$^g$)$^w$, underestimates the calculated mass values. 

2) Averaging over the effective mass values of single-cation oxides in hypothetical phases with five-fold coordination gives better agreement with the calculated values. This may appear to be counterintuitive: since the band gap in hypothetical oxides is smaller compared to that calculated for the oxides in the ground state phases, Table \ref{tsingleband}, one may expect a smaller electron effective mass, and hence, a worse agreement between the calculated and predicted masses than in the case (1) above. However, according to the {\bf k$\cdot$p} theory, the electron effective mass depends not only on the band gap value, but also on the orbital overlap of the neighboring atoms:

\begin{equation}
\frac{m_e}{m_{ii}^{(c)}}=1+\frac{2}{m_e}\sum_{v \neq c} \frac{|\langle u^{(c)}|\hat{p}_i|u^{(v)}\rangle|^2}{E^{(c)}-E^{(v)}},
\label{kp-eq}
\end{equation}
where $\hat{\bf p}$ is the momentum operator, $|u^{(l)}\rangle$ is the Bloch wave function of the $l$'s band at the $\Gamma$ point
(wave vector {\bf k}=0) and $E^{(l)}$ is its energy. Band labels $v$ and $c$ represent the valence and conduction bands, respectively.
The smallest denominator corresponds to $E^{(c)} - E^{(v)} \approx E_g$, and thus, the smaller the band gap, the smaller the electron effective mass. The numerator represents the overlap between the orbitals in the valence band (oxygen $p$ states) and in the conduction band (metal states).   
Because the overlap is greater in the higher-symmetry  phases (with octahedral coordination for CaO, CdO, MgO, tetrahedral for ZnO, etc), the effective mass is smaller in the ground state phases as compared to the hypothetical structures.

3) With the exception for Sc-containing compounds, the equal-weight average provides a better match between the predicted and calculated mass values than the weighted average. For the latter, the respective weights are obtained based on the contributions to the charge density in an energy range at the bottom of the conduction band (see section V.B and Fig. \ref{fcontrib}). The energy range corresponds to a Fermi level shift of 0.7 eV - 1.5 eV (see section V.B.) However, it appears to be insufficient, and the states which are located deep in the conduction band -- such as the states of lighter metals -- play an important role in determining the electron effective mass of multicomponent oxides. Therefore, the corresponding light metal oxide constituents should be given a greater weight. 

The above results suggest that the electron effective mass in multicomponent oxides is highly sensitive to the presence of all oxide constituents independent of their band gap value, i.e., both the semiconductor-like post-transition metal oxides and the insulator light-metal oxides play an equal role in the formation of the conduction band curvature. The local structural peculiarities, i.e., the five-fold coordination of A and M atoms, are of less significance here because of the opposite effect of a reduced orbital overlap and a smaller band gap associated with low symmetry of oxygen polyhedra on the resulting electron effective mass of multicomponent oxides.

\subsection{VI. Conclusions}
In conclusion, the structural and compositional complexity of the considered multicomponent oxides with layered structure RAMO$_4$
allowed us to address two fundamental questions: (1) how the local atomic coordination affects their electronic properties such as the band gap, the electron effective mass and the nature of the conduction band; and (2) how the optical properties and the electron conduction paths of layered multicomponent oxide hosts vary with the chemical composition.

Most significantly, we demonstrate that the unusual five-fold coordination of the A$^{3+}$ and M$^{2+}$ metal atoms stabilized in RAMO$_4$ compounds, results in the electronic band structure of the complex oxides that differs from the one expected based on the electronic properties of the single-cation oxide constituents in their lowest-energy (ground state) phases. In particular, we find:

-- The band gap in oxides shows strong dependence on the atomic coordination. High-symmetry octahedral (six-fold) coordination provides the largest overlap between the metal and oxygen orbitals giving rise to a large band gap. Other coordinations result in a smaller orbital overlap and, hence, the optical band gap is reduced. In multicomponent oxides, the band gap is determined not only by the oxide constituent with the smallest band gap but by {\it all} constituent oxides, although those of lighter metals (Al, Ca, Mg) have smaller contribution to the band gap average compared to the oxides of post-transition metals (In, Cd, Zn). The respective weights of the oxide constituents to the band gap average correlate with the calculated percent atomic contributions to the charge density in the conduction band.

-- The electron effective mass in oxides does not follow the trend expected from the variation in the band gap: we find that the structures with five-coordinated metals exhibit smaller band gaps but larger electron effective masses as compared to their six-coordinated counterparts. This finding is explained based on the {\bf k$\cdot$p} theory. In multicomponent oxides, all oxide constituents give equal contributions to the electron effective mass  
average. 

-- The unusual five-fold coordination of the A and M atoms in InAMO$_4$ compounds promotes a hybrid $s$-like conduction band making isotropic charge transport possible in this layered materials. The calculated charge density distribution shows that the light metal elements, such as Al, Ca, and Mg, contribute their $s$-states to the hybrid conduction band of complex oxides whereas the contributions from their $p$ or $d$-states which are known to cause electron localization in the corresponding single-cation oxides, are significantly reduced.  

Thus, the above results highlight the advantages of incorporating light main group metals in multicomponent oxides, which is highly attractive for lighter-weight, less-expensive, and environmentally-friendly devices. 
Further investigations of how the structural peculiarities and composition affect the formation of native defects in complex oxides are warrant in order to understand their role in carrier generation and transport in doped and/or non-stoichiometric oxides. 

\subsection{Acknowledgement}
This work was supported by the NSF grant DMR-0705626. Computational resources are provided by the NSF supported XSEDE/TeraGrid.

\bibliography{mybib}

\begin{thebibliography}{30}
\expandafter\ifx\csname natexlab\endcsname\relax\def\natexlab#1{#1}\fi
\expandafter\ifx\csname bibnamefont\endcsname\relax
  \def\bibnamefont#1{#1}\fi
\expandafter\ifx\csname bibfnamefont\endcsname\relax
  \def\bibfnamefont#1{#1}\fi
\expandafter\ifx\csname citenamefont\endcsname\relax
  \def\citenamefont#1{#1}\fi
\expandafter\ifx\csname url\endcsname\relax
  \def\url#1{\texttt{#1}}\fi
\expandafter\ifx\csname urlprefix\endcsname\relax\def\urlprefix{URL }\fi
\providecommand{\bibinfo}[2]{#2}
\providecommand{\eprint}[2][]{\url{#2}}

\bibitem[{\citenamefont{Chopra et~al.}(1983)\citenamefont{Chopra, Major, and
  Pandya}}]{Chopra}
\bibinfo{author}{\bibfnamefont{K.~L.} \bibnamefont{Chopra}},
  \bibinfo{author}{\bibfnamefont{S.}~\bibnamefont{Major}}, \bibnamefont{and}
  \bibinfo{author}{\bibfnamefont{D.~K.} \bibnamefont{Pandya}},
  \bibinfo{journal}{Thin Solid Films} \textbf{\bibinfo{volume}{102}},
  \bibinfo{pages}{1} (\bibinfo{year}{1983}).

\bibitem[{\citenamefont{Thomas}(1997)}]{Thomas}
\bibinfo{author}{\bibfnamefont{G.}~\bibnamefont{Thomas}},
  \bibinfo{journal}{Nature} \textbf{\bibinfo{volume}{389}}, \bibinfo{pages}{907
  } (\bibinfo{year}{1997}).

\bibitem[{\citenamefont{Ginley and (Editors)}(2000)}]{MRS}
\bibinfo{author}{\bibfnamefont{D.~S.} \bibnamefont{Ginley}} \bibnamefont{and}
  \bibinfo{author}{\bibfnamefont{C.~B.} \bibnamefont{(Editors)}},
  \bibinfo{journal}{MRS Bull.} \textbf{\bibinfo{volume}{25}}
  (\bibinfo{year}{2000}).

\bibitem[{\citenamefont{Fortunato et~al.}(2007)\citenamefont{Fortunato, Ginley,
  Honoso, and Paine}}]{TCO-optoelectr}
\bibinfo{author}{\bibfnamefont{E.}~\bibnamefont{Fortunato}},
  \bibinfo{author}{\bibfnamefont{D.}~\bibnamefont{Ginley}},
  \bibinfo{author}{\bibfnamefont{H.}~\bibnamefont{Honoso}}, \bibnamefont{and}
  \bibinfo{author}{\bibfnamefont{D.~C.} \bibnamefont{Paine}},
  \bibinfo{journal}{MRS Bulletin} \textbf{\bibinfo{volume}{32}},
  \bibinfo{pages}{242} (\bibinfo{year}{2007}).

\bibitem[{\citenamefont{Edwards et~al.}(2004)\citenamefont{Edwards, Porch,
  Jones, Morgan, and Perks}}]{Edwards}
\bibinfo{author}{\bibfnamefont{P.~P.} \bibnamefont{Edwards}},
  \bibinfo{author}{\bibfnamefont{A.}~\bibnamefont{Porch}},
  \bibinfo{author}{\bibfnamefont{M.~O.} \bibnamefont{Jones}},
  \bibinfo{author}{\bibfnamefont{D.~V.} \bibnamefont{Morgan}},
  \bibnamefont{and} \bibinfo{author}{\bibfnamefont{R.~M.} \bibnamefont{Perks}},
  \bibinfo{journal}{Dalton Trans} \textbf{\bibinfo{volume}{19}},
  \bibinfo{pages}{2995} (\bibinfo{year}{2004}).

\bibitem[{\citenamefont{Facchetti and Marks}(2010)}]{FMbook}
\bibinfo{editor}{\bibfnamefont{A.}~\bibnamefont{Facchetti}} \bibnamefont{and}
  \bibinfo{editor}{\bibfnamefont{T.}~\bibnamefont{Marks}}, eds.,
  \emph{\bibinfo{title}{Transparent Electronics: From Synthesis to
  Applications}} (\bibinfo{publisher}{John Wiley \& Sons},
  \bibinfo{year}{2010}).

\bibitem[{\citenamefont{Ginley et~al.}(2011)\citenamefont{Ginley, Hosono, and
  Paine}}]{TCOhandbook}
\bibinfo{editor}{\bibfnamefont{D.~S.} \bibnamefont{Ginley}},
  \bibinfo{editor}{\bibfnamefont{H.}~\bibnamefont{Hosono}}, \bibnamefont{and}
  \bibinfo{editor}{\bibfnamefont{D.~C.} \bibnamefont{Paine}}, eds.,
  \emph{\bibinfo{title}{Handbook of Transparent Conductors}}
  (\bibinfo{publisher}{Springer}, \bibinfo{year}{2011}).

\bibitem[{\citenamefont{Shannon et~al.}(1977)\citenamefont{Shannon, Gillson,
  and Bouchard}}]{Shannon}
\bibinfo{author}{\bibfnamefont{R.~D.} \bibnamefont{Shannon}},
  \bibinfo{author}{\bibfnamefont{J.~L.} \bibnamefont{Gillson}},
  \bibnamefont{and} \bibinfo{author}{\bibfnamefont{R.~J.}
  \bibnamefont{Bouchard}}, \bibinfo{journal}{J. Phys. Chem. Solids}
  \textbf{\bibinfo{volume}{38}}, \bibinfo{pages}{877} (\bibinfo{year}{1977}).

\bibitem[{\citenamefont{Dawar and Joshi}(1984)}]{Dawar}
\bibinfo{author}{\bibfnamefont{A.~L.} \bibnamefont{Dawar}} \bibnamefont{and}
  \bibinfo{author}{\bibfnamefont{J.~C.} \bibnamefont{Joshi}},
  \bibinfo{journal}{J. Mater. Sci} \textbf{\bibinfo{volume}{19}},
  \bibinfo{pages}{1} (\bibinfo{year}{1984}).

\bibitem[{\citenamefont{Un'no et~al.}(1993)\citenamefont{Un'no, Hikuma, Omata,
  Ueda, Hashimoto, and Kawazoe}}]{Unno}
\bibinfo{author}{\bibfnamefont{H.}~\bibnamefont{Un'no}},
  \bibinfo{author}{\bibfnamefont{N.}~\bibnamefont{Hikuma}},
  \bibinfo{author}{\bibfnamefont{T.}~\bibnamefont{Omata}},
  \bibinfo{author}{\bibfnamefont{N.}~\bibnamefont{Ueda}},
  \bibinfo{author}{\bibfnamefont{T.}~\bibnamefont{Hashimoto}},
  \bibnamefont{and} \bibinfo{author}{\bibfnamefont{H.}~\bibnamefont{Kawazoe}},
  \bibinfo{journal}{Jpn. J. Appl. Phys} \textbf{\bibinfo{volume}{32}},
  \bibinfo{pages}{L1260} (\bibinfo{year}{1993}).

\bibitem[{\citenamefont{Phillips et~al.}(1994)\citenamefont{Phillips, Kwo, and
  Thomas}}]{Phillips}
\bibinfo{author}{\bibfnamefont{J.~M.} \bibnamefont{Phillips}},
  \bibinfo{author}{\bibfnamefont{J.}~\bibnamefont{Kwo}}, \bibnamefont{and}
  \bibinfo{author}{\bibfnamefont{G.~A.} \bibnamefont{Thomas}},
  \bibinfo{journal}{Appl. Phys. Lett} \textbf{\bibinfo{volume}{65}},
  \bibinfo{pages}{115} (\bibinfo{year}{1994}).

\bibitem[{\citenamefont{Kawazoe and Ueda}(1999)}]{spinel-review}
\bibinfo{author}{\bibfnamefont{H.}~\bibnamefont{Kawazoe}} \bibnamefont{and}
  \bibinfo{author}{\bibfnamefont{K.}~\bibnamefont{Ueda}}, \bibinfo{journal}{J.
  Amer. Ceram. Soc} \textbf{\bibinfo{volume}{82}}, \bibinfo{pages}{3330}
  (\bibinfo{year}{1999}).

\bibitem[{\citenamefont{Freeman et~al.}(2000)\citenamefont{Freeman,
  Poeppelmeier, Mason, Chang, and Marks}}]{Freeman}
\bibinfo{author}{\bibfnamefont{A.~J.} \bibnamefont{Freeman}},
  \bibinfo{author}{\bibfnamefont{K.~R.} \bibnamefont{Poeppelmeier}},
  \bibinfo{author}{\bibfnamefont{T.~O.} \bibnamefont{Mason}},
  \bibinfo{author}{\bibfnamefont{R.~P.} \bibnamefont{Chang}}, \bibnamefont{and}
  \bibinfo{author}{\bibfnamefont{T.~J.} \bibnamefont{Marks}},
  \bibinfo{journal}{MRS Bull} \textbf{\bibinfo{volume}{25}},
  \bibinfo{pages}{45} (\bibinfo{year}{2000}).

\bibitem[{\citenamefont{Ingram et~al.}(2004)\citenamefont{Ingram, Gonzalez,
  Kammler, Bertoni, and Mason}}]{Mason-review}
\bibinfo{author}{\bibfnamefont{B.~J.} \bibnamefont{Ingram}},
  \bibinfo{author}{\bibfnamefont{G.~B.} \bibnamefont{Gonzalez}},
  \bibinfo{author}{\bibfnamefont{D.~R.} \bibnamefont{Kammler}},
  \bibinfo{author}{\bibfnamefont{M.~I.} \bibnamefont{Bertoni}},
  \bibnamefont{and} \bibinfo{author}{\bibfnamefont{T.~O.} \bibnamefont{Mason}},
  \bibinfo{journal}{J. Electroceram.} \textbf{\bibinfo{volume}{13}},
  \bibinfo{pages}{167} (\bibinfo{year}{2004}).

\bibitem[{\citenamefont{Walsh et~al.}(2011)\citenamefont{Walsh, Silva, and
  Wei}}]{Walsh}
\bibinfo{author}{\bibfnamefont{A.}~\bibnamefont{Walsh}},
  \bibinfo{author}{\bibfnamefont{J.~D.} \bibnamefont{Silva}}, \bibnamefont{and}
  \bibinfo{author}{\bibfnamefont{S.-H.} \bibnamefont{Wei}},
  \bibinfo{journal}{J. Phys.: Condens. Matter} \textbf{\bibinfo{volume}{23}},
  \bibinfo{pages}{334210} (\bibinfo{year}{2011}).

\bibitem[{\citenamefont{Medvedeva}(2010)}]{JMbook}
\bibinfo{author}{\bibfnamefont{J.~E.} \bibnamefont{Medvedeva}},
  \emph{\bibinfo{title}{in Transparent Electronics: From Synthesis to
  Applications}} (\bibinfo{publisher}{John Wiley \& Sons},
  \bibinfo{year}{2010}), pp. \bibinfo{pages}{1--29}.

\bibitem[{\citenamefont{Medvedeva and Hettiarachchi}(2010)}]{JMprb}
\bibinfo{author}{\bibfnamefont{J.~E.} \bibnamefont{Medvedeva}}
  \bibnamefont{and} \bibinfo{author}{\bibfnamefont{C.~L.}
  \bibnamefont{Hettiarachchi}}, \bibinfo{journal}{Physical Review B}
  \textbf{\bibinfo{volume}{81}}, \bibinfo{pages}{125116}
  (\bibinfo{year}{2010}).

\bibitem[{\citenamefont{Kato et~al.}(1975)\citenamefont{Kato, Kawada, Kimizuka,
  and Katsura}}]{str}
\bibinfo{author}{\bibfnamefont{V.~K.} \bibnamefont{Kato}},
  \bibinfo{author}{\bibfnamefont{I.}~\bibnamefont{Kawada}},
  \bibinfo{author}{\bibfnamefont{N.}~\bibnamefont{Kimizuka}}, \bibnamefont{and}
  \bibinfo{author}{\bibfnamefont{T.}~\bibnamefont{Katsura}},
  \bibinfo{journal}{Z. Krist} \textbf{\bibinfo{volume}{141}},
  \bibinfo{pages}{314} (\bibinfo{year}{1975}).

\bibitem[{\citenamefont{Kimizuka and Mohri}(1985)}]{str-all}
\bibinfo{author}{\bibfnamefont{N.}~\bibnamefont{Kimizuka}} \bibnamefont{and}
  \bibinfo{author}{\bibfnamefont{T.}~\bibnamefont{Mohri}}, \bibinfo{journal}{J.
  Solid State Chem} \textbf{\bibinfo{volume}{60}}, \bibinfo{pages}{382}
  (\bibinfo{year}{1985}).

\bibitem[{\citenamefont{Bylander and Kleinman}(1990)}]{sxlda}
\bibinfo{author}{\bibfnamefont{D.~M.} \bibnamefont{Bylander}} \bibnamefont{and}
  \bibinfo{author}{\bibfnamefont{L.}~\bibnamefont{Kleinman}},
  \bibinfo{journal}{Phys. Rev. B} \textbf{\bibinfo{volume}{41}},
  \bibinfo{pages}{7868} (\bibinfo{year}{1990}).

\bibitem[{\citenamefont{Wimmer et~al.}(1981)\citenamefont{Wimmer, Krakauer,
  Weinert, and Freeman}}]{FLAPW}
\bibinfo{author}{\bibfnamefont{E.}~\bibnamefont{Wimmer}},
  \bibinfo{author}{\bibfnamefont{H.}~\bibnamefont{Krakauer}},
  \bibinfo{author}{\bibfnamefont{M.}~\bibnamefont{Weinert}}, \bibnamefont{and}
  \bibinfo{author}{\bibfnamefont{A.~J.} \bibnamefont{Freeman}},
  \bibinfo{journal}{Phys. Rev. B} \textbf{\bibinfo{volume}{24}},
  \bibinfo{pages}{864} (\bibinfo{year}{1981}).

\bibitem[{\citenamefont{Weinert et~al.}(1982)\citenamefont{Weinert, Wimmer, and
  Freeman}}]{FLAPW1}
\bibinfo{author}{\bibfnamefont{M.}~\bibnamefont{Weinert}},
  \bibinfo{author}{\bibfnamefont{E.}~\bibnamefont{Wimmer}}, \bibnamefont{and}
  \bibinfo{author}{\bibfnamefont{A.~J.} \bibnamefont{Freeman}},
  \bibinfo{journal}{Phys. Rev. B} \textbf{\bibinfo{volume}{26}},
  \bibinfo{pages}{4571} (\bibinfo{year}{1982}).

\bibitem[{\citenamefont{Kimizuka and Mohri}(1989)}]{kimizukaRAM}
\bibinfo{author}{\bibfnamefont{N.}~\bibnamefont{Kimizuka}} \bibnamefont{and}
  \bibinfo{author}{\bibfnamefont{T.}~\bibnamefont{Mohri}}, \bibinfo{journal}{J.
  Solid State Chem} \textbf{\bibinfo{volume}{78}}, \bibinfo{pages}{98}
  (\bibinfo{year}{1989}).

\bibitem[{\citenamefont{Seidl et~al.}(1996)\citenamefont{Seidl, G{\"o}rling,
  Vogl, Majewski, and Levy}}]{sxLDA1}
\bibinfo{author}{\bibfnamefont{A.}~\bibnamefont{Seidl}},
  \bibinfo{author}{\bibfnamefont{A.}~\bibnamefont{G{\"o}rling}},
  \bibinfo{author}{\bibfnamefont{P.}~\bibnamefont{Vogl}},
  \bibinfo{author}{\bibfnamefont{J.~A.} \bibnamefont{Majewski}},
  \bibnamefont{and} \bibinfo{author}{\bibfnamefont{M.}~\bibnamefont{Levy}},
  \bibinfo{journal}{Phys. Rev. B} \textbf{\bibinfo{volume}{53}},
  \bibinfo{pages}{3764} (\bibinfo{year}{1996}).

\bibitem[{\citenamefont{Asahi et~al.}(1999)\citenamefont{Asahi, Mannstadt, and
  Freeman}}]{sx-Asahi}
\bibinfo{author}{\bibfnamefont{R.}~\bibnamefont{Asahi}},
  \bibinfo{author}{\bibfnamefont{W.}~\bibnamefont{Mannstadt}},
  \bibnamefont{and} \bibinfo{author}{\bibfnamefont{A.~J.}
  \bibnamefont{Freeman}}, \bibinfo{journal}{Phys. Rev. B}
  \textbf{\bibinfo{volume}{59}}, \bibinfo{pages}{7486} (\bibinfo{year}{1999}).

\bibitem[{\citenamefont{Geller et~al.}(2001)\citenamefont{Geller, Wolf,
  Picozzi, Continenza, Asahi, Mannstadt, Freeman, and Wimmer}}]{sx-paper}
\bibinfo{author}{\bibfnamefont{C.~B.} \bibnamefont{Geller}},
  \bibinfo{author}{\bibfnamefont{W.}~\bibnamefont{Wolf}},
  \bibinfo{author}{\bibfnamefont{S.}~\bibnamefont{Picozzi}},
  \bibinfo{author}{\bibfnamefont{A.}~\bibnamefont{Continenza}},
  \bibinfo{author}{\bibfnamefont{R.}~\bibnamefont{Asahi}},
  \bibinfo{author}{\bibfnamefont{W.}~\bibnamefont{Mannstadt}},
  \bibinfo{author}{\bibfnamefont{A.~J.} \bibnamefont{Freeman}},
  \bibnamefont{and} \bibinfo{author}{\bibfnamefont{E.}~\bibnamefont{Wimmer}},
  \bibinfo{journal}{Appl. Phys. Lett.} \textbf{\bibinfo{volume}{79}},
  \bibinfo{pages}{368} (\bibinfo{year}{2001}).

\bibitem[{\citenamefont{Kim et~al.}(2002)\citenamefont{Kim, Asahi, and
  Freeman}}]{sx-Kim}
\bibinfo{author}{\bibfnamefont{M.~Y.} \bibnamefont{Kim}},
  \bibinfo{author}{\bibfnamefont{R.}~\bibnamefont{Asahi}}, \bibnamefont{and}
  \bibinfo{author}{\bibfnamefont{A.~J.} \bibnamefont{Freeman}},
  \bibinfo{journal}{J. Comput.-Aided Mater. Des.} \textbf{\bibinfo{volume}{9}},
  \bibinfo{pages}{173} (\bibinfo{year}{2002}).

\bibitem[{\citenamefont{Kimizuka et~al.}(1988)\citenamefont{Kimizuka, Mohri,
  and Matsui}}]{KimizukaRAM2}
\bibinfo{author}{\bibfnamefont{N.}~\bibnamefont{Kimizuka}},
  \bibinfo{author}{\bibfnamefont{T.}~\bibnamefont{Mohri}}, \bibnamefont{and}
  \bibinfo{author}{\bibfnamefont{Y.}~\bibnamefont{Matsui}},
  \bibinfo{journal}{J. Solid State Chem.} \textbf{\bibinfo{volume}{74}},
  \bibinfo{pages}{98} (\bibinfo{year}{1988}).

\bibitem[{\citenamefont{Li et~al.}(1997)\citenamefont{Li, Bando, Nakamura, and
  Kimizuka}}]{Li}
\bibinfo{author}{\bibfnamefont{C.}~\bibnamefont{Li}},
  \bibinfo{author}{\bibfnamefont{Y.}~\bibnamefont{Bando}},
  \bibinfo{author}{\bibfnamefont{M.}~\bibnamefont{Nakamura}}, \bibnamefont{and}
  \bibinfo{author}{\bibfnamefont{M.}~\bibnamefont{Kimizuka}},
  \bibinfo{journal}{J. Electron Microsc} \textbf{\bibinfo{volume}{46}},
  \bibinfo{pages}{119} (\bibinfo{year}{1997}).

\bibitem[{\citenamefont{Medvedeva}(2007)}]{my2epl}
\bibinfo{author}{\bibfnamefont{J.~E.} \bibnamefont{Medvedeva}},
  \bibinfo{journal}{Europhys. Lett.} \textbf{\bibinfo{volume}{78}},
  \bibinfo{pages}{57004} (\bibinfo{year}{2007}).

\end{thebibliography}
\bibliographystyle{apsrev}

\end{document}